\documentclass[conference,10pt]{IEEEtran}

\ifCLASSINFOpdf
\else
\fi

\usepackage{textcomp,amssymb}
\usepackage{hyperref}
\usepackage{setspace}
\usepackage{latexsym,fancyhdr,url}
\usepackage{enumerate}
\usepackage[linesnumbered,ruled]{algorithm2e}
\usepackage{algpseudocode}
\usepackage{graphics}
\usepackage{xparse} %
\usepackage{xspace}
\usepackage{multirow}
\usepackage{csvsimple}
\usepackage{balance}
\usepackage{booktabs}
\usepackage{colortbl}
\usepackage{fancyhdr}
\usepackage{makecell}
\usepackage{xcolor, eucal}
\usepackage{array}
\usepackage{subfigure}
\usepackage{graphicx}
\usepackage{caption}
\usepackage{tabularx}
\usepackage{dsfont}
\usepackage{enumitem}
\usepackage{float}
\usepackage{amsmath}
\usepackage[available,functional]{ndssbadges}

\IEEEoverridecommandlockouts

\definecolor{gray0}{gray}{0.9}

\newtheorem{theorem}{Theorem}
\newtheorem{definition}{Definition}

\usepackage[most]{tcolorbox}
\newcommand\rev[1]{{\color{black}  #1}}

\def \toolname{PrivORL\xspace}
\def \toolnametran{PrivORL-n\xspace}
\def \toolnametraj{PrivORL-j\xspace}

\begin{document}
\title{PrivORL: Differentially Private Synthetic \\ Dataset for Offline Reinforcement Learning}

\author{\IEEEauthorblockN{Chen Gong\IEEEauthorrefmark{1},
Zheng Liu\IEEEauthorrefmark{1},
Kecen Li, and
Tianhao Wang}
\IEEEauthorblockA{University of Virginia}
\thanks{\IEEEauthorrefmark{1}Equal Contributions. Zheng and Kecen work as independent researchers and remote interns at UVA.}}

\IEEEoverridecommandlockouts
\makeatletter\def\@IEEEpubidpullup{6.5\baselineskip}\makeatother
\IEEEpubid{\parbox{\columnwidth}{
    Network and Distributed System Security (NDSS) Symposium 2026 \\
    23-27 February 2026, San Diego, CA, USA \\
    ISBN ISBN 979-8-9919276-8-0 \\
    https://dx.doi.org/10.14722/ndss.2026.[23$|$24]149 \\
    www.ndss-symposium.org \\
}
\hspace{\columnsep}\makebox[\columnwidth]{}}

\maketitle

\begin{abstract}
Recently, offline reinforcement learning (RL) has become a popular RL paradigm. In offline RL, data providers share pre-collected datasets---either as individual transitions or sequences of transitions forming trajectories---to enable the training of RL models (also called agents) without direct interaction with the environments. Offline RL saves interactions with environments compared to traditional RL, and has been effective in critical areas, such as navigation tasks. Meanwhile, concerns about privacy leakage from offline RL datasets have emerged. 

To safeguard private information in offline RL datasets, we propose the first differential privacy (DP) offline dataset synthesis method, \toolname, which leverages a diffusion model and diffusion transformer to synthesize \textit{transitions and trajectories}, respectively, under DP. The synthetic dataset can then be securely released for downstream analysis and research. \toolname adopts the popular approach of pre-training a synthesizer on public datasets, and then fine-tuning on sensitive datasets using DP Stochastic Gradient Descent (DP-SGD).
Additionally, \toolname introduces curiosity-driven pre-training, which uses feedback from the curiosity module to diversify the synthetic dataset and thus can generate diverse synthetic transitions and trajectories that closely resemble the sensitive dataset.
Extensive experiments on five sensitive offline RL datasets show that our method achieves better utility and fidelity in both DP transition and trajectory synthesis compared to baselines. The replication package is available at the GitHub repository.\footnote{{https://github.com/2019ChenGong/PrivORL}}
\end{abstract}

\section{Introduction}

Recent studies highlight that privacy leakage risks are prevalent in RL systems, like using membership inference attacks (MIAs) to infer the environment information or training data~\cite{pan2019you,gomrokchi2022membership}. 
Pan et al.~\cite{pan2019you} present that attackers can use MIAs to steal map information from RL agents. 
Reinforcement Learning from Human Feedback method trains models with human evaluations~\cite{chaudhari2024rlhf}. Human feedback data, such as ratings or preference labels, can hold sensitive user information~\cite{rlhfmia}. 
Similarly, offline RL faces comparable privacy leakage challenges. Du et al.~\cite{du2023orl} propose ORL-Auditor, which infers the training trajectories of agents, an approach that can also be considered a form of MIA.

\rev{Privacy-preserving synthetic data generates artificial datasets that retain key characteristics of real data, enabling secure sharing while reducing privacy risks~\cite{hu2023sok}.} Differential Privacy (DP) dataset synthesis~\cite{yue-etal-2023-synthetic,sun2024netdpsyn} offers a theoretical guarantee for quantifying privacy leakage from real data through the use of synthetic datasets. It protects an individual's data privacy within a dataset throughout the data training. 

The offline RL dataset comprises either \textit{transitions} or \textit{trajectories}. Transitions refer to individual steps in the RL process, capturing the movement from one state to another based on an action taken by an agent, along with the associated reward. In contrast, trajectories encompass complete sequences of such transitions, providing a holistic view of an agent’s behavior over time, including all states, actions, and rewards. Both representations are fundamental to offline RL datasets, as transitions offer granular insights into decision-making, while trajectories enable analysis of long-term patterns and dependencies. 
In practice, as introduced in Section~\ref{sub:dp}, it is common for users to contribute transitions or trajectories to the offline RL datasets. Therefore, it is necessary to consider both transition-level and trajectory-level privacy protection.

\vspace{1.2mm}
\noindent \textbf{Existing Methods.} Previous works have proposed synthesizing offline RL dataset~\cite{lu2023synthetic,he2023diffusion,zhu2024trajsynmadiff}. However, these methods do not provide formal privacy guarantees (i.e., no DP) for the synthetic datasets. The original data still faces the risk of privacy leakage~\cite{2023whithmiadiffusion,carlini2023extracting}. Meanwhile, DP data synthesis methods have been increasingly developed for other modalities, e.g., images~\cite{li2023meticulously,liew2022pearl}, tabular data~\cite{zhang2021privsyn,PGM}, text data~\cite{yue-etal-2023-synthetic}, network records~\cite{sun2024netdpsyn}, and so on, but these are not tailored to offline RL datasets, which pose unique challenges due to their sequential and dynamic nature, which are introduced as follows.

\vspace{1.2mm}
\noindent \textbf{Our Proposal.} \rev{We focus on presenting the feasibility of applying DP to offline RL dataset synthesis and on leveraging existing DP primitives to address challenges unique to this setting. To this end,} we propose \toolname (Differentially \textbf{Priv}ate Dataset Synthesizer for \textbf{O}ffline \textbf{R}einforcement \textbf{L}earning), the first DP synthesizer for offline RL datasets. \toolname comprises \toolnametran and \toolnametraj, which perform \textit{transition}-level and \textit{trajectory}-level DP dataset synthesis, respectively.

We start with adapting existing DP diffusion models, as diffusion models achieve remarkable synthetic performance in complex synthesis tasks~\cite{yang2023diffusion,diffusionText1,Wu_2023_ICCV,ddpm}. However, directly applying diffusion models to DP offline RL dataset synthesis has a couple of challenges: (1) DP noise must be introduced during synthesizer training, which introduces instability into the training process~\cite{dpsgd,li2023meticulously}. (2) The success of agents trained in offline RL is highly dependent on the diversity of the dataset (we elaborate more in Section~\ref{motivation:curiosity}). 
(3) Trajectory-level DP synthesis presents unique challenges, such as high dimensionality and temporal dependencies in the dataset. We provide more discussions in Appendix~\ref{supsubsec:uniqueness}.  
Besides, how to effectively create a diverse DP synthetic dataset remains an open question~\cite{lu2023synthetic}.

We then introduce how \toolnametraj addresses the third unique challenge. Both \toolnametran and \toolnametraj leverage the same paradigm for tackling the first and second common challenges, which we discuss below. 

\begin{itemize}[leftmargin=*]
    \item \rev{First, we adopt the popular paradigm~\cite{li2023meticulously,dp-diffusion} of pre-training with public datasets and only fine-tuning on sensitive data under DP. It is important to use a public dataset during the pre-training phase to achieve fast convergence and generate reasonable synthetic datasets. }
    \vspace{1mm}
    \item To solve the second challenge, inspired by the random network distillation (RND) concept in RL and bug detection~\cite{hong2024curiositydriven,he2024curiosity}, we propose using the curiosity module. The curiosity module quantifies the `novelty' of synthetic data. However, unlike works~\cite{hong2024curiositydriven,he2024curiosity}, diffusion model training lacks a reward mechanism that integrates novelty feedback (as described in Section~\ref{subsec:diffusion_model}). We propose replacing a portion of real data with high-novelty synthetic data, encouraging synthesizers to capture underlying characteristics of high-novelty data for more diverse synthesis.
    
    In addition, we propose using the curiosity module during the pre-training phase instead of the fine-tuning phase. 
    The high-level motivation is that pre-training offers greater flexibility and tolerance for instability.
    \vspace{1mm}
    \item To address the challenges of high dimensionality, \toolnametraj manages long trajectories by splitting trajectories into fragments and using a conditional synthesizer to capture the relationship between fragments, enabling the synthesis of fragments that can be seamlessly stitched into trajectories. To capture long-range temporal dependencies in trajectory-level DP, \toolnametraj extends \toolnametran by integrating a transformer~\cite{diffusiontransformer} into its diffusion model used in \toolnametran, modeling complex temporal relationships in trajectories. 
\end{itemize}

We elaborate on differences in synthesizing transition-level versus trajectory-level datasets in Section~\ref{sec:diffrence}.

\vspace{1.2mm}
\noindent \textbf{Evaluations}. We conduct experiments on five types of sensitive datasets, three {\tt Maze2D}, one {\tt Kitchen}, and one {\tt Mujoco} datasets, to present the effectiveness of \toolnametran and \toolnametraj mainly from the following two perspectives. 

\noindent \textbf{\textit{Utility}}: For DP transition synthesis under privacy budgets $\epsilon = \{1,10\}$, in {\tt Maze2D} domain, agents trained on the synthetic dataset generated by \toolnametran, using three prominent offline RL algorithms, achieve an average normalized return of 51.9 and 69.3 across studied sensitive datasets. This performance exceeds the baseline returns of 11.7, 3.4, 2.0, and 3.2 at $\epsilon=1$, and 18.9, 3.9, 5.1, and 7.2 at $\epsilon=10$, achieved by PGM~\cite{PGM}, PrivSyn~\cite{zhang2021privsyn}, PATE-GAN~\cite{PATE-GAN}, and PrePATE-GAN. In ablation studies, without the curiosity module and pre-training, \toolnametran reduces to the pre-training DP diffusion~\cite{dp-diffusion} and DPDM~\cite{dpdm} proposed in DP image synthesis. Removing pre-training and the curiosity module from \toolnametran reduced the performance, with average normalized returns of trained agents dropping by 25.9\% and 16.3\% in {\tt Maze2D-umaze} dataset under $\epsilon=10$. For DP trajectory synthesis at
$\epsilon = \{1,10\}$, \toolnametraj achieves average 15.0 and 10.4 higher returns than DP-Transformer~\cite{dptransformer}, in the {\tt Maze2D} domain.

\noindent \textbf{\textit{Fidelity}}: In transition synthesis, \toolnametran outperforms the baseline models in both marginal and correlation statistics~\cite{fieller1957tests}. Regarding trajectory synthesis, under $\epsilon=10$, \toolnametraj achieves an average of 15.9\% improvements of TrajScores compared to DP-Transformer~\cite{dptransformer} across studied datasets. The fidelity metrics are introduced in Section~\ref{subsec:metrics}. 

\rev{Besides, Section~\ref{subsec:defending} shows that synthetic transitions generated by \toolnametran exhibit strong resistance to an advanced white-box membership inference attack (MIA) method~\cite{2023whithmiadiffusion}, outperforming synthesis without DP protection.}

\vspace{1mm}
\noindent \textbf{Contributions.} In summary, our contributions are:
\begin{itemize}[leftmargin=*]
    \item We introduce \toolname, a DP offline RL dataset synthesizer for both transition and trajectory.  It facilitates the sharing of datasets and promotes privacy protection.
    \item We introduce the curiosity module to synthesizer training, which enables \toolname to generate synthetic data that is both more diverse and of higher utility.
    \item We conduct a comprehensive evaluation of \toolname. The results show that \toolname outperforms the baselines by synthesizing transitions or trajectories of higher utility and fidelity across datasets from five tasks in three domains. 
\end{itemize}

\section{Backgrounds}
\subsection{Reinforcement Learning}
\label{subsec:rl}

Reinforcement Learning (RL) aims to train a policy, denoted as $\pi$ (also referred to as an agent~\cite{offline_survey}), to solve sequential decision-making tasks. The sequential decision-making tasks can be formulated as a five-tuple Markov Decision Processes (MDP)~\cite{sutton2018reinforcement}, $\langle \mathcal{S}, \mathcal{A}, \mathcal{R}, \mathcal{P}, \gamma \rangle$, where $\mathcal{S}$ and $\mathcal{A}$ represent the state space and action space, $\mathcal{R}: \mathcal{S} \times \mathcal{A} \to \mathbb{R}$ indicates the reward received, and the transition function $\mathcal{P}: \mathcal{S} \times \mathcal{A} \to \mathcal{S}$. $\gamma \in (0,1)$ is the discount factor when computing accumulated rewards. At each timestep $t$, the agent $\pi$ takes an action $a_t$ at the state $s_t$. Then, the agent obtains a reward $r_t \sim \mathcal{R}(s_t,a_t)$ from the reward function, and the MDP transitions to the next state $s_{t+1}$. RL algorithms aim to train the agent $\pi^{\ast}$ to maximize the expected cumulative reward for the specific task, $\pi^{\ast} = \mathop{\arg\max}_{\pi} \mathbb{E}_{a_t\sim \pi} \left[ \sum_{t=0}^\infty \gamma^t \mathcal{R}(s_t, a_t)\right]$~\cite{he2023diffusion}. The agent learns through a trial-and-error paradigm by interacting with the environment.

\vspace{1.2mm}
\noindent \textbf{Offline Reinforcement Learning.} In certain scenarios, such as healthcare~\cite{RL4Treatment, offline_rl_medicial}, online RL is unsuitable. It is impractical to conduct trial-and-error experiments on patients. During training, offline RL relies on learning from a static dataset $D$, which is collected from various users and can be composed in two ways: one where a single user contributes an {\it entire trajectory} or one where a single user contributes a {\it single transition}~\cite{offlinetreatment,rlfortreatment}. Specifically, a trajectory-based dataset can be represented as,
$$D = \left\{ \left. \left( s_0^i, a_0^i, r_0^i, s_1^i, \cdots, s_{|\tau|}^i, a_{|\tau|}^i, r_{|\tau|}^i \right) \right| i = 1, \dots, N \right\}.$$ 
Alternatively, a transition dataset consists of a series of four-tuples, which is defined as follows,
$$D = \{(s_t, a_t, r_t, s_{t+1})\}_{t=1}^N.$$
where $N$ is the offline dataset size. The dataset $D$ in offline RL is typically collected by various strategies~\cite{levine2020offline}. The rationality of these two datasets can be illustrated with examples. A single user contributing a transition, such as a doctor observing a diabetic patient’s blood sugar level, adjusting the insulin dosage, and recording the resulting change and patient response, is practical for analyzing the immediate impact of isolated decisions, offering a focused way to optimize specific actions~\cite{rlfortreatment}. Conversely, a user contributing an entire trajectory, like a driver documenting a full trip from home to work—including every turn, acceleration, and stop along with feedback like fuel efficiency or safety—is reasonable for long-term outcomes, capturing how actions interplay over time~\cite{cql}. These two offline RL datasets reflect realistic data collection scenarios and serve distinct purposes~\cite{levine2020offline}. Therefore, the transition and trajectory level DP protection are both necessary.  \rev{We discuss the uniqueness of offline RL trajectories in Appendix~\ref{supsubsec:uniqueness}.}

\subsection{Differential Privacy}
\label{sub:dp}

Differential privacy (DP)~\cite{dp} protects an individual's data privacy within a dataset throughout the data processing phase. We present the concept of DP as follows.

\begin{definition}[Differential Privacy~\cite{dp}]
     A randomized mechanism $\mathcal{Q}$ satisfies ($\varepsilon, \delta$)-differential privacy (DP) ($\varepsilon > 0$ and $\delta > 0$), if and only if, for any two neighboring datasets $D$, $D'$ and any $\mathcal{O}$, the following is satisfied,
\begin{equation}\label{eq:dp}
    \Pr[\mathcal{Q}(D) \in \mathcal{O}] \leq e^\varepsilon \Pr[\mathcal{Q}(D') \in \mathcal{O}] + \delta.
\end{equation}
where $\mathcal{O}$ denotes the set of all possible outputs from $\mathcal{Q}$.
\end{definition}
The privacy budget parameters, $\varepsilon$ and $\delta$, are both non-negative and measure the privacy loss in the data. A lower $\varepsilon$ value indicates better privacy protections, while a smaller $\delta$ reduces the likelihood that the privacy guarantees provided by $\varepsilon$ will be compromised. Two datasets ${D, D'}$ differing by a single \textit{transition} or \textit{trajectory} are considered neighbors. This can be interpreted as transition-level or trajectory-level DP.

\vspace{2pt}

\vspace{1.5mm}

\noindent \textbf{Differentially Private Stochastic Gradient Descent.} In machine learning, the most popular way to train the model to satisfy DP is DP-SGD~\cite{dpsgd}. This method modifies the standard SGD, which computes gradients based on Poisson sampling of mini-batches, clips the original gradients of the model's parameters, and adds random Gaussian noise to the clipped gradient throughout training. We first denote the `Clip' operation as,
$
 \text{Clip}_{C}\left(\mathbf{g}\right) = \text{min}\left\{1,\frac{C}{||\mathbf{g}||_2}\right\}\mathbf{g},
$
where $\mathbf{g}$ is the original gradient and $C$ is a hyper-parameter. The `Clip' operation scales the norm of the gradient down to less than $C$, and the model parameters $\theta$ are updated via,
\begin{align}
\small
\eta \mathbb{E}_{x_i \in x} \left[ \text{Clip}_{C}\left(\nabla {\mathcal{L}}(\theta, x_i)\right) + C \mathcal{N}(0, \sigma^2 \mathbb{I}) \right],
\label{eq:dpsgd}
\end{align}
where $\eta$ is the learning rate; $L$ is the loss function; $x$ indicates a mini-batch Poisson sampled with a sample rate $q$ (and we denote the size of it by $|x|$); $\nabla {\mathcal{L}}(\theta, x_i)$ represents the gradient for $x_i$. \( \mathcal{N}(0, \sigma^2 \mathbb{I}) \) is the Gaussian noise with the variance $\sigma$, and $B$ is the batch size. 
When there are many iterations of DP-SGD in our method, composition theorems (privacy accounting) can be used to derive the final values of $(\epsilon,\delta)$. We defer those derivations to Appendix~\ref{app:supp_dp}.

\subsection{Diffusion Model}
\label{subsec:diffusion_model}

Diffusion models~\cite{ddpm} are a class of likelihood-based generative models that present excellent generative capabilities in various fields~\cite{he2023diffusion,diffusiontransformer,dpimagebench}. \rev{Prior works~\cite{hu2023sok,li2023meticulously} show that diffusion models provide more stable training dynamics and consistently outperform GAN- and VAE-based approaches in terms of synthesis quality. Most state-of-the-art DP dataset synthesis methods adopt diffusion models as their primary synthesizers~\cite{hu2023sok,dpimagebench}. Following this trend, we use diffusion models as the DP synthesizers. Table~\ref{tab:downstream_offline_rl} shows that diffusion-based methods achieve superior downstream performance compared to GAN-based methods.}

\vspace{1mm}
\noindent \textbf{Traditional diffusion models.} Diffusion models~\cite{ddpm} consist of two processes: (1) The \emph{forward} process that progressively corrupts clean data $x_0$ by adding Gaussian noise, which outputs a noisier data sequence: $\{{x^1}, \ldots, {x^T}\}$, and $T$ is the number of noising steps. As the number of steps increases, the data becomes noisier, gradually resembling Gaussian noise more closely and deviating further from the original data sample with each step. (2) The \emph{reverse} process progressively denoises a noise to clean data using a trainable neural network. The forward process between adjacent noisy data, i.e., $x^{t-1}$ and $x^t$, follows a Gaussian distribution. Then, the forward process of diffusion models is defined as, $p\left( {{x^t}\left| {{x^{t - 1}}} \right.} \right) = \mathcal{N}\left( {{x^t};\sqrt {1 - {\beta _t}} {x^{t - 1}},{\beta _t}\mathbb{I}} \right),$ where ${\beta _t}$ regulates the magnitude of the added noise at each step and is usually pre-defined by users. We note ${{\bar \alpha }_t}: = \prod\nolimits_{s = 1}^t {\left( {1 - {\beta _s}} \right)} $. The likelihood between the clean data $x_0$ and noisy data in step $t$ is, $
    p\left( {{x^t}\left| {{x^0}} \right.} \right) = \mathcal{N}\left( {{x^t};\sqrt {{{\bar \alpha }_t}} {x^0},\left( {1 - {{\bar \alpha }_t}} \right)\mathbb{I}} \right).$
Therefore, we sample ${x^t}$ directly from $x^0$ in closed form instead of adding noise $t$ times as,
$ {x^t} = \sqrt {{{\bar \alpha }_t}} {x^0} + e\sqrt {1 - {{\bar \alpha }_t}},\; e\sim \mathcal{N}\left( {0,\mathbb{I}} \right). $
The final objective of diffusion models is defined as,
\begin{equation}
\label{eq:L_DM}
{\mathcal{L}(\theta, x)} = {\mathbb{E}_{t \sim \mathcal{U} \left( {1,T} \right),{x^t} \sim p\left( {{x^t} \left| {{x^0 = x}} \right.} \right)}}{\left\| {e - {e_\theta }\left( {{x^t},t} \right)} \right\|^2},
\end{equation}
where $e \sim \mathcal{N}\left( {0,\mathbb{I}} \right)$ and ${e_\theta }$ is a neural network parameterized with $\theta$ that is updated to minimize Equation~\eqref{eq:L_DM}. $\mathcal{U}(1,T)$ is the uniform distribution ranging from 1 to $T$. Thus, ${e_\theta }$ learns to predict the noise $e$ of the noisy data at any step $t$. After being trained well, we apply ${e_\theta }$ to gradually denoise a random Gaussian noise to clean the data. 

\rev{ We note that the noise prediction network serves as the core component of diffusion models. Following prior work~\cite{li2023meticulously}, we use $e_\theta$ as a shorthand to represent the entire diffusion model for simplicity. This notation includes the noise prediction network and, when applicable, any additional modules.}

\vspace{1mm}
\noindent \textbf{Conditional diffusion models.} Conditional diffusion models extend diffusion models by incorporating conditional inputs to guide the generation process. They are widely used for tasks such as sequence synthesis. Section~\ref{subsec:traj_synthesis} presents that, for DP trajectory synthesis, conditional inputs enable us to capture temporal dependencies among different transitions in sequences. The objective function is reformulated as ${\mathcal{L}(\theta, x, c)}$, where $c$ represents conditional inputs, and the noise prediction network at step $t$ is defined as $e_\theta(x^t, t, c)$. 

\vspace{1mm}
\noindent \textbf{Diffusion Transformer.} Diffusion transformers leverage transformer-based architecture~\cite{attention} for noise prediction. They excel at modeling sequential data with temporal dependencies, such as text~\cite{yue-etal-2023-synthetic} or structured sequences~\cite{diffusiontransformer}, making them suitable for trajectory-level DP synthesis, where temporal dependencies in sequences must be preserved. 
Since transformers lack inherent sequential awareness, positional encoding is essential to capture the order of inputs~\cite{attention}.

\vspace{1mm}
\rev{Following prior works~\cite{lu2023synthetic,he2023diffusion}, we adopt the Elucidated Diffusion Model (EDM)~\cite{edm} for transition synthesis and the Diffusion Transformer~\cite{diffusiontransformer} for trajectory synthesis. These models use an MLP and a Transformer, respectively, as the noise prediction network. We present more details of architectural designs in Appendix~\ref{subsec:dm}.}

\begin{figure}[!t]
    \centering
    \setlength{\abovecaptionskip}{0pt}
    \includegraphics[width=0.99\linewidth]{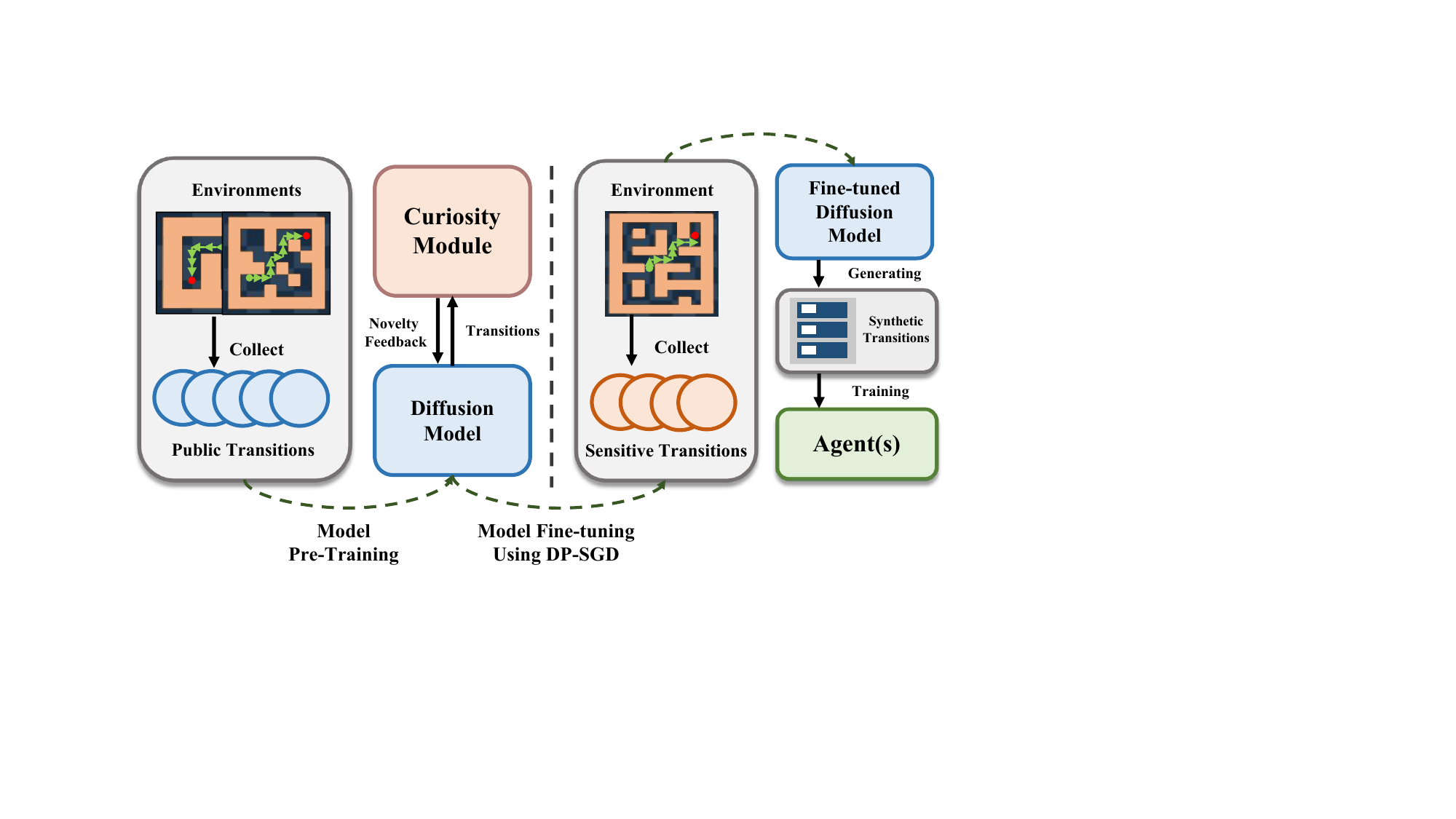}
    \caption{High-level illustration of the overall workflow of \toolname. Initially, \toolname{} pre-trains synthesizers on public datasets, guided by a curiosity module. It then fine-tunes the model on sensitive datasets using DP-SGD. Finally, the fine-tuned model generates synthetic datasets for agent training.}
    \label{fig:privtranr}
    \vspace{-3mm}
\end{figure}

\section{PROBLEM SETUP AND PRELIMINARIES}

\subsection{\rev{Threat Model}}
\label{subsec:threat}

\rev{We assume that the data provider holds a highly sensitive dataset, such as medical records, which can be used for offline RL agent training. Directly sharing such datasets poses significant privacy risks. To mitigate these risks, several approaches advocate generating synthetic datasets as substitutes, while adversaries can still infer the sensitive dataset information~\cite{2023whithmiadiffusion} through the synthetic datasets.}

\rev{DP dataset synthesis addresses this challenge by providing formal, mathematically rigorous guarantees that limit the influence of any individual record on the generated data. This offers general protection and guarantees that, even with auxiliary knowledge, an attacker’s ability to infer specific private details about individual transitions or trajectories remains strictly bounded. DP-based synthesis has gained traction across multiple domains, including image~\cite{li2023meticulously}, text data~\cite{yue-etal-2023-synthetic}, and tabular data~\cite{zhang2021privsyn}, making it a promising paradigm for privacy-preserving offline RL datasets. }

\subsection{Problem Statement}
\label{subsec:ps}
We aim to generate new transition-based and trajectory-based datasets that statistically mirror the original dataset under DP.
Specifically, we possess a set of sensitive offline RL data denoted by $D_s$, and generate a set of synthetic data individuals, $\hat{D}$. Agents trained on the sensitive dataset exhibit a similar level of performance compared to those trained on the synthetic dataset. Besides, the synthetic dataset has statistical characteristics similar to those of the original dataset.

\subsection{Adapting Existing Methods}
\label{subsec:strawman}

\vspace{1mm}
\noindent{\textbf{Marginal-based Solutions.}}
We first consider the marginal-based solutions~\cite{zhang2021privsyn}, notably PrivSyn~\cite{zhang2021privsyn}. 
In transition synthesis, we can treat the data as a table, where each row is a transition, and each column is one element of the transition (state, action, reward, and state can take multiple columns as they can be multi-dimensional). The challenge with marginal-based methods is their difficulty in managing data with large dimensions. These methods are notably slow when processing large-scale sensitive datasets. Section~\ref{sec:utility} empirically shows that directly adapting those methods does not work well. In trajectory synthesis, trajectory lengths vary rather than being fixed, rendering traditional DP tabular synthesis unsuitable.

\vspace{1mm}
\noindent{\textbf{ML-based Solutions.}}
Another approach is to adapt more complex machine-learning models, such as Generative Adversarial Networks (GANs) and diffusion models. For GAN-based methods, we adopt PATE-GAN~\cite{PATE-GAN}. As presented in previous works~\cite{pmlr-v70-arjovsky17a}, GAN suffers from problems of unstable training and sometimes fails to fit the distribution of sensitive datasets~\cite{yin2022practical}.  {Diffusion-based methods encounter similar challenges when processing complex datasets, such as images~\cite{dpdm}.

\vspace{1mm}
\noindent{\textbf{Leveraging Pre-training.}}
Diffusion models achieve excellent synthesis performance in various fields~\cite{dp-diffusion,he2023diffusion,li2023meticulously,ddpm,li2025easy}, but one challenge particular to training diffusion models with DP is that diffusion models are typically larger, and thus need more training. In the era of `large models,' it is increasingly common to begin with a pre-trained model and then proceed to fine-tune it for better performance~\cite{dp-diffusion,Wu_2023_ICCV}. Thus, we adopt this setting and adapt the pre-training method in Ghalebikesab et al.~\cite{dp-diffusion}. 
However, it is less clear how to best use pre-training to enhance marginal-based solutions (because the lower-dimensional marginals are obtained in one-shot with DP, and there is no convergence issue).

\section{Introducing \toolname}
\label{sec:comparing}
We introduce \toolname with two variants \toolnametran and \toolnametraj for handling transition-level and trajectory-level DP definitions. 
\toolname has a unified synthesizer training the synthesis paradigm. As presented in Figure~\ref{fig:privtranr}, \toolname first pre-trains synthesizers on datasets without privacy concerns (i.e., public datasets), guided by feedback from the curiosity module. \toolname then fine-tunes the pre-trained model on sensitive datasets using DP-SGD. Finally, the fine-tuned synthesizer generates synthetic datasets for agent training.

\subsection{Differences Between \toolnametran and \toolnametraj}
\label{sec:diffrence}

\rev{However, there are some differences between \toolnametran and \toolnametraj, which are introduced as follows.}

\begin{itemize}[leftmargin=*]
    \item \textit{Data structure.} \toolnametran models each transition independently, whereas \toolnametraj models the entire trajectory, capturing temporal dependencies (e.g., state transition function). Therefore, \toolnametran models each transition independently using diffusion models, whereas \toolnametraj should model to capture temporal dependencies with sequence generation models like diffusion transformers.
    \item \textit{Data Dimensionality.} Trajectories have significantly larger dimensions than transitions. In \toolnametran, each transition is generated independently, processing small-scale data. Conversely, \toolnametraj handles long sequences, risking a memory explosion. As detailed in Section~\ref{subsec:traj_synthesis}, \toolnametraj handles long trajectories by dividing them into fragments and using a conditional synthesizer to model their interconnections, allowing for the generation of fragments that seamlessly combine into cohesive trajectories.
    \item \textit{Data generation.} \toolnametran generates each transition independently, whereas \toolnametraj synthesizes trajectory fragments and stitches them into a complete trajectory, requiring consideration of inter-fragment dependencies.
\end{itemize}

\subsection{\rev{Design Overall}}
\label{sec:diffrence}

\rev{We use different architectures, training, and synthesis to match the characteristics of transition and trajectory.}

\begin{itemize}[leftmargin=*]
\item 
\rev{Following prior work~\cite{lu2023synthetic,he2023diffusion}, we use the Elucidated Diffusion Model (EDM)~\cite{edm} for \toolnametran (unconditional generation) and the Diffusion Transformer~\cite{diffusiontransformer} for \toolnametraj, conditioned on the link transition to capture fragment relationships, as introduced in Section~\ref{subsec:dataset_process}. EDM and Diffusion Transformer use an MLP and a Transformer for noise prediction. Architectural details appear in Appendix~\ref{subsec:dm}.}

\vspace{0.5mm}

\item \rev{\textit{Training.} Section~\ref{subsec:diffusion_model} details the training process of diffusion models. The input of the noise prediction network ${e_\theta}$ is noisy data, and the output is the prediction of added noise. Equation~(\ref{eq:L_DM}) minimizes the output of ${e_\theta}$ and real added noise $e$. Thus, ${e_\theta}$ learns to predict the noise $e$ added to noisy data at any step $t$. For \toolnametran, $e_\theta$ takes as input noisy transitions and outputs the corresponding noise. For \toolnametraj, $e_\theta$ receives noisy trajectory fragments with the link transition, and outputs the noise. Section~\ref{subsec:traj_synthesis} explains how link transitions are incorporated into the training. }

\vspace{0.5mm}

\item \rev{\textit{Synthesis.} As introduced in Section~\ref{subsec:diffusion_model}, once trained, synthesizers use $e_\theta$ to generate data through denosing the noise sampled from a Gaussian distribution. \toolnametran uses $e_\theta$ to denoise Gaussian noises and generate transitions. As shown in Section~\ref{subsec:fragment_synthesis}, \toolnametraj generates fragments sequentially and stitches fragments to full trajectories. }
\end{itemize}

\rev{We elaborate on the technical details of \toolnametran and \toolnametraj in Section~\ref{sec:methodology} and Section~\ref{sec:privorl-j}, respectively.}

\subsection{\rev{Motivation of Leveraging Curiosity-Driven Pre-training}}
\label{motivation:curiosity}

\rev{Offline RL requires the algorithm to understand the dynamics of the environment's MDP from datasets~\cite{levine2020offline}. Then, agents are trained to achieve the maximum possible cumulative reward when interacting with the environment. To better understand the environment's MDP, a diverse training dataset is necessary, which helps ensure that the agent encounters a comprehensive spectrum of states, actions, and rewards, representing a wide range of environmental scenarios~\cite{levine2020offline,d4rl}. Thus, the success of offline RL agents relies heavily on the breadth and diversity of the datasets.}

\rev{Inspired by the random network distillation in RL and bug detection~\cite{hong2024curiositydriven,he2024curiosity}, we propose using the curiosity module, which quantifies the `novelty' of synthetic data individuals (as detailed in Section~\ref{subsec:curiosity_module}). To incorporate the novelty feedback into synthesizer pre-training, we replace a portion of real data with high-novelty synthetic data, encouraging synthesizers to capture underlying characteristics of high-novelty data. The high-level motivation of curiosity-driven during pre-training instead of fine-tuning phase is that we have more flexibility and can tolerate more instability in pre-training phase, as supported by Table~\ref{tab:FineCurPrivTranR}. Technical details appear in Section~\ref{subsec:curiosity_module}.}

\section{Transition-Level DP: \toolnametran}
\label{sec:methodology}
This section focuses on transition-level DP synthesis and introduces the curiosity module.

\subsection{Curiosity Scores}
\label{subsec:curiosity_module}

We propose a ``curiosity module'' to assess the diversity (or novelty) of the synthetic dataset and promote diverse data synthesis. Our approach is based on random network distillation (RND)~\cite{rnd}.
This section details how to measure the novelty (whether a transition is frequently or rarely generated by the synthesizer)
of synthetic data. RND uses prediction errors to represent the difference between the outputs of a fixed target network and a trainable predictor network. Therefore, when synthetic data is rarely generated, and the predictor predicts that it is unfamiliar, the predictor network’s output deviates 
from the target network’s output, resulting in a higher error.

RND uses two randomly initialized networks: a fixed \textit{target} network $f$ and a \textit{predictor} network $\hat{f}$ that aims to learn the output of $f$. During pre-training, synthetic data $x$ is input to both $f$ and $\hat{f}$. The target network outputs a random vector $f(x) \to \mathbb{R}^d$ ($d$ is the vector dimension), which $\hat{f}$ tries to match. 
Specifically, let the $f$ and $\hat{f}$ be parameterized by $\phi$ and $\hat{\phi}$. The objective is to minimize the following objective,
\begin{equation}
    c(x) = \left\|\hat{f}_{\hat{\phi}}(x) - f_{\phi}(x)\right\|^2_{\ell_2}.
    \label{eq:curi_score}
\end{equation}
This error is also the \textit{curiosity score} $c(x)$ to quantify the `novelty' of a synthetic data 
$x$ and should be higher for `novel' synthetic data than for those previously generated repeatedly. 

This section details how to incorporate the novelty feedback into the diffusion model pre-training.

\subsection{Curiosity-driven Updating}
This section explains how curiosity scores guide synthesizer updates. As described in Section~\ref{subsec:diffusion_model}, diffusion model training lacks a reward mechanism that integrates novelty feedback. In fact, diffusion models fit data distributions by capturing statistical characteristics of training datasets~\cite{ddpm}. To address this challenge, we include high-curiosity synthetic data in the training set, enabling the model to learn its traits and generate diverse data. In particular, in one iteration, we sample a batch of data $X$ from the training dataset with a batch size of $B$. We then generate an equivalent number of synthetic data using the synthesizers, denoted as $\hat{X}$. The curiosity module measures the curiosity scores for synthetic data. We rank these scores and select the top-$p$ synthetic data from $\hat{X}$ to replace the same number of data in the training batch dataset $X$. The $p$ is the \textit{curiosity rate} ranging from 0 to 1, which controls how strongly curiosity is applied. Then, the modified dataset $X_r$ is used to train synthesizers. \rev{Appendix~\ref{supsubsec:theoretical_analysis} presents why our method can benefit the diversity of synthetic datasets. }

\vspace{1.0mm}
\noindent{\textbf{Transitions Synthesis.} For each transition, we first sample a Gaussian noise $x^T$ ($T$ is the number of noising steps), and we use the diffusion model to denoise $x^T$ to the less noisy $x^{T-1}$. For any $t$ in range of 1 to $T$, this denoising entails the computation of the estimated noise mean $\mu$ for $x^{t-1}$~\cite{ddpm},
\begin{equation}
    \mu = \frac{1}{\sqrt{\alpha_t}} \left( x^t - \frac{1-\alpha_t}{\sqrt{1-\beta_t}} e_\theta\left(x^t,t\right)\right),
    \label{eq:calmu}
\end{equation}
where $\alpha_t$ and $\beta_t$ are hyper-parameters as introduced in Section~\ref{subsec:dm}, and $t$ is the current step. The $x^{t-1}$ can be obtained as~\cite{ddpm},
$
    x^{t-1} = \mu + \sigma_t e,\; e \sim \mathcal{N}(0, \mathbb{I}).
    \label{eq:calxt}
$
The $\sigma_t$ regulates the magnitude of the added noise and is pre-defined by users. Repeating this denoising process until $t=0$, the $x^0$ indicates our final synthetic transition.

\subsection{Standard DP-SGD Fine-Tuning.}
We leverage DP-SGD (as described in Section~\ref{sub:dp}) to fine-tune the pre-trained diffusion model on the sensitive transitions to satisfy DP. Algorithm~\ref{alg:privORL-n} presents the training of \toolnametran.

\section{Trajectory-Level DP: \toolnametraj}
\label{sec:privorl-j}

Moving from transitions to trajectories, we face two questions: (1) How to support generate high-dimensional long trajectories? (2) How can we capture long-range temporal dependencies in trajectory-level DP synthesis?

To tackle high-dimensionality challenges, \toolnametraj divides long trajectories into fragments and employs a conditional diffusion transformer to model their interconnections, enabling the generation of fragments that seamlessly combine into cohesive trajectories. To capture long-range temporal dependencies in trajectory-level DP synthesis, \toolnametraj enhances \toolnametran by incorporating a transformer~\cite{diffusiontransformer} into its diffusion model, effectively modeling intricate temporal relationships in trajectories.  
We introduce \toolnametraj as follows.

\begin{algorithm}[!t]
    \caption{Workflow of \toolnametran}
    \label{alg:privORL-n}
    \textbf{Input}: The public and sensitive transition set: $D$ and $D_s$; Synthesizer $e_\theta$ parameterized with $\theta$; Target and predict networks $f_{\phi}$, $f_{\hat{\phi}}$ parameterized with $\phi$ and $\hat{\phi}$; The curiosity rate: $p$. \\
    \tcp{\color{black} Curiosity-Driven Pre-training}
    \While{ training epochs $<$ target epochs }{
            $ X \gets $ Randomly select $B$ data from the $D$;  \\
            $ \hat{X} \gets $ Generate $B$ trajectory fragments using $e_\theta$; \\
            \For{$x \in \hat{X}$ }{
                $c(x) = \left\| \hat{f}_{\hat{\phi}}(x) - f_{\phi}(x) \right\|^2_{\ell_2}$ \\
                Update $\hat{\phi}$ by minimizing $c(x)$ \;
            }
            Sort set $\hat{X}$ according to $c$ and get the top-$p$ sorted set $X_o$\label{privalgo:sort}\;
            \(X_r \gets \) Replace a subset of \(X\) with  from \(X_o\) \label{privalgo:replace}\;
            Train $e_\theta$ on $X_r$ using Eq.~\eqref{eq:L_DM}\;
        }
         \tcp{\color{black} Private Fine-tuning}
        Fine-tune $e_\theta$ on $D_s$ with DP-SGD (using Eq.~\eqref{eq:L_DM}) \label{privalgo:fine-tuning}\;
        \textbf{Output}: The well-trained diffusion model $e_\theta$. \\
    \end{algorithm}

\subsection{Dataset Proprocess} 
\label{subsec:dataset_process}
As previously mentioned, our use of transformers involves working with fragments. To enable this, we preprocess the data into fragments, consisting of consecutive transitions. 
We first segment the trajectory into equal-length segments, referring to processes in text sequences. Any sub-trajectory that does not reach the required length is padded. We should partition the trajectories from both the public dataset $D = \{\tau_n\}_n$ and sensitive dataset $D_s = \{\tau_i\}_i$ into trajectory fragments, resulting in $D' = \{(\tau^s_n, S_n)\}_n$ and $D'_s = \{(\tau^s_i, S_i)\}_i$, respectively. Here, each trajectory $\tau$ is divided into $N$ fragments, $\tau \to \{\tau^{s_i}\}$. The linking transition $S_i$ connects consecutive fragments: for a given fragment, $S_i$ is defined as the preceding transition of the first transition of the fragment or zero-padded for the initial fragment. For instance,  we consider a complete trajectory $\tau = (\cdots, s_p, a_p, r_p, s_{p+1}, a_{p+1}, r_{p+1}, \cdots)$ that might be randomly segmented into two fragments, such as $\tau_1^s = (\cdots, s_p, a_p, r_p)$ and $\tau_2^s = (s_{p+1}, a_{p+1}, r_{p+1}, \cdots)$, where the linking transition $S_2$ for $ \tau_2^s$ is the last transition of $\tau_1^s$, i.e., $S_2 = (s_p, a_p, r_p, s_{p+1})$.

We denote the number of transitions in a trajectory fragment by $H$ (sometimes we call it \textit{horizon}). Additionally, we introduce a terminal signal $d_t$ for each transition: $d_p = 1$ indicates that $s_p$ is a terminal state, in which case $s_{p+1} = 0$; otherwise, $d_p = 0$. Consequently, a trajectory fragment is formalized as $\tau^s = \left[(s_p, a_p, r_p, s_{p+1}, d_p)_{p=1}^H \right]$.

\subsection{Synthesizer Training}
\label{subsec:traj_synthesis}
As described in Section~\ref{subsec:diffusion_model}, \toolnametraj generates trajectory fragments by progressively denoising Gaussian noise through $T$ timesteps, guided by a conditional input. 
During the forward process of the diffusion model, we incrementally add noise to the trajectory fragments, generating a sequence of noisy fragments, $\{\tau^s_t\}_{t=0}^T$. This entails training a noise prediction network to estimate the noise added at each timestep $t$, using the current noisy trajectory fragment $\tau^s_t$ and conditional input $s$.
As introduced in Section~\ref{subsec:diffusion_model}, \toolnametraj leverages the transformer as the prediction network, which is formulated as $e_\theta(\tau^s_t,t,S)$, and $\theta$ means the network parameters. We elaborate on the training processes of \toolnametraj as follows,
\begin{itemize}[leftmargin=*]
    \item \textit{Curiosity-Driven Pre-training.} This section describes the curiosity module pre-training in \toolnametraj.
    \item \textit{Input Embeddings.} We detail the process of treating the inputs of the noise prediction network as an embedding sequence, which prepares them for transformer training.
    \item \textit{Sequence Processing.} It uses the input embeddings to train the transformer, enabling accurate prediction of the added noise in diffusion processing.
    \item \textit{Private Fine-tuning.} We describe how \toolnametraj fine-tunes the synthesizer on sensitive fragments using DP-SGD.
\end{itemize}

\begin{algorithm}[!t]
    \caption{Workflow of \toolnametraj}
    \label{alg:privorl-j}
    \textbf{Input}: The public and sensitive trajectory set: $D$ and $D_s$; The horizon size: $H$; Diffusion transformer $e_\theta$ parameterized with $\theta$; Target and predict networks $f_{\phi}$, $f_{\hat{\phi}}$ parameterized with $\phi$ and $\hat{\phi}$; The curiosity rate: $p$. \\
        \textbf{Initialization}: $D',D'_s = \varnothing$. \\
        \tcp{\color{black} Dataset Preprocess}
        \For{ $\tau \in D, D_s$ }{
            Split $\tau$ to $N$ fragment $\left\{(\tau^s_i,S_i)_{i=1}^N\right\}$, and each $\tau^s$ has $H$ transitions; \\
            \lIf{$\tau \in D$ }{$D' \cup \left\{(\tau^s_i,S_i)_{i=1}^N\right\}$}
            \lElse{$D'_s \cup \left\{(\tau^s_i,S_i)_{i=1}^N\right\}$}
         }
         \tcp{\color{black} Curiosity-Driven Pre-training}
         \While{ training epochs $<$ target epochs }{
            $ X \gets $ Randomly select $B$ data from the $D'$; \\
            $ \hat{X} \gets $ Generate $B$ trajectory fragments using $e_\theta$; \\
            \For{$\tau^s \in \hat{X}$ }{
                $c(\tau^s) = \left\| \hat{f}_{\hat{\phi}}(\tau^s) - f_{\phi}(\tau^s) \right\|^2_{\ell_2}$ \\
                Update $\hat{\phi}$ by minimizing $c(\tau^s)$ \;
            }
            Sort set $\hat{X}$ according to $c$ and get the top-$p$ sorted set $X_o$\label{privalgo:sort}\;
            \(X_r \gets \) Replace a subset of \(X\) with  from \(X_o\) \label{privalgo:replace}\;
            Train $e_\theta$ on $X_r$ using Eq.~\eqref{eq:L_diff}\;
        }
         \tcp{\color{black} Private Fine-tuning}
         Fine-tune $e_\theta$ on $D'_s$ using Eq.~\eqref{eq:dpsgd-trajectory} with DP-SGD\label{privalgo:fine-tuning} to calculate the aggregated gradient for each trajectory\;
        \textbf{Output}: The well-trained diffusion transformer $e_\theta$. 
    \end{algorithm}

\noindent \textbf{Curiosity-Driven Pre-training.} This process is almost the same as what we introduced in Section~\ref{sec:methodology}. In \toolnametraj, we measure the novelty of trajectory fragments instead of single transitions in \toolnametran.
Then, the modified training dataset is leveraged to pre-train the synthesizer. As shown in Section~\ref{subsec:dm}, the objective of conditional diffusion is,
\begin{equation}
    \mathcal{L}_{\text{diff}} = \mathbb{E}_{(\tau^{s_t},S), t} \left[ \left\| e_\theta\left(\tau^{s_t}, t, S\right) - e^{t} \right\|_1 \right]
    \label{eq:L_diff}
\end{equation}
where  $\left\| \cdot \right\|_1$ denotes the $L_1$ norm, and $\tau^{s_t}$ and $e^{t}$ are the noisy trajectory fragment and true noise at the $t$-th timestep.

\vspace{1mm}
\noindent \textbf{Input Embeddings}: Inputs of the prediction network include three parts: (1) the noisy trajectory fragment $\tau^s_t$; (2) timestep in the forward process $t$; and (3) conditional input $S$. Referring to previous works in the text synthesis~\cite{dptransformer}, the sensitive trajectory fragments dataset $D'_{s}$ should be embedded into a high-dimensional space using multilayer perceptrons (MLPs).

First, we embed the timestep $t$ in the diffusion process~\cite{diffusiontransformer}, and recorded as 
$\text{TimeEmb}(t) \in \mathbb{R}^k$, where $k$ is the dimension size of embedding. Then, we embed the conditional input $S$ (a link transition includes a state, an action, a reward, a next state, and a terminal signal) with a separate MLP, $\text{ConditionEmb}(S) \in \mathbb{R}^{5\times k}$. For embedding the trajectory fragment $\tau^s_t$, we divide $\tau_i^s$ into five components, $$\left[(s_p)_{p=1}^H\right], \left[(a_p)_{p=1}^H\right], \left[(r_p)_{p=1}^H\right], \left[(s_{p+1})_{p=1}^H\right], \left[(d_p)_{p=1}^H\right],$$ 
and embed them with five separate MLPs. Thus, each trajectory fragment yields $5\times H$ embeddings; concatenating one time embedding and five condition‑transition embeddings results in $5\times H+5+1$ embeddings in total. As a result, \toolnametraj gets the input embedding $\mathbf{z}$, $\mathbf{z} = \text{InputEmbed}\left( \tau^s_t, t, S \right) \in \mathbb{R}^{(5\times H + 6) \cdot k}.$
Each embedding is treated as a token for subsequent transformer training, resulting in an input embedding $\mathbf{z}$ of $(5\times H + 6)$ tokens, which refers to the basic units of input data for the transformer~\cite{attention}. Then, we formulate the input embedding as the form of tokens, $\mathbf{z} = \{z_i\}_{i=1}^{5\times H + 6}, z_i \in \mathbb{R}^{k}$.

\noindent \textbf{Sequence Processing.} Offline RL trajectory synthesis requires dynamic coherence, and state transitions must align with environmental dynamics~\cite{d4rl,offline_survey}. The transformer models these temporal relationships, ensuring that the generated trajectory is globally coherent, such as how earlier states in the trajectory influence subsequent state-action pairs. Then, we describe how transformers model the input embeddings, i.e., token sequence $\mathbf{z} = \{z_i\}_{i=1}^{5\times H + 6}, z_i \in \mathbb{R}^{k}$, to generate the predicted noise.

To encode the relative positions of tokens in the sequence $\mathbf{z}$, we introduce \textit{position embeddings}. Since transformers lack inherent sequential awareness, positional encoding is essential to distinguish the order of identical tokens. Drawing on implementation of prior work in text data processing~\cite{attention,dptransformer}, we assume that $i$ is the position index of the token ($ i = \left\{0, 1, \ldots,  5\times H + 5 \right\}$, 
and $ \text{PosEmb}(i)$ denotes the position embedding vector for the token at position $i$, and $\text{PosEmb}(i) \in \mathbb{R}^k$. Then, we add the positional embedding $\text{PosEmb}(i)$ to the token embedding and get the final embedded inputs of the transformer. The outputs of the transformer are the embedding matrix, $\text{Transformer}(\mathbf{\mathbf{z}_\text{final}}) \in \mathbb{R}^{H \times k}$, and we decode them into noise prediction components using MLPs. The output of $e_\theta(\tau^s_t,t,S)$ is the predicted noise, matching the dimensionality of the input trajectory fragment $\tau^s_t$. 

\vspace{1mm}
\noindent{\textbf{Private Fine-Tuning.}} We use DP-SGD to fine-tune the transformer on sensitive trajectories to satisfy DP at the complete trajectory level. As presented in Section~\ref{subsec:dataset_process}, we split each trajectory into fragments of size $H$ transitions between consecutive fragments. Distinguished from the DP-SGD applied in transition-level DP, DP-SGD is applied by aggregating the gradients of all fragments $(\tau^s,S)$ belonging to the same trajectory $\tau$, clipping them to a maximum norm $C$, and adding noise at the trajectory level with a noise multiplier $\sigma$, ensuring that we protect per trajectory rather than per fragment. The parameters of the noise prediction network are updated by,
\begin{align}
\small
\mathbb{E}_{\tau 
} \left[ \mathbb{E}_{(\tau^s,S) \in \tau}\left[ \text{Clip}_{C}\left(\nabla {\mathcal{L}}(\theta, (\tau^s,S))\right) \right] + C \mathcal{N}(0, \sigma^2 \mathbb{I}) \right].
\label{eq:dpsgd-trajectory}
\end{align}
We present the workflow of \toolnametraj in Algorithm~\ref{alg:privorl-j}.

\vspace{-2mm}
\subsection{Dataset Synthesis via \toolnametraj}
\label{subsec:fragment_synthesis}
\toolnametraj just generates trajectory fragments. This section introduces how to generate a complete trajectory. 

For the initial fragments of each trajectory, we first sample a Gaussian noise $\tau^s_T$ ($T$ is the number of noising steps), and the conditional input $s$ should be $0$. Then, we use the diffusion transformer to denoise $\tau^s_T$ to the less noisy $\tau^s_{T-1}$. For any $t$ in range of 1 to $T$, this denoising entails the computation of the estimated noise mean $\mu$ for $\tau^s_{t-1}$, defined as~\cite{ddpm},
\begin{equation}
    \mu = \frac{1}{\sqrt{\alpha_t}} \left( \tau^s_{t} - \frac{1-\alpha_t}{\sqrt{1-\beta_t}} e_\theta \left(\tau^s_{t-1},t,S\right)\right),
    \label{eq:trajecotry_syn_1}
\end{equation}
where $\alpha_t$ and $\beta_t$ are hyper-parameters as introduced in Section~\ref{subsec:diffusion_model}, and $t$ is the current step. The $\tau^s_{t-1}$ is defined as~\cite{ddpm},
\begin{equation}
    \tau^s_{t-1} = \mu + \sigma_t e,\;\; e \sim \mathcal{N}(0, \mathbb{I}).
    \label{eq:trajecotry_syn_2}
\end{equation}
The $\sigma_t$ regulates the magnitude of the added noise and is pre-defined by users. Repeating this process until $t=0$, the $\tau^s_{0}$ indicates the synthetic trajectory fragment. Then, we use the final transition of synthetic trajectory fragments as conditional inputs and iterate the synthesis process until either the terminal state is generated—indicated by the fragment trajectory containing the terminal signal $d=1$—or the predefined maximum trajectory length is reached. Figure~\ref{fig:traj_syn} presents the visualization for trajectory synthesis and fragment stitching.

\begin{figure}[!t]
    \centering
    \setlength{\abovecaptionskip}{0pt}
    \includegraphics[width=0.99\linewidth]{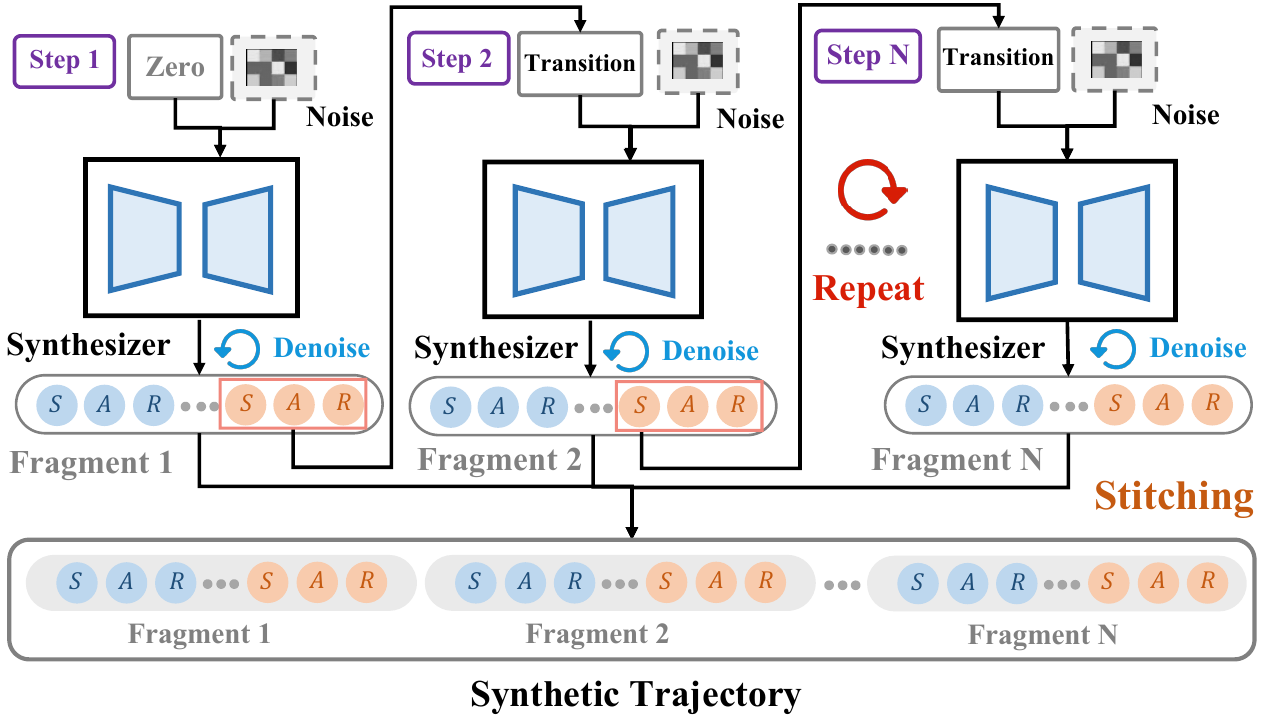}
    \caption{\rev{Visualizations for trajectory synthesis and fragment stitching. The initial condition is zero-padded, and the synthesizer denoises to generate a fragment. We then use the final transition of each fragment as a conditional input and iterate until a terminal state is produced. Finally, the fragments are stitched sequentially to form a complete synthetic trajectory. }}
    \label{fig:traj_syn}
\end{figure}

\vspace{-1mm}
\subsection{\rev{Processes for Discrete Variants}}
\label{subsec:discrete}

\rev{We uniformly represent datasets with real-valued variables to leverage diffusion models’ strengths in handling continuous data~\cite{lu2023synthetic}. For discrete variants, we embed them into the continuous variable using one-hot encoding before the training phase.  During synthesis, we apply an argmax post-processing step to map the sampled synthetic continuous outputs back to valid discrete variants.}

\section{Privacy Analysis}

 Since \toolname leverages DP-SGD to train the synthesizers, its privacy analysis is the same as DP-SGD. Specifically, with the RDP accountant~\cite{rdp}, $K$ finetuning steps of \toolname will cost $\left(\alpha, K\gamma \right)$-RDP, where $\alpha$ and $\gamma$ denote the privacy parameters of RDP, and $\gamma$ is the upper bound of a function. The ultimate DP cost of \toolname is $\left(\gamma + \frac{\log 1/\delta}{\alpha-1}, \delta\right)$-DP. Please refer to Appendix~\ref{app:supp_dp} for more details. Besides, we present the detailed parameters of DP-SGD, like noise scale and sampling probability, in Table~\ref{tab:dpsgd_hyperparams} of the Appendix. For \toolnametran, the sampling probability is defined as the ratio of the batch transition size to the total transition dataset size. For \toolnametraj, although it is trained on trajectory fragments, the sampling probability is calculated as the ratio of the batch trajectory size to the overall trajectory dataset size, ensuring trajectory-level DP protection. \rev{ We use the RDP for fair comparisons with baselines. Appendix~\ref{supsubsec:prv} presents results under Privacy Random Variable (PRV)~\cite{PRV}, which provides a tighter privacy analysis than RDP. }

According to the post-processing theorem~\cite{dp}, if an algorithm satisfies $(\epsilon, \delta)$-DP, then any form of post-processing will not incur additional privacy loss. Therefore, agents trained on DP synthetic datasets (consisting of transitions or trajectories) without increasing the risk of data leakage.

\section{Experimental Setup}
\label{sec:setup}

\subsection{Investigated Tasks and Datasets.}
We conduct the experiments across three domains from D4RL~\cite{d4rl}: {\tt Maze2D}~\cite{d4rl}, {\tt Kitchen}~\cite{gupta2019relay}, and {\tt Mujoco}~\cite{mujoco}, all of which are commonly used in offline RL research~\cite{du2023orl, lu2023synthetic, tarasov2022corl}. D4RL is a benchmark specifically designed for evaluating offline RL algorithms.

 Each environment comprises various tasks, each featuring similar yet distinct map or robot configurations. For example, in {\tt Maze2D}, the agent controls the same robot across different tasks but is required to achieve various goals on different maps, as presented in Figure~\ref{fig:maze2d_env}. To protect the privacy of the real dataset, we select the pre-training and sensitive datasets from the same domain but with different tasks. For instance, if we designate {\tt Maze2D-umaze} as the sensitive dataset requiring protection, {\tt Maze2D-medium} and {\tt Maze2D-large} are designated as pre-training datasets. Further details on the selection of pre-training and sensitive datasets and processing of D4RL for transition and trajectory synthesis to match the real-application requirements are provided in Appendix~\ref{app:task_dataset}.

For the downstream task, we selected three state-of-the-art offline RL algorithms widely used~\cite{tarasov2022corl, lu2023synthetic}, including EDAC~\cite{awac}, IQL~\cite{iql}, and TD3PlusBC~\cite{td3plusbc}.  Our implementations of offline RL algorithms are based on the open-source repository, CoRL~\cite{tarasov2022corl}, with consistent hyper-parameters settings. We refer to codes released in the repository~\cite{lu2023synthetic,he2023diffusion}, implementing the offline RL dataset synthesizers.
We elaborate on the details of hyper-parameters in Appendix~\ref{subsec:dm}. Besides, we discuss the details of the curiosity module in Appendix~\ref{appsubsec:curo}. Unless otherwise specified, the experimental configurations are the same as those in Section~\ref{sec:utility}.

\begin{figure}[!t]
    \centering
    \setlength{\abovecaptionskip}{0pt}
    \includegraphics[width=0.99\linewidth]{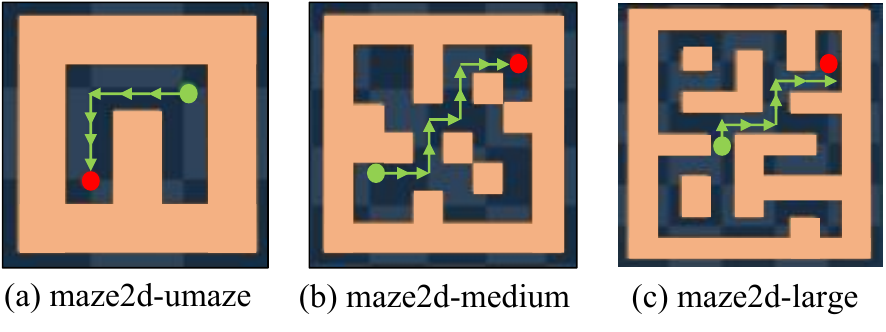}
    \caption{An illustrative example from {\tt Maze2D}. The green and red dots represent the start and end points of the 2D ball. }
    \label{fig:maze2d_env}
    \vspace{-3mm}
\end{figure}

\subsection{Baselines} 
\label{sub:baselines}
We compare \toolnametran with both PATE-GAN and PATE-GAN with pre-training (PrePATE-GAN)~\cite{PATE-GAN}. 
We also consider PGM~\cite{PGM} and PrivSyn~\cite{zhang2021privsyn}, which are two marginal-based DP dataset synthesis methods. They achieve state-of-the-art performance in the `SynMeter' library~\cite{du2024systematic}. These baselines are originally implemented for DP tabular synthesis, and we edit them for DP transition synthesis.

We present variants of \toolnametran evaluated in ablation studies to assess the role of curiosity-driven pre-training.

\begin{itemize}[leftmargin=*]
    \item \textbf{NonPrePrivORL (DPDM~\cite{dpdm}).} This baseline omits pre-training on public datasets for \toolnametran, exploring the significance of pre-training in transition synthesis. 
    \item \textbf{NonCurPrivORL (PDP-Diffusion~\cite{dp-diffusion}).} This baseline excludes the curiosity module in the pre-training. This variant is equivalent to the pre-training DP diffusion method proposed by Ghalebikesab et al.~\cite{dp-diffusion}.
\end{itemize}

For DP trajectory synthesis, we compare \toolnametraj with DP-Transformer~\cite{dptransformer} used in DP text synthesis. 
The second baseline is PrivORL-j-U, which does not consider the temporal relationship in the fragment trajectory. Specifically, we use U-Net as the noise prediction network instead of the transformer, and the other components, like fragment synthesizing, are the same as \toolnametraj. We elaborate on baselines in Appendix~\ref{app:baselines_imp}.

\subsection{Evaluation Metrics}
\label{subsec:metrics}  

We outline the principles of DP dataset synthesis in Appendix~\ref{app:eval_metrics} and introduce the evaluation metrics as follows. We elaborate on the details of these metrics in Appendix~\ref{supsubsec:metrics}

\vspace{1mm}
\noindent \textbf{Averaged Cumulative Return.} This metric assesses the utility of a synthetic dataset by measuring the average total reward a trained agent accumulates over multiple trajectories in real environments. The higher normalized returns, scaled to [0, 100], indicate that the trained agent has better performance and the synthetic dataset is of higher utility~\cite{d4rl}. we use ``\textit{normalized return}'' for simplicity.

\vspace{1mm}
\noindent \textbf{Marginal \& Correlation.} \textit{Marginal} uses the mean Kolmogorov-Smirnov~\cite{massey1951kolmogorov} statistic to measure the maximum distance between empirical cumulative distribution functions of each dimension. \textit{Correlation} measures differences in pairwise Pearson rank correlations~\cite{fieller1957tests}. Scores range from 0 to 1, with higher values indicating greater fidelity.

\vspace{1mm}
\noindent \textbf{TrajScore.} Inspired by BERTScore~\cite{Bertscore}, TrajScore uses pre-trained MLPs in an autoencoder as a trajectory encoder, similar to BERT’s embedding layer~\cite{BERT}, to compute trajectory embeddings. It calculates similarity via cosine similarity between generated and real trajectories.

\begin{table*}[!t]
\small
    \centering
    \caption{The normalized returns of agents trained on synthetic transitions using \toolnametran and baselines under privacy budget $\epsilon = \{1,10\}$ and real datasets. `no-DP' means agents trained on real datasets. We show the mean $\pm$ standard deviation of the performance averaged over five seeds. The best score is marked with the \colorbox{gray0}{gray} color box. }
    \setlength{\tabcolsep}{2.5mm}{
    \resizebox{0.99\textwidth}{!}{
   \begin{tabular}{p{1.2cm}|p{1.4cm}|l|c|c|c|c|c|c|c|c|c}
        \toprule
         \multirow{2}{*}{\textbf{Domains}}   &  \multirow{2}{*}{\begin{minipage}{1.3cm}\textbf{Real} \\ \textbf{Dataset}\end{minipage}}  & \multirow{2}{*}{\textbf{Methods}} & \multicolumn{3}{c|}{\textbf{EDAC}~\cite{an2021uncertainty}} & \multicolumn{3}{c|}{\textbf{IQL}~\cite{iql}} & \multicolumn{3}{c}{\textbf{TD3PLUSBC}~\cite{td3plusbc}} \\
         \cline{4-12}
          &  & & $\epsilon = 1$ & $\epsilon = 10$ & no-DP & $\epsilon = 1$ & $\epsilon = 10$ & no-DP & $\epsilon = 1$ & $\epsilon = 10$ & no-DP  \\
         \midrule
         \toprule
         \multirow{15}{2em}{{\tt Maze2D}}
         &\multirow{5}{*}{umaze} & PGM & 17.2 $\pm$ 10.4  & 32.8 $\pm$ 9.1 & \multirow{5}{*}{70.8 $\pm$ 13.6} & 30.1 $\pm$ 3.0 &  41.6 $\pm$ 0.9  & \multirow{5}{*}{71.0 $\pm$ 2.3}  & -12.4 $\pm$ 3.0 &  -8.5 $\pm$ 2.9 & \multirow{5}{*}{73.6 $\pm$ 3.2} \\
         & & PrivSyn  & 0.1 $\pm$ 10.3 & 1.3 $\pm$ 11.9 &  & 0.0 $\pm$ 1.5 & 2.9 $\pm$ 2.3  & & 3.0 $\pm$ 3.0 & 5.7 $\pm$ 4.0 & \\
         & & PATE-GAN &  -8.2 $\pm$ 12.0 & -16.8 $\pm$ 13.8 & & -10.5 $\pm$ 5.9 & -14.4 $\pm$ 7.8  &  & 1.5 $\pm$ 4.2 & 9.1 $\pm$ 6.2 & \\
         & & PrePATE-GAN & 12.1 $\pm$ 6.0 & 18.4 $\pm$ 4.8 &  & 12.4 $\pm$ 10.3 & 20.2 $\pm$ 4.5  & & 15.3 $\pm$ 6.4 & 26.4 $\pm$ 2.3 & \\
         & & \textbf{\toolnametran}  &   \cellcolor{gray!20}{63.2 $\pm$ 10.1} & \cellcolor{gray!20}{69.1 $\pm$ 14.5} & & \cellcolor{gray!20}{60.1 $\pm$ 8.6} & \cellcolor{gray!20}{70.3 $\pm$ 2.1} & & \cellcolor{gray!20}{30.6 $\pm$ 11.8} & \cellcolor{gray!20}{60.3 $\pm$ 6.3} & \\
         \cline{2-12}
         & \multirow{5}{*}{medium} & PGM & 12.1 $\pm$ 7,7 & 20.3 $\pm$ 8.7 & \multirow{5}{*}{73.0 $\pm$ 10.2} & 35.5 $\pm$ 1.7 & 46.8 $\pm$ 2.4 & \multirow{5}{*}{93.1 $\pm$ 10.7} & 4.3 $\pm$ 2.1 & 7.7 $\pm$ 1.0 & \multirow{5}{*}{53.3 $\pm$ 0.3} \\
         & & PrivSyn & 0.2 $\pm$ 1.4 & -6.3 $\pm$ 5.2 &  & 2.0 $\pm$ 1.5 &  4.0 $\pm$ 2.5 & & 30.0 $\pm$ 12.2 & 31.6 $\pm$ 10.0 \\
         & & PATE-GAN & 10.0 $\pm$ 3.0 & 15.5 $\pm$ 5.9 & & 17.7 $\pm$ 0.3 & 25.7 $\pm$ 0.7  & & 15.3 $\pm$ 12.9 & 20.0 $\pm$ 10.2 \\
         & & PrePATE-GAN & 2.3 $\pm$ 5.1  & 14.3 $\pm$ 6.3 & & -4.2 $\pm$ 3.0 & -7.0 $\pm$ 2.3 & & -2.2 $\pm$ 2.0 & -5.4 $\pm$ 1.1 \\
         & & \textbf{\toolnametran}  & \cellcolor{gray!20}{32.0 $\pm$ 3.0}  & \cellcolor{gray!20}{45.6 $\pm$ 1.9} & & \cellcolor{gray!20}{73.5 $\pm$ 13.2} & \cellcolor{gray!20}{90.7 $\pm$ 8.6}  &  & \cellcolor{gray!20}{33.8 $\pm$ 8.3} & \cellcolor{gray!20}{50.4 $\pm$ 6.4} & \\
         \cline{2-12}
         & \multirow{5}{*}{large} & PGM  & 0.4 $\pm$ 1.0 & 3.1 $\pm$ 3.7 & \multirow{5}{*}{89.9 $\pm$ 6.3} & 16.7 $\pm$ 3.2 &  21.9 $\pm$ 5.4   & \multirow{5}{*}{85.9 $\pm$ 0.2}  & 1.6 $\pm$ 2.0 & 4.0 $\pm$ 1.1 &  \multirow{5}{*}{96.8 $\pm$ 3.2}\\
         & & PrivSyn &  -8.2 $\pm$ 1.2 & -9.9 $\pm$ 0.6  &  & 0.4 $\pm$ -5.2 &  2.5 $\pm$ 1.6 &  & 3.4 $\pm$ 1.1 & 2.9 $\pm$ 2.3  & \\
         & & PATE-GAN &  -14.3 $\pm$ 2.0  & -5.2 $\pm$ 1.8 &  & 2.9 $\pm$ 0.7 & 5.9 $\pm$ 0.2 &  & 3.6 $\pm$ 1.0 & 5.8 $\pm$ 0.0 & \\
         & & PrePATE-GAN & -6.6 $\pm$ 4.3 & -2.4 $\pm$ 2.2 &  & 0.0 $\pm$ 0.0 &  0.4 $\pm$ 1.2 & & 0.0 $\pm$ 0.0 & 0.0 $\pm$ 0.0 & \\
         & & \textbf{\toolnametran}  &  \cellcolor{gray!20}{62.4 $\pm$ 12.2} & \cellcolor{gray!20}{80.6 $\pm$ 14.5} & &  \cellcolor{gray!20}{50.3 $\pm$ 8.1} &  \cellcolor{gray!20}{81.0 $\pm$ 11.8} &  & \cellcolor{gray!20}{61.3 $\pm$ 4.7} &  \cellcolor{gray!20}{75.3 $\pm$ 13.2} &  \\
         \midrule
         \toprule
         \multirow{5}{2em}{{\tt Kitchen}}
         & \multirow{5}{*}{partial} & PGM & 0.0 $\pm$ 0.0  & \cellcolor{gray!20}{3.5 $\pm$ 5.0} & \multirow{5}{*}{10.0 $\pm$ 0.0}  & 2.0 $\pm$ 1.5  &  1.5 $\pm$ 1.7 & \multirow{5}{*}{40.0 $\pm$ 2.5} & 2.0 $\pm$ 0.4 &   2.5 $\pm$ 1.8 & \multirow{5}{*}{18.0 $\pm$ 6.0} \\
         & & PrivSyn &  0.0 $\pm$ 0.0 & 0.0 $\pm$ 0.0 & & 0.0 $\pm$ 0.0 &  0.0 $\pm$ 0.0 & & 0.0 $\pm$ 0.0 & 0.2 $\pm$ 0.3 & \\
         & & PATE-GAN &  0.0 $\pm$ 0.0  & 0.0 $\pm$ 0.0 &  &  0.0 $\pm$ 0.0  & 0.0 $\pm$ 0.0  &  & 1.9 $\pm$ 0.6 & 4.2 $\pm$ 6.5 & \\
         & & PrePATE-GAN &  0.0 $\pm$ 0.0 & 0.0 $\pm$ 0.0 &  & 0.0 $\pm$ 0.0 & 0.0 $\pm$ 0.0  &  & 0.0 $\pm$ 0.0 & 0.8 $\pm$ 1.2 & \\
         & & \textbf{\toolnametran}  & \cellcolor{gray!20}{0.0 $\pm$ 0.0}  & 2.5 $\pm$ 1.5 &  & \cellcolor{gray!20}{ 12.5 $\pm$ 2.5} & \cellcolor{gray!20}{ 25.5 $\pm$ 2.5} & & \cellcolor{gray!20}{7.5 $\pm$ 1.5} & \cellcolor{gray!20}{ 11.5 $\pm$ 0.0} & \\
         \midrule
         \toprule
         \multirow{5}{2em}{{\tt MujoCo}}
         & \multirow{5}{*}{halfcheetah} & PGM &  0.0 $\pm$ 0.0 & 0.2 $\pm$ 0.1 & \multirow{5}{*}{60.8 $\pm$ 1.9} & 1.5 $\pm$ 0.0 & 4.5 $\pm$ 0.5 & \multirow{5}{*}{48.3 $\pm$ 0.5} & 0.0 $\pm$ 4.0 &  1.6 $\pm$ 0.5 & \multirow{5}{*}{48.5 $\pm$ 0.3} \\
         & & PrivSyn & 0.0 $\pm$ 0.0  & 0.0 $\pm$ 0.3 &  & 0.0 $\pm$ 1.6 & 2.4 $\pm$ 0.4  & & 0.4 $\pm$ 1.0 & 1.3 $\pm$ 0.5 & \\
         & & PATE-GAN & -2.8 $\pm$ 0.5  & -3.4 $\pm$ 0.6 &  & 1.8 $\pm$ 3.1 &  1.7 $\pm$ 0.5 & & 4.0 $\pm$ 5.1 & 2.5 $\pm$ 0.6 & \\
         & & PrePATE-GAN & 3.0 $\pm$ 1.3  & 3.0 $\pm$ 0.3 &  & 1.8 $\pm$ 0.9 &  3.6 $\pm$ 0.3 &  & 5.3 $\pm$ 2.3 & 5.6 $\pm$ 1.9 & \\
         & & \textbf{\toolnametran}  & \cellcolor{gray!20}{38.7 $\pm$ 4.9}  & \cellcolor{gray!20}{48.8 $\pm$ 9.7} &  & \cellcolor{gray!20}{25.2 $\pm$ 0.7} & \cellcolor{gray!20}{36.9 $\pm$ 2.4} & & \cellcolor{gray!20}{27.4 $\pm$ 3.2} &  \cellcolor{gray!20}{45.2 $\pm$ 3.2} & \\
        \bottomrule
    \end{tabular}
    }}
    \label{tab:downstream_offline_rl}
    \vspace{-2mm}
\end{table*}

\section{EMPIRICAL EVALUATIONS}
\label{sec:eval}
This section first compares the effectiveness of \toolnametran and \toolnametraj with baselines in downstream tasks. Then, we assess the fidelity between the synthetic and real datasets. \rev{Next, we evaluate the DP-protected synthetic datasets against MIA and analyze the impact of hyperparameters and privacy budgets on our methods.} Finally, we conduct an ablation study to emphasize the importance of curiosity and pre-training.

\subsection{The Utility of Synthetic Datasets}
\label{sec:utility}
\noindent \textbf{Experiment Design}. This experiment evaluates the utility of the DP synthetic dataset with the size of $1 \times 10^6$ transitions for transition-level synthesis and $5 \times 10^3$ trajectories for trajectory-level synthesis using \toolnametran and \toolnametraj. We also compare the utility of synthetic and real datasets. The data synthesizer is pre-trained for ten epochs on a public dataset and then fine-tuned for five epochs on a sensitive dataset, under $\epsilon = \{1, 10\}$. The privacy budget $\delta$ is not sensitive in our analysis and is set to $1 \times 10^{-6}$ across all experiments~\cite{opacus}. An epoch means one complete pass through the entire dataset. The hyper-parameter of the curiosity rate is set at 0.3. We provide further details on the selection of pre-training and sensitive datasets in Appendix~\ref{app:task_dataset}. We train agents for $5 \times 10^5$ timesteps. Appendix~\ref{app:task_dataset} notes the small trajectory size in the {\tt MuJoco} domain dataset, so we exclude it from trajectory synthesis.

\begin{table*}[!t]
\small
    \centering
    \caption{The normalized returns of agents trained on synthetic trajectories using \toolnametraj and baselines under $\epsilon = \{1,10\}$ and real datasets. `PrivORL-j-U' denotes a variant of PrivORL-j that uses a U-Net as the noise prediction network instead of a transformer, while retaining the other components. }
    \setlength{\tabcolsep}{3.0mm}{
    \resizebox{0.99\textwidth}{!}{
   \begin{tabular}{p{1.2cm}|p{0.9cm}|l|c|c|c|c|c|c|c|c|c}
        \toprule
         \multirow{2}{*}{\textbf{Domains}}   &  \multirow{2}{*}{\begin{minipage}{1.3cm}\textbf{Real} \\ \textbf{Dataset}\end{minipage}}  & \multirow{2}{*}{\textbf{Methods}} & \multicolumn{3}{c|}{\textbf{EDAC}~\cite{an2021uncertainty}} & \multicolumn{3}{c|}{\textbf{IQL}~\cite{iql}} & \multicolumn{3}{c}{\textbf{TD3PLUSBC}~\cite{td3plusbc}} \\
         \cline{4-12}
          &  & & $\epsilon = 1$ & $\epsilon = 10$  & no-DP & $\epsilon = 1$ & $\epsilon = 10$  & no-DP & $\epsilon = 1$ & $\epsilon = 10$ & no-DP \\
         \midrule
         \toprule
         \multirow{9}{2em}{{\tt Maze2D}}
         &\multirow{3}{*}{umaze} & PrivORL-j-U & 11.2 $\pm$ 9.0  & 32.0 $\pm$ 6.1 & \multirow{3}{*}{70.8 $\pm$ 13.6} &  24.6 $\pm$ 4.4 & 28.9 $\pm$ 5.0  &   \multirow{3}{*}{71.0 $\pm$ 2.3} & 26.8 $\pm$ 3.7 & 38.8 $\pm$ 2.9 & \multirow{3}{*}{73.6 $\pm$ 3.2} \\
         & & DP-Transformer  &  28.4 $\pm$ 4.1 & 39.8 $\pm$ 5.1 &  &  25.9 $\pm$ 6.5 & 41.2 $\pm$ 3.3  &   & 28.2 $\pm$ 2.9   &  49.3 $\pm$ 2.6 &  \\
         & & \textbf{\toolnametraj}  &  \cellcolor{gray!20}{45.5 $\pm$ 9.1} & \cellcolor{gray!20}{
52.2 $\pm$ 2.6} &  &  \cellcolor{gray!20}{42.1 $\pm$ 2.4} &  \cellcolor{gray!20}{49.8 $\pm$ 6.8} &   & 
\cellcolor{gray!20}{38.7 $\pm$ 6.5} & \cellcolor{gray!20}{49.9 $\pm$ 4.9} &  \\
         \cline{2-12}
         & \multirow{3}{*}{medium} & PrivORL-j-U & 13.5 $\pm$ 1.2 & 16.4 $\pm$ 5.1 & \multirow{3}{*}{73.0 $\pm$ 10.2} & 5.5 $\pm$ 1.0  &  14.1 $\pm$ 2.9 &   \multirow{3}{*}{93.1 $\pm$ 10.7} & 6.7 $\pm$ 0.2 & 10.8 $\pm$ 1.6 & \multirow{3}{*}{53.3 $\pm$ 0.3}\\
         & & DP-Transformer & 10.2 $\pm$ 0.4  &  19.0 $\pm$ 4.8 &  &  18.1 $\pm$ 2.2 & 32.8 $\pm$ 4.4  & & 7.6 $\pm$ 1.0  & 26.6 $\pm$ 0.4 &  \\
         & & \textbf{\toolnametraj}  & \cellcolor{gray!20}{31.5 $\pm$ 1.1}  & \cellcolor{gray!20}{35.0 $\pm$ 3.4} &  & \cellcolor{gray!20}{23.4 $\pm$ 5.2}  &  \cellcolor{gray!20}{49.3 $\pm$ 1.7} &   & \cellcolor{gray!20}{31.1 $\pm$ 10.6}&\cellcolor{gray!20}{ 38.0 $\pm$ 1.6}&  \\
         \cline{2-12}
         & \multirow{3}{*}{large} & PrivORL-j-U  & 10.9 $\pm$ 0.6  & 19.8 $\pm$ 0.1 & \multirow{3}{*}{89.9 $\pm$ 6.3} & 9.8 $\pm$ 0.4  & 31.9 $\pm$ 1.2  &   \multirow{3}{*}{85.9 $\pm$ 0.2} & 9.9 $\pm$ 0.3 & 14.3 $\pm$ 1.2 & \multirow{3}{*}{96.8 $\pm$ 3.2} \\
         & & DP-Transformer & 7.6 $\pm$ 3.5  & 21.8 $\pm$ 2.4 &  & 4.9 $\pm$ 0.8  & 23.4 $\pm$ 3.1  &   & 6.3 $\pm$ 0.2 & 28.3 $\pm$ 3.2 &  \\
         & & \textbf{\toolnametraj}  & \cellcolor{gray!20}{20.5 $\pm$ 0.2} & \cellcolor{gray!20}{30.5 $\pm$ 6.0}  &  &  \cellcolor{gray!20}{9.9 $\pm$ 0.3}& \cellcolor{gray!20}{37.7 $\pm$ 7.0}  &  & \cellcolor{gray!20}{29.5 $\pm$ 6.6} & \cellcolor{gray!20}{35.6 $\pm$ 2.6} &  \\
         \midrule
         \toprule
         \multirow{3}{2em}{{\tt Kitchen}}
         & \multirow{3}{*}{partial} & PrivORL-j-U &  0.0 $\pm$ 0.0  &  0.0 $\pm$ 0.0 & \multirow{3}{*}{10.0 $\pm$ 0.0} &  0.0 $\pm$ 0.0  &   0.0 $\pm$ 0.0 &   \multirow{3}{*}{40.0 $\pm$ 2.5} & 0.0 $\pm$ 0.0 & 0.0 $\pm$ 0.0 & \multirow{3}{*}{18.0 $\pm$ 6.0} \\
         & & DP-Transformer &   0.0 $\pm$ 0.0  &  0.0 $\pm$ 0.0  &  &  0.0 $\pm$ 0.0   & 2.5 $\pm$ 1.8   &   &  0.0 $\pm$ 0.0 &  0.0 $\pm$ 0.0 &  \\
         & & \textbf{\toolnametraj}  &  \cellcolor{gray!20}{
0.0 $\pm$ 0.0}  &  \cellcolor{gray!20}{0.0 $\pm$ 0.0} &   &  \cellcolor{gray!20}{7.5 $\pm$ 1.0}   & \cellcolor{gray!20}{13.8 $\pm$ 7.5}  &   & \cellcolor{gray!20}{5.0 $\pm$ 1.8} &\cellcolor{gray!20}{8.3 $\pm$ 2.5}&  \\
        \bottomrule
    \end{tabular}
    }}
    \label{tab:trajecotory_offline_rl}
\end{table*}

\vspace{1mm}
\noindent \textbf{Result Analysis.} For transition-level DP synthesis, Table~\ref{tab:downstream_offline_rl} presents that agents trained on synthetic transitions using \toolnametran achieve the highest normalized returns across all tasks. In the {\tt Maze2D} domain, under $\epsilon = \{1,10\}$, agents trained on synthetic transitions from \toolnametran achieve an average normalized return of 51.9 $(=(63.2+32.0+62.4+60.1+73.5+50.3+30.6+33.8+61.3)/9)$ 
and 69.3 $(=(69.1+45.6+80.6+70.3+90.7+81.0+60.3+50.4+75.3)/9)$ 
across three types of sensitive datasets. This performance surpasses the baseline returns of 11.7, 3.4, 2.0, 3.2, and 18.9, 3.9, 5.1, 7.2 (calculated similarly to that of \toolnametran), from PGM, PrivSyn, PATE-GAN, and PrePATE-GAN, under $\epsilon = \{1,10\}$. In {\tt Maze2D}, where the transition dimension is low at just 11, baselines perform well in certain scenarios; for example, agents trained on synthetic transitions from PGM using the IQL in {\tt Maze2D-medium} achieve a normalized return of 41.6. In {\tt Kitchen} and {\tt MujoCo}, where the transition dimensions are 130, the baseline methods struggle to perform effectively. \toolnametran achieves  average return scores of 13.1 $(=(2.5+25.5+11.5)/3)$ and 43.6 $(=(48.8+36.9+45.2)/3)$, under $\epsilon = 10$, outperforming baselines.

\begin{table*}[!t]
\small
\vspace{4.5mm}
    \centering
    \caption{Comparison of fidelity metrics of the synthetic transitions using \toolnametran and baselines ($\epsilon = 10$) to real datasets. The highest scores are highlighted in bold font. `Margin.' and `Correlat.' are abbreviations for `Marginal' and `Correlation.' }
    \setlength{\tabcolsep}{3.8mm}{
    \resizebox{0.99\textwidth}{!}{
   \begin{tabular}{l|c|cc|cc|cc|cc|cc}
        \toprule
            \multirow{2}{*}{\textbf{Domains}} & \multirow{2}{*}{\begin{minipage}{1.3cm}\centering\textbf{Real} \\ \textbf{Dataset}\end{minipage}}  & \multicolumn{2}{c|}{\textbf{PGM}~\cite{PGM}} & \multicolumn{2}{c|}{\textbf{PrivSyn}~\cite{zhang2021privsyn}} & \multicolumn{2}{c|}{\textbf{PATE-GAN}~\cite{PATE-GAN}} & \multicolumn{2}{c|}{\textbf{PrePATE-GAN}} & \multicolumn{2}{c}{\textbf{\toolnametran}} \\
         \cline{3-12}
          & &  Margin. & Correlat.  & Margin. & Correlat.   & Margin. & Correlat. & Margin. & Correlat. & Margin. & Correlat. \\
         \midrule
         \multirow{3}{*}{{\tt Maze2D}} & umaze & 0.793 & 0.983 & 0.784 & 0.969   & 0.672  &  0.844 & 0.784 &  0.969& \textbf{0.948} & \textbf{0.994}\\
         & medium  & 0.801 & \textbf{0.995} & 0.763 &  0.973  & 0.721  &  0.911 & 0.763 & 0.793 & \textbf{0.947} & 0.983 \\
         & large & 0.803 & 0.982 & 0.832 &  0.997  & 0.589  & 0.912  & 0.713 & 0.974 & \textbf{0.937} & \textbf{0.997}\\
         \midrule
         \multirow{1}{*}{{\tt Kitchen}} & partial & 0.783 & 0.697 & 0.737 &   0.806 & 0.647  & 0.788  & 0.695 & 0.819 & \textbf{0.861} & \textbf{0.901} \\
         \midrule
         \multirow{1}{*}{{\tt Mujoco}} & halfcheetah & 0.776 &  0.921 & 0.862 &  0.946  & 0.768  &  0.937 & 0.849 & 0.945 &\textbf{0.949} & \textbf{0.982} \\
         \hline
        \multicolumn{2}{c|}{ \textbf{Average}}  & 0.791   &  0.916 &  0.796 &  0.938  &  0.679 &  0.878 &  0.761 &  0.900 &  \textbf{0.928} &  \textbf{0.971} \\
        \bottomrule
    \end{tabular}
     }
     }
    \label{tab:mar_and_corr}
    \vspace{-2mm}
\end{table*}

\rev{Table~\ref{tab:pretrain} presents the performance of synthesizers only pre-trained on public datasets, without fine-tuning on sensitive datasets. We observe that fine-tuning on sensitive datasets significantly improves the synthetic quality. Synthesizers trained only on the pre-training datasets of {\tt Maze2D-umanze} achieve only 6.6 downstream agent performance. Although public datasets may resemble sensitive datasets, the training objectives differ substantially. For instance, the goal position and the layout map in {\tt Maze2D} vary across datasets. Thus, a synthesizer pretrained solely on public data may not generalize well to sensitive datasets, as the optimal trajectories and MDP (introduced in Section~\ref{subsec:rl}) are task-specific. }

For trajectory-level synthesis, Table~\ref{tab:trajecotory_offline_rl} shows that \toolnametraj achieves better performance than baselines. In the {\tt Maze2D}, under $\epsilon = \{1,10\}$, agents trained on synthetic trajectories from \toolnametraj achieve average normalized returns of 30.2 and 41.8, surpassing 13.2 and 23.0 obtained by PrivORL-j-U, and DP-Transformer’s 15.2 and 31.4. Thus, \toolnametraj achieves average 15.0 and 10.4 higher returns than DP-Transformer. In {\tt Kitchen}, \toolnametraj still outperforms baselines, presenting its better ability to handle complex trajectories.

Agents trained on real datasets achieve average returns of 78.6, 22.6, and 52.5 across {\tt Maze2D}, {\tt Kitchen}, and {\tt MujoCo}, while \toolnametran obtains 69.3, 13.0, and 43.5 at $\epsilon = 10$. The synthetic dataset exhibits comparable utility to the real dataset.  
We find that in {\tt Kitchen}, the agents trained using EDAC have 0.0 averaged returns.
Prior works~\cite{d4rl,tarasov2022corl,d3rlpy} show that no single existing offline RL algorithm excels across all datasets due to the inherent instability training problem in offline RL. Thus, it is usual for certain methods to appear ineffective for specific tasks. These results present that \toolnametran and \toolnametraj both achieve better synthetic utility than baselines.

\begin{table}[!t]
\small
    \centering
    \caption{\rev{Average normalized returns of agents trained on synthetic transitions ($\epsilon = 10$) using \toolnametran, with and without fine-tuning on sensitive datasets under IQL algorithm.}}
    \setlength{\tabcolsep}{4.0mm}{
    \resizebox{0.49\textwidth}{!}{
   \begin{tabular}{p{1.3cm}|c|cc}
        \toprule
            \multirow{1}{*}{\textbf{Domains}} & \multirow{1}{*}{\textbf{Dataset}}  & \textbf{Only pretraining}  & \textbf{\toolnametran} \\
         \midrule
         \multirow{3}{*}{{\tt Maze2D}} & umaze & 6.6 $\pm$ 0.0 & 70.3 $\pm$ 2.1  \\
         & medium & 6.6 $\pm$ 1.1 & 90.7 $\pm$ 8.6 \\
         & large & 7.8 $\pm$ 4.7 & 81.0 $\pm$ 11.8  \\
         \midrule
         \multirow{1}{*}{{\tt Kitchen}} & partial & 0.0 $\pm$ 0.0 & 25.5 $\pm$ 2.5 \\
         \midrule
         \multirow{1}{*}{{\tt Mujoco}} & halfcheetah & 14.4 $\pm$ 3.7 & 36.9 $\pm$ 2.4  \\
        \bottomrule
    \end{tabular}
    }}
    \label{tab:pretrain}
    \vspace{-1mm}
\end{table}

\subsection{The Fidelity of Synthetic Datasets}

\noindent \textbf{Experiment Design.} This section investigates whether \toolnametran and \toolnametraj can generate transitions and trajectories with greater fidelity than baselines under privacy budget $\epsilon = 10$. To visualize the synthetic distribution, we sample 500 transitions from each synthetic dataset and use t-SNE~\cite{tsne} to visualize them in a two-dimensional space.

\vspace{1mm}
\noindent \textbf{Result Analysis.} Table~\ref{tab:mar_and_corr} compares the marginal and correlation statistics of the synthetic transitions using \toolnametran under $\epsilon = 10$ to real datasets. We observe that \toolnametran outperforms the baselines in both marginal and correlation statistics. 
For correlation statistics, \toolnametran similarly excels with an average score of 0.971, exceeding baseline scores of 0.916, 0.938, 0.878, and 0.900. Marginal provides insights into the accuracy of single-variable distributions, while correlation effectively introduces synthetic data and preserves the relational structure observed in real data~\cite{massey1951kolmogorov,fieller1957tests}.
Based on these results, we conclude that while all methods effectively synthesize transitions with strong pairwise relationships between variables as observed in real transitions, \toolnametran achieves higher similarity in the distribution of individual variables between the synthetic and real transitions than the baseline methods. Table~\ref{tab:trajscores} in Appendix~\ref{supsubsec:fidelity_traj} presents that \toolnametraj achieves an average TrajScore of 0.902, which is 0.124
higher than DP-Transformer (0.778), under $\epsilon = 10$.

Figure~\ref{fig:tsne} shows the distribution of synthetic transitions from \toolnametran and baselines compared to real transitions. These results highlight the strength of \toolnametran strength in handling high-dimensional transitions. The number of transition dimensions is 11 and 130 in the {\tt Maze2D-medium} and {\tt Kitchen-partial} datasets. Baselines match the real distribution well in {\tt Maze2D-medium} but fail in {\tt Kitchen-partial}, while \toolnametran consistently generates transitions resembling the real dataset.

\begin{figure}[!t]
    \centering
    \setlength{\abovecaptionskip}{0pt}
    \includegraphics[width=0.98\linewidth]{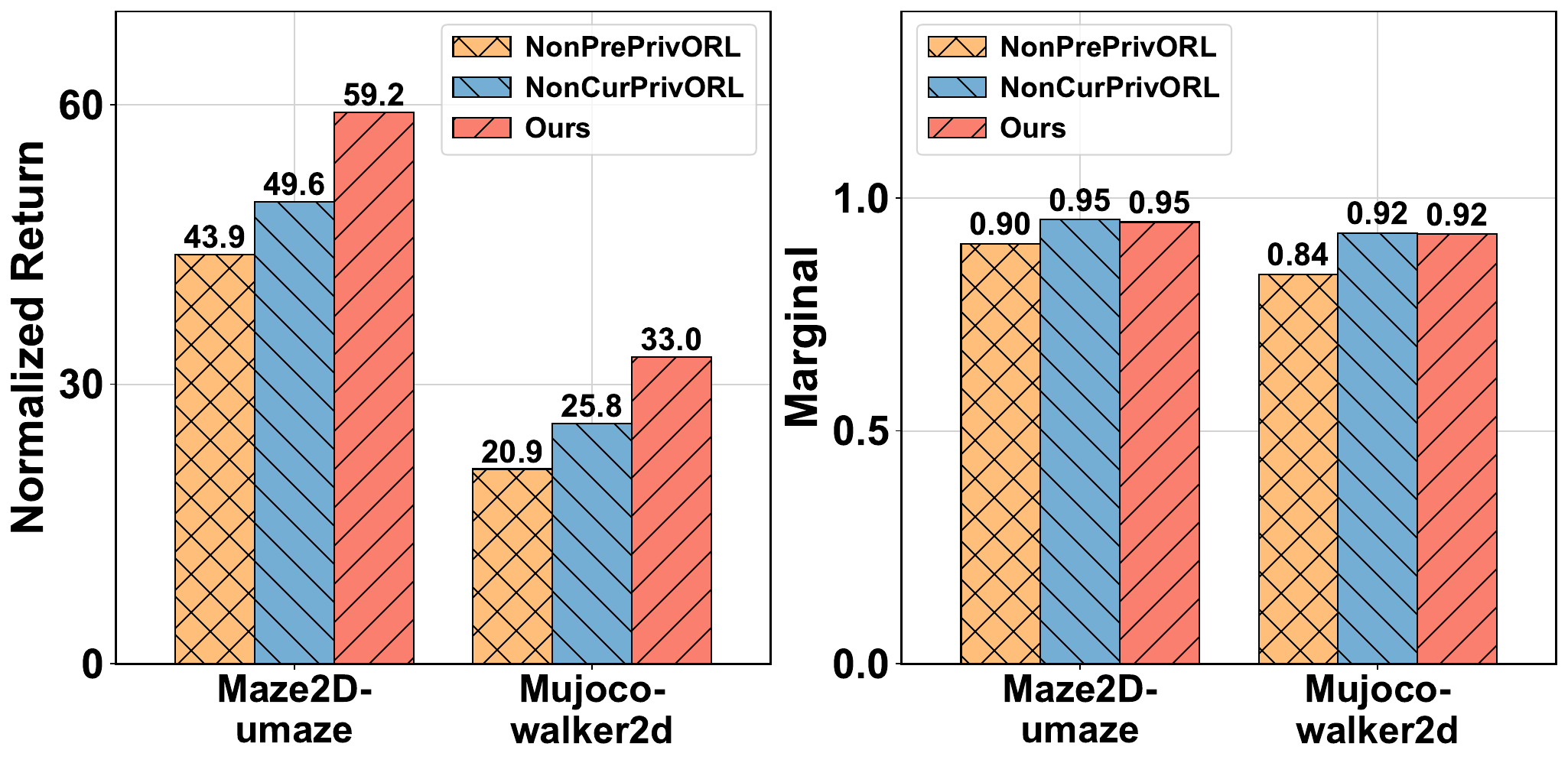}
    \caption{ The averaged normalized return across three RL algorithms under $\epsilon=10$ and marginal as achieved by (1) \toolnametran (Ours), (2) NonPrePrivORL, and (3) NonCurPrivORL.}
    \label{fig:abalation_performace}
    \vspace{-3mm}
\end{figure}

\subsection{Ablation Study}
\label{subsec:ablation}

\begin{figure*}[!t]
    \centering
    \includegraphics[width=0.98\linewidth]{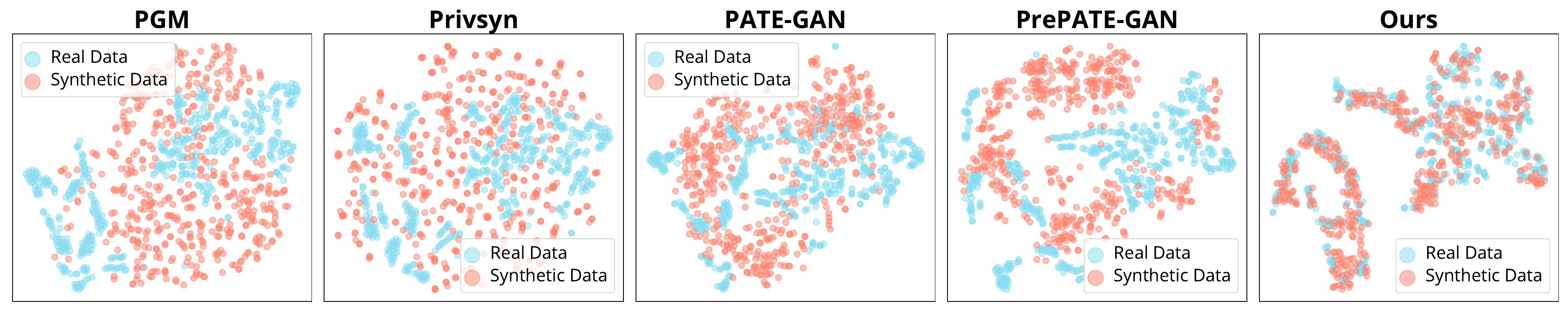}
    \caption{The t-SNE visualizations of data distribution for {\tt Kitchen-partial}. We synthesize the dataset using \toolnametran and baselines under $\epsilon=10$. We show the t-SNE visualizations for {\tt Maze2D-medium} in Figure~\ref{fig:tsne_maze2d} of Appendix~\ref{appsubsec:tsne}.}
    \label{fig:tsne}
    \vspace{-3mm}
\end{figure*}

\begin{figure*}[!t]
    \centering
    \setlength{\abovecaptionskip}{0pt}
    \includegraphics[width=1.0\linewidth]{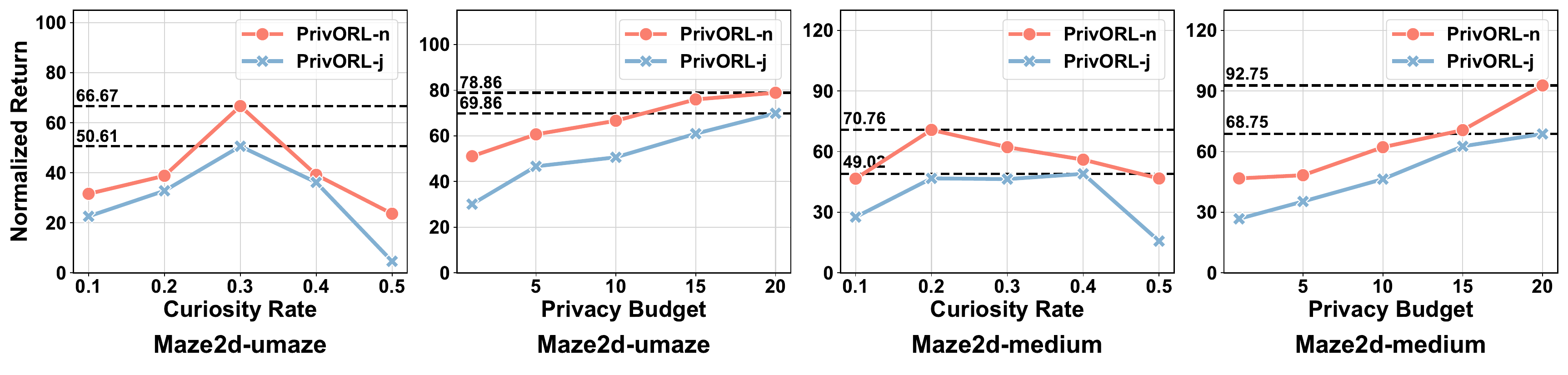}
    \caption{The average normalized returns of agents trained using three offline algorithms on synthetic transitions and trajectories using \toolnametran and \toolnametraj. These experiments and trajectories vary with different curiosity rates and privacy budgets. }
    \label{fig:hyper-parameter}
    \vspace{-4mm}
\end{figure*}

\noindent \textbf{Experiment Design.} These studies aim to explore the importance of pre-training and curiosity modules and how the \toolname performs while we introduce the curiosity module in the fine-tuning stage. We conduct experiments on transition synthesis. We conduct experiments under $\epsilon=10$. As mentioned in Section~\ref{sub:baselines}, `NonCurPrivORL' and `NonPrePrivORL' are equal to applying DPDM and PDP-Diffusion.

\noindent \textbf{Result Analysis.} Figure~\ref{fig:abalation_performace} shows the utility and fidelity of synthetic transitions using various methods. Notably, in {\tt Maze2D-umaze}, the removal of the pre-training stage and the curiosity module from \toolnametran leads to a decrease in average normalized returns of trained agents—by 25.9\% $(1 -  43.9/59.2) \times 100\%$ and 16.3\% $(1 -  49.6/59.2) \times 100\%$, and  Besides, 36.7\% $=(1 -  20.9/33.0) \times 100\%$ and 21.9\% $=(1 -  25.8/33.0) \times 100\%$ for {\tt Mujoco-walker2d}. Table~\ref{tab:FineCurPrivTranR} shows that using FineCurPrivORL leads to a degradation of the utility of synthetic transitions. Specifically, the average returns decrease from 70.3, 90.7, 81.0, 25.5, 36.9 to 54.1, 65.9, 54.1, 5.2, 18.7. The curiosity module adds novelty feedback to the training, increasing variability, which destabilizes fine-tuning.

\subsection{\rev{Defending against MIAs}}
\label{subsec:defending}

\begin{table}[!t]
\small
    \centering
    \caption{The normalized returns of agents trained on a DP synthetic dataset using IQL ($\epsilon=10$). }
    \resizebox{0.49\textwidth}{!}{
   \begin{tabular}{p{2.3cm}|ccc|c|c}
        \toprule
            \multirow{2}{*}{\textbf{Method}}   & \multicolumn{3}{c|}{\textbf{Maze}} & \multicolumn{1}{c|}{\textbf{Kitchen}} & \multicolumn{1}{c}{\textbf{Mujoco}} \\
         \cline{2-6}
          &  umaze & medium & large &  umaze & halfcheetah  \\
         \midrule
         FineCurPrivTranR & 54.1 & 65.9  & 54.1 & 5.2 & 18.7   \\
        Ours & 70.3 & 90.7 & 81.0 & 25.5 & 36.9  \\
        \bottomrule
    \end{tabular}
     }
    \label{tab:FineCurPrivTranR}
    \vspace{-6mm}
\end{table}

\rev{\noindent \textbf{Experiment Design.} We explore whether \toolnametran can defend against MIAs under DP protection. 
We use a white-box MIA~\cite{2023whithmiadiffusion} to attack synthesizers, aiming to identify members of a sensitive dataset using the synthetic dataset. We use True Positive Rate@10\%False Positive Rate (TPR@10\%FPR) and TPR@1\%FPR to evaluate the attacker's performance,
and a higher metric means a higher attack success rate. 
Please refer to more details of implementations of MIA in~\cite{2023whithmiadiffusion}. Following prior research~\cite{2023whithmiadiffusion,li2023meticulously}, the fixed FPR is set as a low rate, e.g., TPR@1\%FPR indicates the FPR threshold at 1\%. We introduce more about this metric in Appendix~\ref{supsubsec:metrics}.}

\rev{\noindent \textbf{Result Analysis.} We present the TPR@10\%FPR and TPR@1\%FPR of the MIA for diffusion models~\cite{2023whithmiadiffusion} under $\epsilon = \{1,10,\infty\}$ in Table~\ref{tab:mia}, where ``$\infty$'' means without DP protection. From this table, we observe that the MIA is ineffective against \toolnametran, approximating the effectiveness of random guessing. These results show that \toolnametran can synthesize transitions without privacy leakage of real sensitive transitions. Setting the privacy budget at 10 provides an effective defense with negligible differences compared to a budget set at 1. Although as $\epsilon$ increases, \toolnametran presents a reduction in synthetic transitions utility. We still advise using smaller $\epsilon$ values, such as 10, as recommended by previous research~\cite{li2023meticulously,dp-diffusion,dockhorn2023differentially}, to defend against unknown attacks. }

\subsection{Hyper-parameter and Privacy Budget}
\label{subsec:hyper}

\noindent \textbf{Experiment Design.} This experiment studies how \toolnametran and \toolnametraj perform under different hyper-parameter settings: (1) curiosity rate, $p = \{0.1,0.2,0.3,0.4,0.5\}$, (2) privacy budget, $\epsilon = \{1, 5, 10, 15, 20\}$, (3) horizon for \toolnametraj, $H=\{8,16,32,64,128\}$ (presented in Appendix~\ref{supsubsec:hyper}).

\begin{table}[!t]
\small
    \centering
    \caption{\rev{The TPR@10\%FPR / TPR@1\%FPR (\%) of MIA~\cite{2023whithmiadiffusion} under different $\epsilon$. The privacy budget ``$\infty$'' means training \toolnametran without DP protection.}}
    \setlength{\tabcolsep}{3.5mm}{
    \resizebox{0.49\textwidth}{!}{
   \begin{tabular}{p{1.3cm}|c|ccc}
        \toprule
            \multirow{2}{*}{\textbf{Domains}} & \multirow{2}{*}{\begin{minipage}{1.3cm}\centering\textbf{Real} \\ \textbf{Dataset}\end{minipage}}  & \multicolumn{3}{c}{\textbf{$\epsilon$}} \\
         \cline{3-5}
          & &  1 & 10 & $\infty$ \\
         \midrule
         \multirow{3}{*}{{\tt Maze2D}} & umaze & 11.9 / 1.5 & 11.4 / 1.9 & 31.0 / 12.8 \\
         & medium & 9.2 / 0.7 & 9.7 / 0.8 & 30.9 / 9.2\\
         & large & 8.9 / 0.7 & 10.9 / 0.6 & 27.6 / 8.9 \\
         \midrule
         \multirow{1}{*}{{\tt Kitchen}} & partial & 10.6 / 1.6 & 10.1 / 1.4 & 96.4 / 86.0 \\
         \midrule
         \multirow{1}{*}{{\tt Mujoco}} & halfcheetah & 10.8 / 1.7 & 11.0 / 1.6 & 95.9 / 86.0\\
        \bottomrule
    \end{tabular}
    }}
    \label{tab:mia}
    \vspace{-2mm}
\end{table}

\noindent \textbf{Result Analysis.} Figure~\ref{fig:hyper-parameter} presents the average normalized returns of agents trained using the studied algorithms. We observe that the performance of \toolnametran and \toolnametraj both initially increase but then decrease as curiosity rates continue to rise. ``There is no such thing as a free lunch.'' Excessively pursuing diversity can also increase the risk of generating erroneous data and affect the training. This highlights the need for a moderate curiosity rate for best results. Optimal curiosity rates vary across different tasks to maximize synthesis performance, e.g., for {\tt Maze2D-umaze} and {\tt Maze2D-medium}, the optimal curiosity rates for \toolnametran are 0.3 and 0.2, and for \toolnametraj are 0.3 and 0.3. We observe that with an increase in the privacy budget $\epsilon$, there is a corresponding upward trend in the performance of the synthetic data. As the privacy budget $\epsilon$ increases, introducing less noise during the training of synthesizers, the quality of synthetic datasets shows improvements~\cite{li2023meticulously,zhang2021privsyn}. 

\rev{Selecting an appropriate curiosity rate for different datasets is crucial. However, tuning the curiosity rate requires repeated access to the sensitive dataset~\cite{dp}, which can degrade the overall privacy guarantee and negatively impact the quality of the synthetic data. A practical mitigation strategy is to tune the curiosity rate on a public or non-sensitive dataset and then apply the optimal value to the sensitive data. Figure~\ref{fig:hyper-parameter} shows that a curiosity rate of approximately 0.3 achieves strong performance across datasets. Therefore, we adopt this fixed value without further tuning on sensitive data.}

\begin{table}[!t]
\small
    \centering
    \caption{The normalized return and marginal of synthetic transitions as achieved by (1) our method ($\epsilon=10$), \toolnametran, and (2) NonPrivORL ($\epsilon=\infty$). `Difference' means the results of NonPrivORL minus that of \toolnametran. }
    \resizebox{0.48\textwidth}{!}{
   \begin{tabular}{p{1.6cm}|cc|c|cc|c}
        \toprule
         \multirow{3}{*}{\textbf{Method}} & \multicolumn{3}{c|}{\textbf{Normalized return}} & \multicolumn{3}{c}{\textbf{Marginal}} \\
        \cline{2-7}
           & \multicolumn{2}{c|}{\textbf{Maze}} & \multicolumn{1}{c|}{\textbf{Mujoco}} & \multicolumn{2}{c|}{\textbf{Maze}} & \multicolumn{1}{c}{\textbf{Mujoco}} \\
         \cline{2-7}
          &  medium & large &  halfcheetah &  medium & large &  halfcheetah  \\
         \midrule
        NonPrivORL & 74.8 & 80.5  & 54.9 & 0.98 & 0.97  & 0.97 \\
        Ours & 66.7 & 79.0 & 43.6 & 0.95 & 0.94  & 0.95 \\
        \hline
        \rowcolor{gray0} Difference  &  8.1 & 1.5 & 11.3 & 0.03 & 0.03  & 0.02  \\
        \bottomrule
    \end{tabular}
     }
    \label{tab:7.1}
    \vspace{-1mm}
\end{table}

\subsection{\toolname without Privacy Protection}
\label{supsubsec:without_protection}
\noindent \textbf{Experiment Design.} This experiment focuses on DP transition synthesis. We explore how the DP harms the synthetic performance of \toolnametran ($\epsilon=10$). We compare \toolnametran with the method `NonPrivORL', which fine-tunes diffusion models on sensitive datasets without DP protection (i.e., $\epsilon=\infty$).

\vspace{1mm}
\noindent \textbf{Result Analysis.} Table~\ref{tab:7.1} illustrates that \toolnametran, on average, leads to a 7.0 $(=(8.1+1.5+11.3)/3)$ reduction in normalized returns for agents trained using four different algorithms, to meet the requirements of the DP constraint. However, in terms of high-level statistics, \toolnametran results in only a 0.03 $(=(0.03+0.03+0.02)/3)$ decrease in the marginal. It is understood that incorporating the DP framework may damage transition synthesis performance, as observed in various prior studies~\cite{li2023meticulously,dp-diffusion}. For example, in DP image synthesis, PrivImage also experiences an approximate 11.6\% performance reduction in downstream tasks under $\epsilon = 10$. Therefore, a reduction in performance is deemed acceptable for privacy protection. These results indicate that further development is necessary to mitigate the performance reduction caused by DP. \rev{Besides, Appendix~\ref{subapp:curve} shows learning curves of the downstream agent over training steps. }

We discuss the efficiency of \toolnametran compared to baselines and the application scenario of offline RL dataset synthesis in Appendix~\ref{app:eff} and~\ref{app:app_sce}, respectively. \rev{Besides, Appendix~\ref{supsubsec:scaling_up} shows how scaling up improves the performance of synthetic transitions generated by \toolnametran. We discuss the limitations of our methods in Appendix~\ref{supsubsec:limitation}.}

\vspace{-1mm}
\section{Related Work}
\label{sec:related}

We discuss related work briefly here, and we provide a more comprehensive discussion of related works in Appendix~\ref{app:related}.

\vspace{1mm}
\noindent \textbf{DP Dataset Synthesis.} DP dataset synthesis methods fall into two categories: 
(1) \textit{Marginal-based approaches}, which replicate marginal and joint distributions of variables~\cite{zhang2021privsyn,PGM,privmrf}. These methods work well for small tabular data but struggle with discrete values, high-dimensional attributes, and large datasets~\cite{zhang2021privsyn}. 
(2) \textit{Generative model-based approaches}, which train DP synthesizers such as GANs~\cite{PATE-GAN,yin2022practical} and diffusion models~\cite{ddpm,dp-diffusion} using DP-SGD~\cite{dpsgd}. While PATE-GAN~\cite{PATE-GAN} and DP-CGAN~\cite{DPCGAN} perform well on {\tt MNIST}~\cite{mnist}, they degrade on larger datasets. Recent work shows diffusion models outperform GANs in DP synthesis~\cite{dockhorn2023differentially}. Sabra et al.~\cite{dp-diffusion} propose to pre-training DP-Diffusion on public data and fine-tune on sensitive data.

\vspace{1mm}
\noindent \textbf{Offline RL with Synthetic Data.} Diffusion models augment datasets with synthetic data, widely used in computer vision~\cite{li2023meticulously,yuan2024realfake} and suitable for offline RL to address data scarcity~\cite{offline_survey}. Recent works~\cite{yu2023scaling,chen2023genaug} augmented robotic observations with text-guided diffusion models. SynthER~\cite{lu2023synthetic} and MTDIFF~\cite{he2023diffusion} generate transitions with novel actions. Zhao et al.~\cite{zhao2024trajsyn} used transformers for trajectory synthesis. Prior work~\cite{zhu2024trajsynmadiff} explored multi-agent dataset synthesis. However, generative models still risk MIAs, a concern from images~\cite{2023whithmiadiffusion,carlini2023extracting} to offline RL datasets~\cite{du2023orl,pan2019you,gomrokchi2022membership,ye2023RLunlearning}.

\vspace{1mm}
\noindent \textbf{Applications of Offline RL.} Offline RL excels in healthcare~\cite{RL4Treatment,offline_rl_medicial}, energy management~\cite{RL4Energy,zhang2023mutual}, autonomous driving~\cite{RL4AutonomousVehicles,RL4AutonomousVehicles2}, and recommendation systems~\cite{RL4Recommender,RL4Recommender2}. In healthcare, ethical constraints limit online RL. Mila et al.~\cite{RL4Treatment} optimized diabetes treatment policies, while Emerson et al.~\cite{RL4BGC} determined insulin doses using offline RL. Offline RL enhances data efficiency in costly data collection scenarios like autonomous driving~\cite{RL4AutonomousVehicles,RL4AutonomousVehicles2}.

\section{Conclusions}
\label{sec:con}
This paper is the first to propose offline RL dataset synthesis under the DP framework and introduces \toolname, which allows users to create datasets from both transition and trajectory levels that closely resemble sensitive datasets while safeguarding their privacy under DP. We identify the importance of synthesizing diverse offline RL datasets and propose the curiosity module to improve the diversity of synthetic datasets.  \toolname initially pre-trains diffusion models on public datasets, using feedback from the curiosity module to diversify the synthetic datasets. The pre-trained model is then fine-tuned on sensitive datasets with DP-SGD. Finally, the fine-tuned diffusion model generates synthetic datasets. For trajectory-level DP synthesis, we introduce a transformer to the diffusion model to capture long-range temporal dependencies. To handle the high dimensionality of trajectories, we split trajectories into fragments and use a conditional synthesizer to model fragment relationships, enabling seamless trajectory stitching. Experiments show that \toolname achieves higher utility for various downstream agents' training in both transition and trajectory synthesis, compared to baselines. This work aims to advance the privacy-preserving sharing of offline RL datasets and further the progress of open offline RL research.

\bibliographystyle{IEEEtranS}
\bibliography{bib.bib}

@misc{d4rl,
      title={D4RL: Datasets for Deep Data-Driven Reinforcement Learning}, 
      author={Justin Fu and Aviral Kumar and Ofir Nachum and George Tucker and Sergey Levine},
      year={2021},
      eprint={2004.07219},
      archivePrefix={arXiv}
}

@misc{levine2020offline,
      title={Offline Reinforcement Learning: Tutorial, Review, and Perspectives on Open Problems}, 
      author={Sergey Levine and Aviral Kumar and George Tucker and Justin Fu},
      year={2020},
      eprint={2005.01643},
      archivePrefix={arXiv},
      primaryClass={cs.LG}
}

@book{sutton2018reinforcement,
  title     = {Reinforcement learning: An introduction},
  author    = {Sutton, Richard S and Barto, Andrew G},
  year      = {2018},
  publisher = {MIT press}
}

@inproceedings{RL4Treatment,
  author       = {Mila Nambiar and
                  Supriyo Ghosh and
                  Priscilla Ong and
                  others},
  title        = {Deep Offline Reinforcement Learning for Real-world Treatment Optimization
                  Applications},
  booktitle    = {Proceedings of the 29th {ACM} {SIGKDD} Conference on Knowledge Discovery
                  and Data Mining},
  year         = {2023},
}

@inproceedings{cql,
  author       = {Aviral Kumar and
                  Aurick Zhou and
                  George Tucker and
                  Sergey Levine},
  title        = {Conservative Q-Learning for Offline Reinforcement Learning},
  booktitle    = {NeurIPS},
  year         = {2020},
}

@article{RL4BGC,
  author       = {Harry Emerson and
                  Matthew Guy and
                  Ryan McConville},
  title        = {Offline reinforcement learning for safer blood glucose control in
                  people with type 1 diabetes},
  journal      = {J. Biomed. Informatics},
  year         = {2023},
}

@inproceedings{RL4Energy,
  author       = {Xianyuan Zhan and
                  Haoran Xu and
                  Yue Zhang and
                  others},
  title        = {DeepThermal: Combustion Optimization for Thermal Power Generating
                  Units Using Offline Reinforcement Learning},
  booktitle    = {AAAI},
  year         = {2022},
}

@article{RL4AutonomousVehicles,
  author       = {Lixian Zhang and
                  Ruixian Zhang and
                  Tong Wu and
                  others},
  title        = {Safe Reinforcement Learning With Stability Guarantee for Motion Planning of Autonomous Vehicles},
  journal      = {{IEEE} Trans. Neural Networks Learn. Syst.},
  year         = {2021},
}

@article{offline_survey,
	year = 2023,
	publisher = {Institute of Electrical and Electronics Engineers ({IEEE})},
	author = {Rafael Figueiredo Prudencio and Marcos R. O. A. Maximo and others},
	title = {A Survey on Offline Reinforcement Learning: Taxonomy, Review, and Open Problems},
	journal = {{IEEE} Transactions on Neural Networks and Learning Systems}
}

@inproceedings{offline_rl_medicial,
  author       = {Mehdi Fatemi and
                  Taylor W. Killian and
                  Jayakumar Subramanian and
                  others},
  title        = {Medical Dead-ends and Learning to Identify High-Risk States and Treatments},
  booktitle    = {Advances in Neural Information Processing Systems},
  pages        = {4856--4870},
  year         = {2021}
}

@inproceedings{RL4AutonomousVehicles2,
  author       = {Daniel Graves and
                  Nhat M. Nguyen and
                  Kimia Hassanzadeh and
                  others},
  title        = {Learning robust driving policies without online exploration},
  booktitle    = {{IEEE} International Conference on Robotics and Automation},
  publisher    = {{IEEE}},
  year         = {2021},
}

@misc{RL4Recommender,
      title={Causal Decision Transformer for Recommender Systems via Offline Reinforcement Learning}, 
      author={Siyu Wang and Xiaocong Chen and Dietmar Jannach and others},
      year={2023},
      eprint={2304.07920},
      archivePrefix={arXiv},
      primaryClass={cs.IR}
}

@inproceedings{RL4Recommender2,
  author       = {Qihua Zhang and
                  Junning Liu and
                  Yuzhuo Dai and
                  others},
  title        = {Multi-Task Fusion via Reinforcement Learning for Long-Term User Satisfaction
                  in Recommender Systems},
  booktitle    = {The 28th {ACM} {SIGKDD} Conference on Knowledge Discovery
                  and Data Mining},
  year         = {2022},
}

@article{d3rlpy,
  author  = {Takuma Seno and Michita Imai},
  title   = {d3rlpy: An Offline Deep Reinforcement Learning Library},
  journal = {Journal of Machine Learning Research},
  year    = {2022},
  volume  = {23}
}

@inproceedings{iql,
title={Offline Reinforcement Learning with Implicit Q-Learning},
author={Ilya Kostrikov and Ashvin Nair and Sergey Levine},
booktitle={International Conference on Learning Representations},
year={2022}
}

@inproceedings{td3plusbc,
 author = {Fujimoto, Scott and Gu, Shixiang (Shane)},
 booktitle = {Advances in Neural Information Processing Systems},
 pages = {20132--20145},
 publisher = {Curran Associates, Inc.},
 title = {A Minimalist Approach to Offline Reinforcement Learning},
 volume = {34},
 year = {2021}
}

@inproceedings{mujoco,
  title        = {Mujoco: A physics engine for model-based control},
  author       = {Todorov, Emanuel and Erez, Tom and Tassa, Yuval},
  booktitle    = {2012 IEEE/RSJ International Conference on Intelligent Robots and Systems},
  pages        = {5026--5033},
  year         = {2012},
  organization = {IEEE}
}

@misc{gomrokchi2022membership,
      title={Membership Inference Attacks Against Temporally Correlated Data in Deep Reinforcement Learning}, 
      author={Maziar Gomrokchi and Susan Amin and Hossein Aboutalebi and Alexander Wong and Doina Precup},
      year={2022},
      eprint={2109.03975},
      archivePrefix={arXiv},
      primaryClass={cs.LG}
}

@inproceedings{pan2019you,
  title={How You Act Tells a Lot: Privacy-Leaking Attack on Deep Reinforcement Learning.},
  author={Pan, Xinlei and Wang, Weiyao and Zhang, Xiaoshuai and others},
  booktitle={AAMAS},
  volume={19},
  number={2019},
  year={2019}
}

@article{ye2023RLunlearning,
  title={Reinforcement Unlearning},
  author={Ye, Dayong and Zhu, Tianqing and Zhu, Congcong and Wang, Derui and Shen, Sheng and Zhou, Wanlei and others},
  journal={arXiv:2312.15910},
  year={2023}
}

@article{zhang2023mutual,
  title={Mutual Information as Intrinsic Reward of Reinforcement Learning Agents for On-demand Ride Pooling},
  author={Zhang, Xianjie and Sun, Jiahao and Gong, Chen and others},
  journal={arXiv preprint arXiv:2312.15195},
  year={2023}
}

@inproceedings{li2023meticulously,
  title={$\{$PrivImage$\}$: Differentially private synthetic image generation using diffusion models with $\{$Semantic-Aware$\}$ pretraining},
  author={Li, Kecen and Gong, Chen and Li, Zhixiang and Zhao, Yuzhong and Hou, Xinwen and Wang, Tianhao},
  booktitle={33rd USENIX Security Symposium (USENIX Security 24)},
  pages={4837--4854},
  year={2024}
}

@article{
dockhorn2023differentially,
title={Differentially Private Diffusion Models},
author={Tim Dockhorn and Tianshi Cao and Arash Vahdat and others},
journal={Transactions on Machine Learning Research},
issn={2835-8856},
year={2023},
note={}
}

@inproceedings{yue-etal-2023-synthetic,
    title = "Synthetic Text Generation with Differential Privacy: A Simple and Practical Recipe",
    author = "Yue, Xiang  and
      Inan, Huseyin  and
      Li, Xuechen  and
      others",
    booktitle = "Proceedings of the 61st Annual Meeting of the Association for Computational Linguistics",
    month = jul,
    year = "2023",
}

@inproceedings{dpsgd,
  author       = {Mart{\'{\i}}n Abadi and Andy Chu and Ian J. Goodfellow and et al.},
  title        = {Deep Learning with Differential Privacy},
  booktitle    = {Proceedings of the 2016 {ACM} {SIGSAC} Conference on Computer and
                  Communications Security},
  pages        = {308-318},
  doi          = {10.1145/2976749.2978318},
}

@article{rdp,
  author       = {Ilya Mironov},
  title        = {Renyi Differential Privacy},
  journal      = {CoRR},
  volume       = {abs/1702.07476},
  year         = {2017},
}

@article{sgm,
  author       = {Ilya Mironov and
                  Kunal Talwar and
                  Li Zhang},
  title        = {R{\'{e}}nyi Differential Privacy of the Sampled Gaussian Mechanism},
  journal      = {CoRR},
  volume       = {abs/1908.10530},
  year         = {2019},
  eprinttype    = {arXiv},
  eprint       = {1908.10530},
}

@inproceedings{diffusionText1,
  author       = {Xiang Li and
                  John Thickstun and
                  Ishaan Gulrajani and
                  et al.},
  title        = {Diffusion-LM Improves Controllable Text Generation},
  booktitle    = {NeurIPS},
  year         = {2022}
}

@inproceedings{ddpm,
  author       = {Jonathan Ho and
                  Ajay Jain and
                  Pieter Abbeel},
  title        = {Denoising Diffusion Probabilistic Models},
  booktitle    = {Advances in Neural Information Processing Systems},
  year         = {2020},
}

@article{dp-diffusion,
  author       = {Sahra Ghalebikesabi and
                  Leonard Berrada and
                  Sven Gowal and
                  others},
  title        = {Differentially Private Diffusion Models Generate Useful Synthetic
                  Images},
  journal      = {CoRR},
  volume       = {abs/2302.13861},
  year         = {2023},
}

@inproceedings{hu2023sok,
  title={SoK: Privacy-Preserving Data Synthesis},
  author={Hu, Yuzheng and Wu, Fan and Li, Qinbin and Long, Yunhui and Garrido, Gonzalo and Ge, Chang and Ding, Bolin and Forsyth, David and Li, Bo and Song, Dawn},
  booktitle={2024 IEEE Symposium on Security and Privacy (SP)},
  pages={2--2}
}

@inproceedings{liew2022pearl,
title={{PEARL}: Data Synthesis via Private Embeddings and Adversarial Reconstruction Learning},
author={Seng Pei Liew and Tsubasa Takahashi and Michihiko Ueno},
booktitle={International Conference on Learning Representations},
year={2022}
}

@article{privmrf,
author = {Cai, Kuntai and Lei, Xiaoyu and Wei, Jianxin and Xiao, Xiaokui},
title = {Data Synthesis via Differentially Private Markov Random Fields},
year = {2021},
issue_date = {July 2021},
publisher = {VLDB Endowment},
volume = {14},
number = {11},
journal = {Proc. VLDB Endow.},
pages = {2190–2202},
numpages = {13}
}

@inproceedings{zhang2021privsyn,
  title={$\{$PrivSyn$\}$: Differentially Private Data Synthesis},
  author={Zhang, Zhikun and Wang, Tianhao and Li, Ninghui and others},
  booktitle={30th USENIX Security Symposium},
  pages={929--946},
  year={2021}
}

@inproceedings{dp,
  title={Calibrating noise to sensitivity in private data analysis},
  author={Dwork, Cynthia and McSherry, Frank and Nissim, Kobbi and Smith, Adam},
  booktitle={Theory of Cryptography: Third Theory of Cryptography Conference,},
  pages={265--284},
  year={2006}
}

@article{awac,
  author    = {Ashvin Nair and
               Murtaza Dalal and
               Abhishek Gupta and
               Sergey Levine},
  title     = {Accelerating Online Reinforcement Learning with Offline Datasets},
  journal   = {CoRR},
  eprinttype = {arXiv}
}

@inproceedings{kumar2023offlinepre,
  author       = {Aviral Kumar and
                  Rishabh Agarwal and
                  Xinyang Geng and
                  others},
  title        = {Offline Q-learning on Diverse Multi-Task Data Both Scales And Generalizes},
  booktitle    = {The Eleventh International Conference on Learning Representations},
  year         = {2023}
}

@inproceedings{PGM,
  author       = {Ryan McKenna and
                  Daniel Sheldon and
                  Gerome Miklau},
  title        = {Graphical-model based estimation and inference for differential privacy},
  booktitle    = {Proceedings of the 36th International Conference on Machine Learning,
                  {ICML}},
  volume       = {97},
  pages        = {4435--4444},
  publisher    = {{PMLR}},
  year         = {2019},
}

@inproceedings{
PATE-GAN,
title={{PATE}-{GAN}: Generating Synthetic Data with Differential Privacy Guarantees},
author={Jinsung Yoon and James Jordon and Mihaela van der Schaar},
booktitle={International Conference on Learning Representations},
year={2019}
}

@inproceedings{2023whithmiadiffusion,
  author       = {Tomoya Matsumoto and
                  Takayuki Miura and
                  Naoto Yanai},
  title        = {Membership Inference Attacks against Diffusion Models},
  booktitle    = {2023 {IEEE} Security and Privacy Workshops (SPW)},
  pages        = {77-83},
}

@inproceedings{
tarasov2022corl,
  title={{CORL}: Research-oriented Deep Offline Reinforcement Learning Library},
  author={Denis Tarasov and Alexander Nikulin and Dmitry Akimov and others},
  booktitle={3rd Offline RL Workshop: Offline RL as a ''Launchpad''},
  year={2022}
}

@inproceedings{
lu2023synthetic,
title={Synthetic Experience Replay},
author={Cong Lu and Philip J. Ball and Yee Whye Teh and Jack Parker-Holder},
booktitle={Thirty-seventh Conference on Neural Information Processing Systems},
year={2023}
}

@inproceedings{edm,
  author       = {Tero Karras and
                  Miika Aittala and
                  Timo Aila and
                  Samuli Laine},
  title        = {Elucidating the Design Space of Diffusion-Based Generative Models},
  booktitle    = {Advances in Neural Information Processing Systems},
  year         = {2022},
}

@inproceedings{mlp_mixer,
  author       = {Ilya O. Tolstikhin and
                  Neil Houlsby and
                  Alexander Kolesnikov and
                  et al.},
  title        = {MLP-Mixer: An all-MLP Architecture for Vision},
  booktitle    = {Advances in Neural Information Processing Systems},
  pages        = {24261--24272},
  year         = {2021},
}

@article{gupta2019relay,
  title={Relay policy learning: Solving long-horizon tasks via imitation and reinforcement learning},
  author={Gupta, Abhishek and Kumar, Vikash and Lynch, Corey and Levine, Sergey and Hausman, Karol},
  journal={arXiv preprint arXiv:1910.11956},
  year={2019}
}

@inproceedings{
yuan2024realfake,
title={Real-Fake: Effective Training Data Synthesis Through Distribution Matching},
author={Jianhao Yuan and Jie Zhang and Shuyang Sun and Philip Torr and Bo Zhao},
booktitle={The Twelfth International Conference on Learning Representations},
year={2024}
}

@article{massey1951kolmogorov,
  title={The Kolmogorov-Smirnov test for goodness of fit},
  author={Massey Jr, Frank J},
  journal={Journal of the American statistical Association},
  pages={68--78},
  year={1951},
  publisher={Taylor \& Francis}
}

@article{fieller1957tests,
  title={Tests for rank correlation coefficients. I},
  author={Fieller, Edgar C and Hartley, Herman O and Pearson, Egon S},
  journal={Biometrika},
  volume={44},
  pages={470--481},
  year={1957},
  publisher={JSTOR}
}

@article{yu2023scaling,
  title={Scaling robot learning with semantically imagined experience},
  author={Yu, Tianhe and Xiao, Ted and Stone, Austin and others},
  journal={arXiv:2302.11550},
  year={2023}
}

@article{chen2023genaug,
  title={Genaug: Retargeting behaviors to unseen situations via generative augmentation},
  author={Chen, Zoey and Kiami, Sho and Gupta, Abhishek and Kumar, Vikash},
  journal={arXiv preprint arXiv:2302.06671},
  year={2023}
}

@inproceedings{
he2023diffusion,
title={Diffusion Model is an Effective Planner and Data Synthesizer for Multi-Task Reinforcement Learning},
author={Haoran He and Chenjia Bai and Kang Xu and others},
booktitle={Thirty-seventh Conference on Neural Information Processing Systems},
year={2023},
url={https://openreview.net/forum?id=fAdMly4ki5}
}

@inproceedings{carlini2023extracting,
  title={Extracting training data from diffusion models},
  author={Carlini, Nicolas and Hayes, Jamie and Nasr, Milad and others},
  booktitle={32nd USENIX Security Symposium},
  pages={5253--5270},
  year={2023}
}

@misc{chaudhari2024rlhf,
      title={RLHF Deciphered: A Critical Analysis of Reinforcement Learning from Human Feedback for LLMs}, 
      author={Shreyas Chaudhari and Pranjal Aggarwal and Vishvak Murahari and others},
      year={2024},
      eprint={2404.08555},
      archivePrefix={arXiv},
      primaryClass={cs.LG}
}

@article{an2021uncertainty,
  title={Uncertainty-based offline reinforcement learning with diversified q-ensemble},
  author={An, Gaon and Moon, Seungyong and Kim, Jang-Hyun and Song, Hyun Oh},
  journal={Advances in neural information processing systems},
  pages={7436--7447},
  year={2021}
}

@misc{du2024systematic,
      title={Systematic Assessment of Tabular Data Synthesis Algorithms}, 
      author={Yuntao Du and Ninghui Li},
      year={2024},
      eprint={2402.06806},
      archivePrefix={arXiv},
      primaryClass={cs.CR}
}

@article{opacus,
  title={Opacus: {U}ser-Friendly Differential Privacy Library in {PyTorch}},
  author={Ashkan Yousefpour and Igor Shilov and Alexandre Sablayrolles and others},
  journal={arXiv:2109.12298},
  year={2021}
}

@inproceedings{yin2022practical,
  title={Practical gan-based synthetic ip header trace generation using netshare},
  author={Yin, Yucheng and Lin, Zinan and Jin, Minhao and Fanti, Giulia and Sekar, Vyas},
  booktitle={Proceedings of the ACM SIGCOMM 2022 Conference},
  pages={458--472},
  year={2022}
}

@inproceedings{rnd,
  title={Exploration by random network distillation},
  author={Burda, Yuri and Edwards, Harrison and Storkey, Amos and Klimov, Oleg},
  booktitle={International Conference on Learning Representations},
  year={2018}
}

@article{tsne,
  title={Visualizing data using t-SNE.},
  author={Van der Maaten, Laurens and Hinton, Geoffrey},
  journal={Journal of machine learning research},
  volume={9},
  number={11},
  year={2008}
}

@inproceedings{DPCGAN,
  author       = {Reihaneh Torkzadehmahani and
                  Peter Kairouz and
                  Benedict Paten},
  title        = {{DP-CGAN:} Differentially Private Synthetic Data and Label Generation},
  booktitle    = {{IEEE} Conference on Computer Vision and Pattern Recognition Workshops,
                  {CVPR} Workshops},
  pages        = {98-104},
  year         = {2019}
}

@article{mnist,
  author       = {Yann LeCun and
                  L{\'{e}}on Bottou and
                  Yoshua Bengio and
                  et al.},
  title        = {Gradient-based learning applied to document recognition},
  journal      = {Proc. {IEEE}},
  volume       = {86},
  number       = {11},
  pages        = {2278-2324},
  year         = {1998},
  url          = {https://doi.org/10.1109/5.726791},
  doi          = {10.1109/5.726791},
}

@inproceedings{he2024curiosity,
  title={Curiosity-Driven Testing for Sequential Decision-Making Process},
  author={He, Junda and Yang, Zhou and Shi, Jieke and others},
  booktitle={Proceedings of the IEEE/ACM 46th International Conference on Software Engineering},
  pages={1--14},
  year={2024}
}

@inproceedings{
hong2024curiositydriven,
title={Curiosity-driven Red-teaming for Large Language Models},
author={Zhang-Wei Hong and Idan Shenfeld and Tsun-Hsuan Wang and others},
booktitle={The Twelfth International Conference on Learning Representations},
year={2024}
}

@InProceedings{Wu_2023_ICCV,
    author    = {Wu, Weijia and Zhao, Yuzhong and Shou, Mike Zheng and Zhou, Hong and Shen, Chunhua},
    title     = {DiffuMask: Synthesizing Images with Pixel-level Annotations for Semantic Segmentation Using Diffusion Models},
    booktitle = {Proceedings of the IEEE/CVF International Conference on Computer Vision},
    year      = {2023},
    pages     = {1206-1217}
}

@InProceedings{pmlr-v70-arjovsky17a,
  title = 	 {{W}asserstein Generative Adversarial Networks},
  author =       {Martin Arjovsky and Soumith Chintala and L{\'e}on Bottou},
  booktitle = 	 {Proceedings of the 34th International Conference on Machine Learning},
  pages = 	 {214--223},
  year = 	 {2017},
  volume = 	 {70},
  month = 	 {06--11 Aug}
}

@article{scikit-learn,
  title={Scikit-learn: Machine Learning in {P}ython},
  author={Pedregosa, F. and Varoquaux, G. and Gramfort, A. and others},
  journal={Journal of Machine Learning Research},
  volume={12},
  pages={2825--2830},
  year={2011}
}

@article{rlhfmia,
  title={Towards label-only membership inference attack against pre-trained large language models},
  author={He, Yu and Li, Boheng and Liu, Liu and Ba, Zhongjie and Dong, Wei and Li, Yiming and Qin, Zhan and Ren, Kui and Chen, Chun},
  journal={arXiv preprint arXiv:2502.18943},
  year={2025}
}

@inproceedings{du2023orl,
  author       = {Linkang Du and
                  Min Chen and
                  Mingyang Sun and
                  others},
  title        = {{ORL-AUDITOR:} Dataset Auditing in Offline Deep Reinforcement Learning},
  booktitle    = {31st Annual Network and Distributed System Security Symposium, {NDSS}},
  year         = {2024}
}

@inproceedings{sun2024netdpsyn,
  title={Netdpsyn: synthesizing network traces under differential privacy},
  author={Sun, Danyu and Chen, Joann Qiongna and Gong, Chen and Wang, Tianhao and Li, Zhou},
  booktitle={Proceedings of the 2024 ACM on Internet Measurement Conference},
  pages={545--554},
  year={2024}
}

@article{zhao2024trajsyn,
  title={Offline Trajectory Generalization for Offline Reinforcement Learning},
  author={Zhao, Ziqi and Ren, Zhaochun and Yang, Liu and others},
  journal={arXiv preprint arXiv:2404.10393},
  year={2024}
}

@article{zhu2024trajsynmadiff,
  title={Madiff: Offline multi-agent learning with diffusion models},
  author={Zhu, Zhengbang and Liu, Minghuan and Mao, Liyuan and Kang, Bingyi and Xu, Minkai and Yu, Yong and Ermon, Stefano and Zhang, Weinan},
  journal={Advances in Neural Information Processing Systems},
  year={2024}
}

@article{dpimagebench,
  author = {Chen Gong and Kecen Li and Zinan Lin and Tianhao Wang},
  title = {DPImageBench: A Unified Benchmark for Differentially Private Image Synthesis},
  year = {2025},
  journal={arXiv preprint arXiv:2503.14681}
}

@inproceedings{diffusiontransformer,
  title={Scalable diffusion models with transformers},
  author={Peebles, William and Xie, Saining},
  booktitle={Proceedings of the IEEE/CVF international conference on computer vision},
  year={2023}
}

@article{rlfortreatment,
  title={Reinforcement learning in healthcare: A survey},
  author={Yu, Chao and Liu, Jiming and Nemati, Shamim and Yin, Guosheng},
  journal={ACM Computing Surveys (CSUR)},
  number={1},
  pages={1--36},
  year={2021},
  publisher={ACM New York, NY}
}

@inproceedings{Bertscore,
  author       = {Tianyi Zhang and
                  Varsha Kishore and
                  Felix Wu and
                  Kilian Q. Weinberger and
                  Yoav Artzi},
  title        = {BERTScore: Evaluating Text Generation with {BERT}},
  booktitle    = {8th International Conference on Learning Representations},
  year         = {2020}
}

@inproceedings{dptransformer,
  title={Differentially private optimization on large model at small cost},
  author={Bu, Zhiqi and Wang, Yu-Xiang and Zha, Sheng and Karypis, George},
  booktitle={International Conference on Machine Learning},
  pages={3192--3218},
  year={2023},
  organization={PMLR}
}

@article{offlinetreatment,
  title={Offline reinforcement learning for safer blood glucose control in people with type 1 diabetes},
  author={Emerson, Harry and Guy, Matthew and McConville, Ryan},
  journal={Journal of Biomedical Informatics},
  volume={142},
  pages={104376},
  year={2023},
  publisher={Elsevier}
}

@inproceedings{attention,
  title={Attention Is All You Need},
  author={Vaswani, Ashish and Shazeer, Noam and Parmar, Niki and Uszkoreit, Jakob and Jones, Llion and Gomez, Aidan N and Kaiser, {\L}ukasz and Polosukhin, Illia},
  booktitle={Advances in Neural Information Processing Systems},
  pages={5998--6008},
  year={2017}
}

@article{
dpdm,
title={Differentially Private Diffusion Models},
author={Tim Dockhorn and Tianshi Cao and Arash Vahdat and others},
journal={Transactions on Machine Learning Research},
issn={2835-8856},
year={2023}
}

@inproceedings{BERT,
title	= {BERT: Pre-training of Deep Bidirectional Transformers for Language Understanding},
author	= {Jacob Devlin and Ming-Wei Chang and Kenton Lee and Kristina N. Toutanova},
year	= {2018},
URL	= {https://arxiv.org/abs/1810.04805}
}

@inproceedings{li2025easy,
  author       = {Kecen Li and
                  Chen Gong and
                  Xiaochen Li and
                  Yuzhong Zhao and
                  Xinwen Hou and
                  Tianhao Wang},
  title        = {From Easy to Hard: Building a Shortcut for Differentially Private
                  Image Synthesis},
  booktitle    = {{IEEE} Symposium on Security and Privacy, {SP}},
  year         = {2025},
}

@inproceedings{dptuning,
  title={Hyperparameter Tuning with Renyi Differential Privacy},
  author={Papernot, Nicolas and Steinke, Thomas},
  booktitle={International Conference on Learning Representations}
}

@inproceedings{PRV,
  author       = {Sivakanth Gopi and
                  Yin Tat Lee and
                  Lukas Wutschitz},
  title        = {Numerical Composition of Differential Privacy},
  booktitle    = {Advances in Neural Information Processing Systems},
  year         = {2021},
}

@article{grover2019bias,
  title={Bias correction of learned generative models using likelihood-free importance weighting},
  author={Grover, Aditya and Song, Jiaming and Kapoor, Ashish and others},
  journal={Advances in neural information processing systems},
  volume={32},
  year={2019}
}

@article{lecun2006tutorial,
  title={A tutorial on energy-based learning},
  author={LeCun, Yann and Chopra, Sumit and Hadsell, Raia and Ranzato, M and Huang, Fujie and others},
  journal={Predicting structured data},
  year={2006}
}

@article{yang2023diffusion,
  title={Diffusion models: A comprehensive survey of methods and applications},
  author={Yang, Ling and Zhang, Zhilong and Song, Yang and Hong, Shenda and Xu, Runsheng and Zhao, Yue and Zhang, Wentao and Cui, Bin and Yang, Ming-Hsuan},
  journal={ACM computing surveys},
  volume={56},
  number={4},
  pages={1--39},
  year={2023},
  publisher={ACM New York, NY, USA}
}

@article{Privcode,
      title={PrivCode: When Code Generation Meets Differential Privacy}, 
      author={Zheng Liu and Chen Gong and Terry Yue Zhuo and Kecen Li and Weichen Yu and Matt Fredrikson and Tianhao Wang},
      journal={arXiv preprint arXiv:2512.05459},
      year={2025}
}

\appendices

\setcounter{section}{0}
\setcounter{equation}{0}
\renewcommand\thesection{\Alph{section}}

\section{\rev{Ethical Considerations}}
\label{app:ethical}

\rev{DP dataset synthesis provides strong guarantees against individual data leakage, making it a cornerstone for ethical synthetic data generation. However, DP-based synthesis is not without risks. While privacy is preserved, the injected noise can distort data distributions, potentially introducing bias or reducing fairness in downstream tasks. Furthermore, synthetic data can be misused, for example, to create plausible but biased datasets or to facilitate adversarial attacks such as data poisoning. To mitigate these risks, practitioners should adopt safeguards such as bias auditing, transparency reports, and watermarking synthetic data to trace misuse. Ethical deployment also requires clear communication of limitations and responsible governance to ensure that synthetic data serves its intended purpose without compromising fairness or trust.}

\section{More Details about DP-SGD}
\label{app:supp_dp}

This section introduces how to account for the privacy cost of DP-SGD using the R\'{e}nyi DP, which is defined as follows.

\begin{definition}[\textit{R\'{e}nyi DP}~\cite{sgm}]
     We define the R\'{e}nyi divergence between two probability distributions $Y$ and $N$ as ${D_\alpha }\left( {Y\left\| N \right.} \right) = \frac{1}{{\alpha  - 1}}\ln {\mathbb{E}_{x\sim N}}{\left[ {\frac{{Y\left( x \right)}}{{N\left( x \right)}}} \right]^\alpha }$, where $\alpha>1$ is a real number. A randomized mechanism $\mathcal{A}$ satisfies ($\alpha, \gamma$)-RDP, if ${D_\alpha }\left( {\mathcal{A}(D)\left\| \mathcal{A}(D') \right.} \right) < \gamma$ holds for any neighboring dataset $D$ and $D'$.
\end{definition}

\noindent Given the batch size $B$, dataset size $N$, clip hyper-parameter $C$, and noise variance $\sigma^2$, as described in Section~\ref{sub:dp}, we denote the sampling ratio $q = B / N$. The RDP privacy cost for one training step can be obtained via Theorem~\ref{eq:rdp_gamma}~\cite{sgm}.

\begin{theorem}
\label{eq:rdp_gamma}
      Let $p_0$ and $p_1$ be the probability density function of $\mathcal{N}(0,C^2\sigma^2)$ and $\mathcal{N}(1,C^2\sigma^2)$. One training step with DP-SGD satisfies ($\alpha, \gamma_i$)-RDP for any $\gamma_i$ such that,
\begin{equation}
\gamma_i  \ge {D_\alpha }\left( { {\left( {1 - q} \right){p_0} + q{p_1}\left\| {{p_0}} \right.} } \right).
\end{equation}
\end{theorem}
\noindent The above theorem shows that the privacy bound $\gamma_i$ of one training step can be computed using the term ${D_\alpha }\left( { {\left( {1 - q} \right){p_0} + q{p_1}\left\| {{p_0}} \right.} } \right)$. Based on the RDP composition theorem~\cite{rdp}, we compose RDP costs of multiple training steps through $\gamma = \sum_i \gamma_i$. We convert the RDP privacy cost  $(\alpha,\gamma)$ to the  $(\epsilon,\delta)-$DP privacy cost as follows.

\begin{theorem}[\textit{From $(\alpha,\gamma)$-RDP to $(\epsilon,\delta)$-DP}~\cite{rdp}]
     If $\mathcal{A}$ is an ($\alpha, \gamma$)-RDP
mechanism, it also satisfies ($\epsilon, \delta$)-DP, for any $0 < \delta < 1$, where $\epsilon=\gamma + \frac{\log 1/\delta}{\alpha-1}$.
\end{theorem}

\noindent Therefore, we can use different Gaussian noise variance $\sigma^2$ to calculate the final privacy cost $\gamma + \frac{\log 1/\delta}{\alpha-1}$ until the required privacy budget $\varepsilon$ is reached.

\section{\rev{Theoretical Analysis of Curiosity-Driven Pre-training}}
\label{supsubsec:theoretical_analysis}

\rev{During pre-training, we replace a proportion $p \in [0,1]$ of real samples in each batch with synthetic samples from the current diffusion model distribution \(q_\theta(x)\). However, directly using $q_\theta(x)$ would reinforce existing patterns, limiting diversity. To encourage exploration of under-represented regions, we reweight synthetic samples using an exponential scheme~\cite{lecun2006tutorial}:
\begin{equation}
\begin{split}
    \tilde{p}_{\mathrm{synth}}(x) &= \frac{\exp(\beta c(x))\,q_\theta(x)}{Z_\beta}, \\
    Z_\beta &= \int \exp(\beta c(x'))\,q_\theta(x')\,dx'.
\end{split}
\label{eq:effdist2}
\end{equation}
where $c(x)$ is the curiosity score (Equation~\eqref{eq:curi_score}), and $\beta > 0$ controls the sharpness of curiosity emphasis. This reweighting strategy is motivated by the Importance Sampling View~\cite{grover2019bias}, and the term $\exp(\beta c(x))$ biases the distribution toward novel samples in low-density regions. In practice, we approximate \(\tilde{p}_{\mathrm{synth}}(x)\) by sampling a batch data from \(q_\theta(x)\) and selecting the top-\(k\) samples with the highest curiosity scores \(c(x)\). This avoids computing the intractable normalization constant \(Z_\beta\) while still emphasizing under-represented regions.

Let $p_{\mathrm{real}}(x)$ denote the real data distribution. This batch replacement induces an effective training distribution:
\begin{equation}
    p_{\mathrm{eff}}(x) = (1-p)\,p_{\mathrm{real}}(x) + p\,\tilde{p}_{\mathrm{synth}}(x),
    \label{eq:effdist1}
\end{equation}
where $\tilde{p}_{\mathrm{synth}}(x)$ is the curiosity-weighted synthetic distribution. From a score-matching perspective, diffusion training minimizes,
\begin{equation}
    \mathcal{L}(\theta) = \mathbb{E}_{x \sim p_{\mathrm{eff}},\, t}\Big[\big\|e_\theta(x^t,t) - \varepsilon\big\|^2\Big],
\end{equation}
where $x^t$ is the noisy version of clean data $x_0$ at step $t$. Since $p_{\mathrm{eff}}$ upweights high-curiosity regions, gradient contributions from under-represented modes are amplified, guiding $q_\theta$ toward a higher-coverage distribution and enhancing diversity.}

\section{Investigated Tasks and the Dataset}
\label{app:task_dataset}

We carry out experiments across five tasks in three domains ({\tt Maze2D}, {\tt MuJoCo}, and {\tt Kitchen}) sourced from D4RL~\cite{d4rl}, a benchmark recently introduced and widely studied for evaluating offline RL algorithms. We describe these tasks and their datasets in detail as follows,

\begin{table*}[!t]

\centering
\caption{Information of each task and the dataset.}
\small
\resizebox{1.0\textwidth}{!}{
\begin{tabular}{l|l|l|c|c|c|c|c}
    \toprule
     Domain & Tasks   & Sensitive Datasets &  Observations & Action Shape & Transition Shape & Trajectory Size & Transition Size \\
    \midrule 
    \multirow{3}{*}{Maze2D}& {\tt Maze2D}   & ``maze2d-umaze''  & 4 & 2 &  11 & $1976$ & $1 \times 10^6$ \\
    & {\tt Maze2D}   & ``maze2d-medium''  & 4 & 2 & 11 & $3977$ & $2 \times 10^6$ \\
    & {\tt Maze2D}   & ``maze2d-large''  & 4 & 2 & $11$ & $7967$ & $4 \times 10^6$ \\
    \midrule 
    \multirow{1}{*}{FrankaKitchen}& {\tt Kitchen}   & ``kitchen-partial''  & 60 & 9 & 130 & 601 & 136950  \\
    \midrule 
    MuJoCo& {\tt Half-Cheetah}   & ``halfcheetah-medium-replay''  & 17 & 6 & 41 & 404 & 101000 \\
    \bottomrule
\end{tabular}
}
\label{tab:app_discribe_env}
\end{table*}

\begin{table*}[!t]
\centering
\vspace{3mm}
\caption{Information on pre-training datasets for each dataset we utilize for fine-tuning.}
\small
\resizebox{1.0\textwidth}{!}{
\begin{tabular}{l|l|l|c|c|c|c}
    \toprule
     Domain & Fine-tuning Datasets   & Pre-training Datasets &  Observations & Action Shape & Transition Shape & Transition Size \\
    \midrule 
    \multirow{9}{*}{Maze2D}
    & \multirow{3}{*}{``maze2d-umaze''}&``maze2d-open'' & 4 & 2 & 11 &  $1 \times 10^6$  \\
    &&``maze2d-medium'' & 4 & 2 & 11 &  $2 \times 10^6$  \\
    &&``maze2d-large'' & 4 & 2 & 11 &  $4 \times 10^6$  \\
    \cline{2-7}
    & \multirow{3}{*}{``maze2d-medium''}&``maze2d-open'' & 4 & 2 & 11 &  $1 \times 10^6$  \\
    &&``maze2d-umaze'' & 4 & 2 & 11 &  $1 \times 10^6$  \\
    &&``maze2d-large'' & 4 & 2 & 11 &  $4 \times 10^6$  \\
    \cline{2-7}
    & \multirow{3}{*}{``maze2d-large''}&``maze2d-open'' & 4 & 2 & 11 &  $1 \times 10^6$  \\
    &&``maze2d-umaze'' & 4 & 2 & 11 &  $1 \times 10^6$  \\
    &&``maze2d-medium'' & 4 & 2 &11 &  $2 \times 10^6$  \\
    \midrule 
    
    \multirow{2}{*}{FrankaKitchen}
    & \multirow{2}{*}{``kitchen-partial''}&``kitchen-complete'' & 60 & 9 & 130 &  3680  \\
    &&``kitchen-mixed'' & 60 & 9 & 130 &  136950  \\
    \midrule 

    \multirow{3}{*}{MuJoCo}
    & \multirow{3}{*}{``halfcheetah-medium-replay''}&``walker2d-full-replay'' & 17 & 6 & 41 &  $1 \times 10^6$  \\
    &&``halfcheetah-expert'' & 17 & 6 & 41 &  $1 \times 10^6$  \\
    &&``walker2d-medium'' & 17 & 6 & 41 &  $1 \times 10^6$  \\
    \bottomrule
\end{tabular}
}
\label{tab:app_discribe_pub_sen}
\end{table*}

\begin{itemize}[leftmargin=*]
\item {\tt Maze2D}: We use the ``umaze'', ``medium'', and ``large'' layouts for a navigation task that demands a 2D agent reach a predetermined location. The generation of data involves randomly selecting goal locations, followed by the deployment of a planner that creates sequences of waypoints.
Agents aim to discover the shortest path to the goal. We present the visualization of {\tt Maze2D} map in Figure~\ref{fig:maze2d_env}.

\item {\tt MujoCo}: We select ``halfcheetah'' as the sensitive dataset in this domain. The robot is two-dimensional with nine linkages and eight joints, including two paws. The objective is to apply torque to six active joints—excluding the stationary torso and head—to drive the robot forward as swiftly as possible. Forward movement garners positive rewards based on the distance covered, whereas backward movement incurs a penalty. The active joints connect the front and rear thighs to the torso, the shins to the thighs, and the feet to the shins.

\item {\tt Kitchen}: The  {\tt Kitchen} domain, introduced by Gupta et al.~\cite{gupta2019relay}, features a 9-DoF robot operating in a kitchen setup with common household items such as a microwave, kettle, overhead light, cabinets, and an oven. This domain tests multi-task capabilities. We utilize the ``partial'' datasets as the sensitive dataset, consisting of undirected data, where the robot executes subtasks that may not directly align with the goal configuration. 

\end{itemize}

The D4RL~\cite{d4rl} dataset provides trajectories within its collection. For experiments on transition-level DP, each transition in the offline RL dataset is contributed by distinct users. To construct this dataset, we first decompose the trajectories into individual transitions. These transitions are assumed to be independent, as we randomly sample a subset from the dataset to mitigate potential correlations. We select 80\% of the sensitive transitions for the fine-tuning dataset. This assumption of independence is practical, as in real-world scenarios, users typically contribute transitions independently in the same tasks without coordinated dependencies across their contributions. 

In DP trajectory synthesis, datasets often contain a limited number of trajectories, resulting in substantial DP noise during privatization, which can degrade data utility. To address this, we augment the dataset by segmenting long trajectories into shorter fragments. This approach increases the effective number of data points, diluting the relative impact of DP noise while preserving temporal dependencies within fragments. In real-world applications, where trajectory datasets may be larger, this method remains effective, as it adapts to varying dataset sizes, simulating realistic scenarios.

We introduce the investigated datasets in Table~\ref{tab:app_discribe_env}. As presented in Section~\ref{subsec:rl}, when calculating the dimensions of transitions, we consider a transition that consists of two states: an action and a reward. Besides, we present the division of pre-training and sensitive datasets in our experiments in Table~\ref{tab:app_discribe_pub_sen}.

\begin{table*}[!t]
\centering
\caption{Default DP-SGD hyper-parameters under $\epsilon=10$ and $\delta=1\times10^{-6}$. Sampling rate $q$ is set depending on the dataset size; clipping norm $C$ is set to $1$. We use the Adam optimizer with a learning rate of $3 \times 10^{-4}$, while all other hyper-parameters adhered to the default settings in~\cite{lu2023synthetic}.}
\label{tab:dpsgd_hyperparams}
\resizebox{1.0\textwidth}{!}{
\setlength{\tabcolsep}{6.1mm}{
\begin{tabular}{l|c|c|c|c|c|c}
\toprule
\multirow{2}{*}{\textbf{Method}} & \multirow{2}{*}{\textbf{Parameter}} &  \multicolumn{3}{c|}{\textbf{Maze2D}} & \textbf{Kitchen} & \multicolumn{1}{c}{\textbf{Mujoco}}  \\
\Xcline{3-7}{0.5pt}
& & umaze & medium & large & partial & halfcheetah \\ \hline
\multirow{4}{*}{\textbf{\toolnametran}} & sampling ratio $q$ & $1.28\times 10^{-4}$ & $0.64\times 10^{-4}$ & $0.32\times 10^{-4}$ & $9.35\times 10^{-4}$ & $12.67\times 10^{-4}$ \\
& training steps & 574K & 294K & 433K & 15K & 240K  \\
& noise multiplier $\sigma$ & 0.44 & 0.39 & 0.37 & 0.47 & 0.68\\
& batch size & 128  &  128 &  128 & 128  & 128 \\
\midrule
\multirow{4}{*}
{\textbf{\toolnametraj}} & sampling ratio $q$ & $5.18\times 10^{-2}$ & $2.57\times 10^{-2}$ & $1.29\times 10^{-2}$ & $1.70\times 10^{-1}$ & $2.53\times 10^{-1}$ \\
& training steps & 200K & 200K & 200K & 200K & 200K  \\
& noise multiplier $\sigma$ & 12.1 & 6.3 & 3.2 & 40.4 & 60.9 \\
& batch size &  1024 &  1024 & 1024  &  1024 & 1024 \\
\bottomrule
\end{tabular}
}
}
\end{table*}

\section{Baseline Implementation}
\label{app:baselines_imp} 
These baselines achieve state-of-the-art performance in the `SynMeter' library~\cite{du2024systematic}. For further information, we strongly encourage readers to refer to the `SynMeter'~\cite{du2024systematic}. 

\begin{itemize}[leftmargin=*]
    \item  \textbf{PGM~\cite{PGM}.} In light of our samples possessing higher data dimension, we configure the number of two-way marginals to 50, the number of three-way marginals to 10, and the maximum number of iterations to 4,000. This adjustment facilitates a robust learning of the complex interdependencies among variables. We set the number of bins to 10.
    \item \textbf{PrivSyn~\cite{zhang2021privsyn}.} We set the max bins parameter to 32 to capture our dataset's subtle patterns and complexities. Similarly, we have set the privacy budget to 10.
    \item \textbf{PATE-GAN~\cite{PATE-GAN}.} We set the generator $n$ layers hidden to 3 and the generator $n$ units hidden to 141 to adeptly learn complex data distributions. For the discriminator, we configure the discriminator $n$ layers with hidden to 2 and the discriminator $n$ units with hidden to 113, ensuring effective discrimination between real and generated data. We set $n$ teachers to 15 to enhance data generation diversity. 
    \item \textbf{PrePATE-GAN.} We use identical training hyper-parameters as PATE-GAN and introduce a pre-training phase to enhance the model's training effectiveness. Specifically, we exclude DP protection during pre-training and train directly on the public transitions. We define the epoch for pre-training the same as that of \toolname.
     \item \textbf{DP-Transformer.} We designed an autoregressive trajectory synthesis framework based on Transformer architecture to model and generate decision-making trajectories while ensuring privacy. Trajectories are represented as sequences of state, action, reward, termination, and next-state tokens, processed by a Transformer encoder with self-attention to capture temporal dependencies. For generations, trajectories are iteratively sampled autoregressively from initial states, producing coherent sequences.
\end{itemize}

\section{Hyper-parameter Settings}
\label{supsec:hyper}

\rev{Following prior implementations~\cite{lu2023synthetic,he2023diffusion}, we use the Elucidated Diffusion Model (EDM)~\cite{edm} for transition synthesis and the Diffusion Transformer~\cite{diffusiontransformer} for trajectory synthesis in our synthesizer. }

\subsection{Diffusion Model}
\label{subsec:dm}

\vspace{1.5mm}
\noindent \textbf{Denoising Network.}
We use the Elucidated Diffusion Model (EDM) formulation of the denoising network~\cite{edm}. The denoising network $D_\theta$ is parameterized as a Multi-Layer Perceptron (MLP) with skip connections from the previous layer, following the architecture described in~\cite{mlp_mixer}. We encode the noise level of the diffusion process using a Random Fourier Feature embedding, with dimensions of 16. The base network size adopts a width of 1024 and a depth of 6, resulting in approximately 6 million parameters.

\vspace{1.5mm}
\noindent \textbf{Sampling of Diffusion Model.}
For the diffusion sampling process, we utilize the stochastic SDE sampler introduced by~\cite{edm}, using the default hyper-parameters designed for ImageNet. To enhance sample fidelity, we select a higher number of diffusion timesteps at 128. Our implementation is sourced from the repository\footnote{\url{https://github.com/lucidrains/denoising-diffusion-pytorch}}.

\vspace{1.5mm}
\noindent \textbf{DP-SGD Training.}  We refer to the official repository Opacus\footnote{\url{https://github.com/pytorch/opacus/}} to implement DP-SGD. The hyper-parameters are detailed in Table~\ref{tab:dpsgd_hyperparams}. We calculate the sampling ratio $q$ by dividing the batch size by the number of transitions within the dataset. For all datasets, we set the fine-tuning epoch to 5, and the training step is obtained by dividing the epoch by the sampling ratio. Consequently, the noise multiplier must be adjusted based on the varying number of training steps across different datasets to ensure uniform privacy protection across various training settings. The noise multiplier is calculated using the standard privacy analysis function of Opacus, and the max-grad-norm $C$ is set to 1.0, following the default setting in Opacus. We use the same setting in all experiments for the max-grad-norm $C$, batch size, optimizer, and learning rate. We preprocess the dataset using splitting and random sampling techniques to simulate real-world scenarios, as detailed in Appendix~\ref{app:task_dataset}.

\vspace{1.5mm}
\noindent \textbf{Transformer.} For \toolnametraj, we represent the noise prediction network as the transformer-based architecture proposed by~\cite{he2023diffusion}, which adopts a GPT2-like model. The transformer is configured as six hidden layers and four attention heads. We use $T$ = 200 for diffusion steps, while all other hyper-parameters
adhered to their default settings. 

\rev{
\begin{table*}[!t]
\centering
    \caption{\rev{GPU memory consumption, runtime, and normalized return: comparing \toolnametran and \toolnametraj with baselines on synthetic datasets ({\tt Maze2D-medium}). `h' means hours.}}
    \setlength{\tabcolsep}{4mm}{
    \resizebox{0.99\textwidth}{!}{
    \begin{tabular}{l|c|cccc|ccc}
    \toprule
      \multicolumn{2}{c|}{\textbf{Evaluation Metrics}}& \textbf{PGM} & \textbf{PrivSyn} & \textbf{PrePATE-GAN} & \textbf{PrivORL-n} & \textbf{PrivORL-j-U} & \textbf{DP-Transformer} & \textbf{PrivORL-j} \\
    \midrule
    \multirow{3}{*}{\textbf{Memory}} & Pre-train & -  & - & 1.5GB &0.9GB & 7.45GB & 19.7GB & 13.12GB \\
     & Fine-tune & -  & - & 0.4GB & 21.4GB & 14.59GB & 43.11GB & 39.72GB\\
     & Synthesis& 11.1GB & 9.4GB & 8.1GB &4.3GB & 4.1GB & 9.74GB & 8.15GB \\
     \midrule
    \multirow{3}{*}{\textbf{Time}}& Pre-train & - & - & 12h &1.5h & 0.45h & 1.37h & 0.77h \\
     & Fine-tune  &  - &- & 11h & 2h & 0.96h & 5.18h & 3.03h \\
     & Synthesis &  6h & 8.5h & 0.18h &0.5h & 0.24h & 10.1h & 1.42h \\
    \bottomrule
\end{tabular}
}}
\label{tab:computationalResource}
\vspace{-1mm}
\end{table*}}

\subsection{Curiosity Module}
\label{appsubsec:curo}

The curiosity module enhances the diversity of synthetic data during the pre-training phase. Both the target and prediction networks in our architecture are structured as MLPs. The target network is characterized by greater depth and higher-dimensional modeling capabilities. Throughout each epoch of the pre-training, we first generate a predefined number of synthetic transitions using the current diffusion model. Based on a designated curiosity rate, we identify and select those samples from the generated data with the highest curiosity scores. These selected samples are then replaced with the current batch of training samples to update the synthesizer. We apply the same settings across various training tasks unless differences are specifically noted.  The output dimension of the target network is 32. The default curiosity rate is 0.3.

\section{Efficiency}
\label{app:eff}
\rev{This section highlights the efficiency of \toolnametran and \toolnametraj} relative to the baselines and the notable trade-off between computational resource costs (including the memory and time cost) and synthesis performance. Table~\ref{tab:computationalResource} compares GPU memory consumption and runtime \rev{between \toolnametran, \toolnametraj, and baselines. All experiments are conducted on the server with Python 3.9.18, four NVIDIA GeForce A6000 GPUs, and 512GB of memory.}

For DP transition synthesis, Table~\ref{tab:computationalResource} shows that due to the large parameter size in diffusion models, \toolnametran requires 21.4GB of GPU memory to fine-tune these models using DP-SGD. This characteristic is inherent to DP-SGD~\cite{dpsgd}. During dataset synthesis, \toolnametran and PrePATE-GAN complete the process in just 0.5 hours and 0.18 hours, significantly faster than the 6.0 and 8.5 hours required by PGM and PrivSyn, respectively. The difference in synthesis time arises because PGM and PrivSyn must consider the entire sensitive dataset during processing~\cite{PGM,zhang2021privsyn}. In contrast, generative model-based methods (i.e., PrePATE-GAN and \toolnametran) can directly generate datasets without needing to analyze the entire dataset. Regarding total processing time, \toolnametran is more efficient, requiring 4 hours less than the 6.0, 8.5, and 23.2 hours needed by PGM, PrivSyn, and PrePATE-GAN. \toolnametran consistently delivers competitive normalized returns and the highest marginal utility, indicating superior performance and efficiency compared to baselines.

\rev{For DP trajectory synthesis, as shown in Table~\ref{tab:computationalResource}, the peak of GPU memory consumption for \toolnametraj is 39.72GB, and the time consumptions for pre-train, fine-tune, and dataset synthesis are 0.77h, 3.03h, and 1.42h. Thus, trajectory-level synthesis requires nearly 40GB of GPU memory and completes training in 3.8 hours.  Once the DP synthesizer is well-trained, it can generate unlimited synthetic data without further privacy concerns, making resource requirements manageable. }

\section{Additional Discussions}
\label{app:addi_discussions}

\subsection{\rev{Uniqueness of Offline RL Trajectories.}}  
\label{supsubsec:uniqueness}

\rev{Compared to image synthesis, trajectory generation places a stronger emphasis on preserving temporal consistency across the entire sequence. In text generation, coherence is often captured locally, so models can train on smaller fragments (e.g., sentence turns) independently. In contrast, an offline RL trajectory encodes a complete decision-making process with sparse and delayed rewards, making long-term dependencies critical for policy quality. While our method also segments trajectories, we incorporate conditioning signals to preserve inter-fragment dependencies, ensuring that the synthesized segments collectively capture the global structure and reward dynamics of the original trajectory.}

\subsection{Synthesis Principles}
\label{app:eval_metrics}

To evaluate dataset synthesis in offline RL, drawing from other fields~\cite{yue-etal-2023-synthetic,li2023meticulously,zhang2021privsyn}, we analyze from three perspectives: (1) utility, and (2) fidelity.

\vspace{1mm}
\noindent \textbf{Utility}: A crucial aspect of dataset synthesis is that the synthetic dataset should be capable of accomplishing the downstream task. We hope that training agents with the synthesized dataset can achieve performance comparable to that of agents trained directly on the original dataset. 

\vspace{1mm}
\noindent \textbf{Fidelity}: This principle ensures that the synthetic dataset has statistical characteristics similar to those of the raw dataset.

\subsection{Evaluation Metrics}
\label{supsubsec:metrics}  

We introduce the evaluation metrics as follows.

\vspace{1mm}
\noindent \textbf{Averaged Cumulative Return. } The metric for downstream tasks measures the utility of the synthesized dataset by training agents and evaluating their performance using the \textit{Averaged Cumulative Return}. An agent interacts with the environment over multiple rounds, generating test trajectories,
$
\tau = \left\{ s_1,a_1,r_1, \cdots, s_{|\tau|},a_{|\tau|},r_{|\tau|} \right\}.
$
The cumulative return per trajectory is $R(\tau) = \sum_{i=0}^{|\tau|} r_i$, and the Averaged Cumulative Return is $\frac{1}{|\mathcal{T}|}\sum_{\tau \in \mathcal{T}} R(\tau)$. Higher returns indicate better agent performance and greater dataset utility. Returns are normalized to [0, 100] per D4RL~\cite{d4rl}, and we use ``normalized return'' for simplicity.

\noindent \textbf{Marginal \& Correlation.} These metrics measure the differences between the synthetic and real datasets~\cite{lu2023synthetic}. \textit{Marginal} is the mean Kolmogorov-Smirnov~\cite{massey1951kolmogorov} statistic,
which measures the maximum distance between the empirical cumulative distribution functions of each dimension in the synthetic and real data. \textit{Correlation} represents mean correlation similarity, quantifying
the difference in pairwise Pearson rank correlations~\cite{fieller1957tests} between the synthetic and real data. These scores range from 0 to 1, with a higher value indicating greater fidelity between the synthetic and real datasets.

\vspace{0.5mm}
\noindent \textbf{TrajSocre:} In generating offline RL trajectories, we design a novel metric inspired by BERTScore~\cite{Bertscore}. TrajScore first leverages MLPs, pre-trained in an autoencoder, as a trajectory encoder—similar to the embedding layer in BERT~\cite{BERT}—to compute high-dimensional embeddings of trajectories. It measures similarity by calculating the cosine similarity score between generated and real trajectories.

\vspace{1mm}
\noindent \rev{\textbf{TPR@FPR:} The True Positive Rate (TPR) at a fixed False Positive Rate (FPR) quantifies the attacker's ability to accurately identify sensitive information (true positives) while limiting the misidentification of non-sensitive instances as sensitive (false positives).}

\begin{figure*}[!t]
    \centering
    \includegraphics[width=1.0\linewidth]{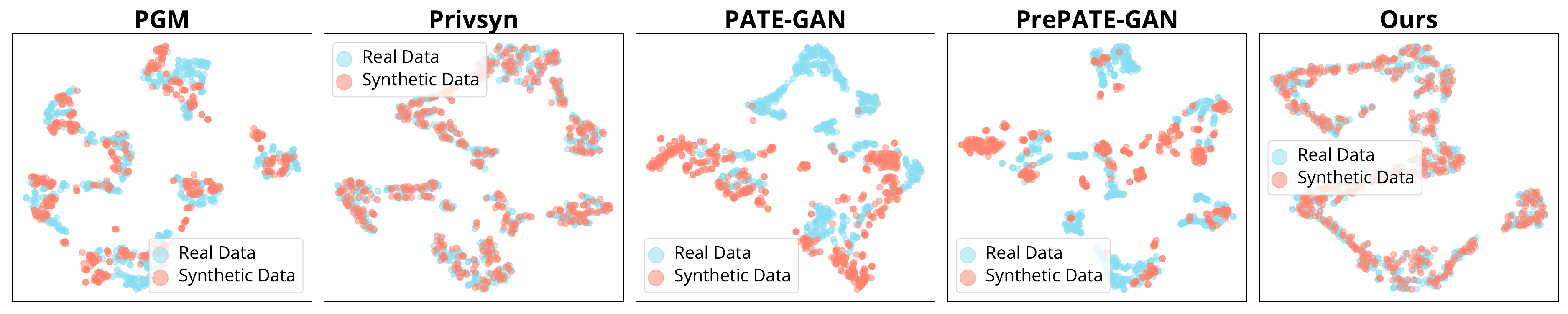}   
    \caption{The t-SNE visualizations of data distribution for {\tt Maze2D-medium}. The t-SNE visualizations illustrate the two-dimensional data distribution of the DP synthetic dataset and the real dataset. We synthesize the dataset using \toolnametran and baseline methods under privacy budget $\epsilon=10$. }
    \label{fig:tsne_maze2d}
\end{figure*}

\subsection{Application Scenario}
\label{app:app_sce}
Offline RL has been widely applied across various fields, such as healthcare~\cite{RL4Treatment,RL4BGC,offline_rl_medicial}, indoor navigation~\cite{pan2019you,levine2020offline}, and recommendation systems~\cite{RL4Recommender,RL4Recommender2}. In these applications, offline RL datasets often contain sensitive information. For instance, in healthcare, this might include patients' treatment records; in indoor navigation, the map information (e.g., room structure); and in recommendation systems, customers' shopping histories. It is a struggle to directly release these transitions to the third party for training agents.

\toolname allows us to train a transition or a trajectory synthesizer under DP. This synthesizer can generate datasets that closely mimic real data while ensuring privacy is maintained. Organizations can then share these synthetic datasets for downstream agent training, reducing privacy concerns. Analogous to the pre-training image synthesizer on some public datasets~\cite{li2023meticulously,dp-diffusion}, we believe that it is also reasonable for offline RL dataset synthesizers to utilize public datasets for pre-training. It is easy to collect transitions from environments that are similar to, yet distinct from, those containing sensitive datasets for both transitions and trajectories. For instance, in healthcare, there are public treatment records available for medical research obtained with the patient's consent. In indoor navigation, we can acquire map information from some environments without privacy concerns.

\begin{table}[!t]
\footnotesize
    \centering
    \caption{The average normalized returns of agents trained on synthetic transitions (with a privacy budget of $\epsilon = 10$) across various dataset sizes using the IQL algorithm. }
    \resizebox{0.48\textwidth}{!}{
   \begin{tabular}{p{1.3cm}|c|ccccc}
        \toprule
            \multirow{2}{*}{\textbf{Domains}} & \multirow{2}{*}{\begin{minipage}{1.3cm}\centering\textbf{Real} \\ \textbf{Dataset}\end{minipage}}  & \multicolumn{5}{c}{\textbf{Sizes}} \\
         \cline{3-7}
          & &  0.1 M & 0.5 M & 1 M & 2 M & 5 M \\
         \midrule
         \multirow{3}{*}{{\tt Maze2D}} & umaze & 35.2 & 72.5 & 70.3 & 70.9 & 81.3\\
         & medium & 42.3 & 42.6 & 90.7 & 89.5 & 89.5 \\
         & large & 45.9 & 77.5 & 81.0 & 84.3 & 98.9 \\
         \midrule
         \multirow{1}{*}{{\tt Kitchen}} & partial & 7.5 & 20.9 & 25.5 & 17.5 & 17.3 \\
         \midrule
         \multirow{1}{*}{{\tt Mujoco}} & halfcheetah & 32.8 & 36.0 & 36.9 & 42.5 & 45.3\\
         \midrule
         \multicolumn{2}{c|}{\cellcolor{gray!20} \textbf{Average}} \cellcolor{gray!20}  & \cellcolor{gray!20} 32.7 & \cellcolor{gray!20} 49.9 & \cellcolor{gray!20} 60.9 & 60.9 \cellcolor{gray!20} & \cellcolor{gray!20} 66.5 \\
        \bottomrule
    \end{tabular}
     }
    \label{tab:data_size}
\end{table}

\subsection{T-SNE visualizations}
\label{appsubsec:tsne}

We use the public package from `sklearn'~\cite{scikit-learn}, with default settings, to implement T-SNE. The T-SNE figure can reflect the distribution of the dataset to some extent. Figure~\ref{fig:tsne_maze2d} visualizes the distribution of synthetic transitions generated by \toolnametran and the baselines compared to the real sensitive transitions in the {\tt Maze2D-medium} environment.

\subsection{Limitations}
\label{supsubsec:limitation}

\rev{We discuss the limitations of our work as follows.}

\begin{itemize}[leftmargin=*]
\item This work relies on public datasets for pre-training synthesizers. Without access to a suitable public dataset, \toolname may reduce the performance of generating complex datasets. 

\vspace{0.5mm}
\item Optimal curiosity rates vary across different sensitive datasets. 
\rev{Section~\ref{subsec:hyper} introduces that tuning the curiosity rate introduces additional privacy budget. A potential solution is incorporating DP hyper-parameter tuning~\cite{dptuning}, which provides a formal framework for selecting hyper-parameters under DP. This approach typically leverages techniques such as privacy-preserving grid search or adaptive tuning under a fixed privacy budget. Although DP hyper-parameter tuning introduces additional complexity, it offers a principled way to balance utility and privacy during parameter selection.}

\vspace{0.5mm}
\item DP-SGD performs well with large datasets, but when the dataset is small, the added noise is significant, leading to poor synthetic quality. 

\vspace{0.5mm}
\item \rev{Besides, our method faces challenges with high-dimensional datasets. The curse of dimensionality amplifies the noise introduced by DP, making it harder to preserve utility while ensuring privacy. High-dimensional feature spaces often require larger models and more iterations, which further increase computational costs and exacerbate the trade-off between privacy and data fidelity.} 

\end{itemize}

\rev{Future work aims to develop high-quality datasets without relying on public datasets and explore methods that can adaptively search for the optimal curiosity rate across different tasks under the DP constraint.}

\begin{table}[!t]
\small
    \centering
    \caption{Comparison of TrajSocres of synthetic trajectories using \toolnametraj and baselines ($\epsilon = 10$) to the real dataset. `DP-Trans.' is the abbreviation for `DP-Transformer.' }
    \setlength{\tabcolsep}{3.0mm}{
    \resizebox{0.48\textwidth}{!}{
   \begin{tabular}{p{1.2cm}|c|ccc}
        \toprule
            \multirow{1}{*}{\textbf{Domains}} & \multirow{1}{*}{\textbf{Dataset}}  & {\textbf{\toolnametran}} & {\textbf{DP-Trans.}}  & {\textbf{\toolnametraj}} \\
         \midrule
         \multirow{3}{*}{{\tt Maze2D}} & umaze & 0.761 & 0.830  & 0.946 \\
         & medium  & 0.602 & 0.705 & 0.922 \\
         & large & 0.739 & 0.866 & 0.952  \\
         \midrule
         \multirow{1}{*}{{\tt Kitchen}} & partial & 0.629 & 0.712 & 0.787 \\
         \hline
         \multicolumn{2}{c|}{\cellcolor{gray!20}\textbf{Average}}  &  \ 0.683 \cellcolor{gray!20} & \ 0.778 \cellcolor{gray!20}  & \cellcolor{gray!20} 0.902 \\
        \bottomrule
    \end{tabular}
     }
     }
    \label{tab:trajscores}
\end{table}

\section{Additional Experimental Results}

\subsection{Fidelity Evaluation of Trajectory-level Synthesis} 
\label{supsubsec:fidelity_traj}

Table~\ref{tab:trajscores} presents that the synthetic trajectories from \toolnametraj are more similar to the trajectories in real datasets compared with baselines. Considering all datasets, \toolnametraj achieves an average TrajScore of 0.902, which is 0.124 higher than DP-Transformer (0.778), corresponding to a 15.9\% relative improvement under $\epsilon=10$.

\subsection{Scaling Up Synthetic Offline Dataset} 
\label{supsubsec:scaling_up}

This section explores whether \toolnametran can scale up the synthetic offline dataset and assesses agent performance that is trained on the datasets varying with the sizes of transition. 
The transitions in the synthetic dataset are synthesized across sizes of $\{0.1\text{M}, 0.5\text{M}, 1\text{M}, 2\text{M}, 5\text{M}\}$.

\begin{table*}[!t]
\vspace{3mm}
\footnotesize
    \centering
    \caption{\rev{The average normalized returns of agents trained on synthetic datasets ($\epsilon = 10$) under two privacy accounting methods (RDP and PRV), evaluated with IQL and TD3PlusBC algorithms. The values in parentheses are the relative changes. } }
    \resizebox{1.0\textwidth}{!}{
   \begin{tabular}{p{1.3cm}|c|cc|cc|cc|cc}
        \toprule
            \multirow{3}{*}{\textbf{Domains}} & \multirow{3}{*}{\textbf{Datasets}}  & \multicolumn{4}{c|}{\textbf{IQL}}  & \multicolumn{4}{c}{\textbf{TD3PlusBC}}\\
         \cline{3-10}
         & &  \multicolumn{2}{c|}{\toolnametran} & \multicolumn{2}{c|}{\toolnametraj} &  \multicolumn{2}{c|}{\toolnametran} & \multicolumn{2}{c}{\toolnametraj} \\
         \cline{3-10}
          & &  RDP & PRV & RDP & PRV &  RDP & PRV & RDP & PRV \\
         \midrule
         \multirow{3}{*}{{\tt Maze2D}} & umaze & 70.3 $\pm$ 2.1 & 72.5 $\pm$ 3.6 ({\color{blue} $\uparrow$2.2})  & 49.8 $\pm$ 6.8 & 55.5 $\pm$ 7.8 ({\color{blue} $\uparrow$5.7}) & 60.3 $\pm$ 6.3 & 63.1 $\pm$ 5.8 ({\color{blue} $\uparrow$2.8}) & 49.9 $\pm$ 4.9 & 52.9 $\pm$ 5.1 ({\color{blue} $\uparrow$3.0}) \\
         & medium & 90.7 $\pm$ 8.6 & 92.1 $\pm$ 4.7 ({\color{blue} $\uparrow$1.4}) & 49.3 $\pm$ 1.7 & 41.4 $\pm$ 6.8 ({\color{red} $\downarrow$7.9}) & 50.4 $\pm$ 6.4 & 51.3 $\pm$ 6.6 ({\color{blue} $\uparrow$0.9}) & 38.0 $\pm$ 1.6 & 38.9 $\pm$ 1.9 ({\color{blue} $\uparrow$0.9}) \\
         & large & 81.0 $\pm$ 11.8 & 82.0 $\pm$ 4.6 ({\color{blue} $\uparrow$1.0}) & 37.7 $\pm$ 7.0 & 40.5 $\pm$ 8.1 ({\color{blue} $\uparrow$2.8}) & 75.3 $\pm$ 13.2 & 72.1 $\pm$ 6.3 ({\color{red} $\downarrow$2.8}) & 35.6 $\pm$ 2.6 & 40.9 $\pm$ 3.6 ({\color{blue} $\uparrow$5.3}) \\
         \midrule
         \multirow{1}{*}{{\tt Kitchen}} & partial & 25.5 $\pm$ 2.5 & 24.0 $\pm$ 1.5 ({\color{red} $\downarrow$1.5}) & 13.8 $\pm$ 7.5 & 15.0 $\pm$ 6.5 ({\color{blue} $\uparrow$1.2}) & 11.5 $\pm$ 0.0 & 13.0 $\pm$ 1.5 ({\color{blue} $\uparrow$1.5}) & 8.3 $\pm$ 2.5 & 8.0 $\pm$ 2.5 ({\color{red} $\downarrow$0.3}) \\
         \midrule
         \multicolumn{2}{c|}{\cellcolor{gray!20}\textbf{Average}} 
         & \cellcolor{gray!20} $66.9 \pm 6.3$
         & \cellcolor{gray!20} $67.7 \pm 3.6$ ({\color{blue} $\uparrow$0.6})
         & \cellcolor{gray!20} $41.9 \pm 5.7$
         & \cellcolor{gray!20} $42.4 \pm 7.3$ ({\color{blue} $\uparrow$0.5})
         & \cellcolor{gray!20} $49.4 \pm 6.5$
         & \cellcolor{gray!20} $49.9 \pm 5.1$ ({\color{blue} $\uparrow$0.5})
         & \cellcolor{gray!20} $33.0 \pm 2.9$
         & \cellcolor{gray!20} $35.2 \pm 3.3$ ({\color{blue} $\uparrow$2.2}) \\
        \bottomrule
    \end{tabular}
     }
    \label{tab:prv}
\end{table*}

Table~\ref{tab:data_size} shows the average normalized returns of IQL agents trained on DP synthetic transitions with varying sizes. This table shows that the agents' performance gradually improves as the number of synthetic transitions increases. Specifically, the average performance of trained agents increases from 32.7 to 66.5 as the size of synthetic transitions grows from 0.1M to 5M. These results also show that \toolnametran synthesizes transitions with high diversity, so increasing the number of synthetic transitions assists the dataset's utility. However, these improvements are not unlimited, and they eventually reach a plateau, e.g., in {\tt Maze2D-medium}, the performance of the trained agents converges to approximately 90. In {\tt Kitchen-partial}, the utility of synthetic transitions decreases somewhat due to the real dataset involving only about 136,950 transitions (approximately 0.1M), which is smaller than others. Preferring a diverse generation may lead to unexpected transitions.

\begin{figure}[!t]
\vspace{-1.5mm}
    \centering
    \setlength{\abovecaptionskip}{0pt}
    \includegraphics[width=1.0\linewidth]{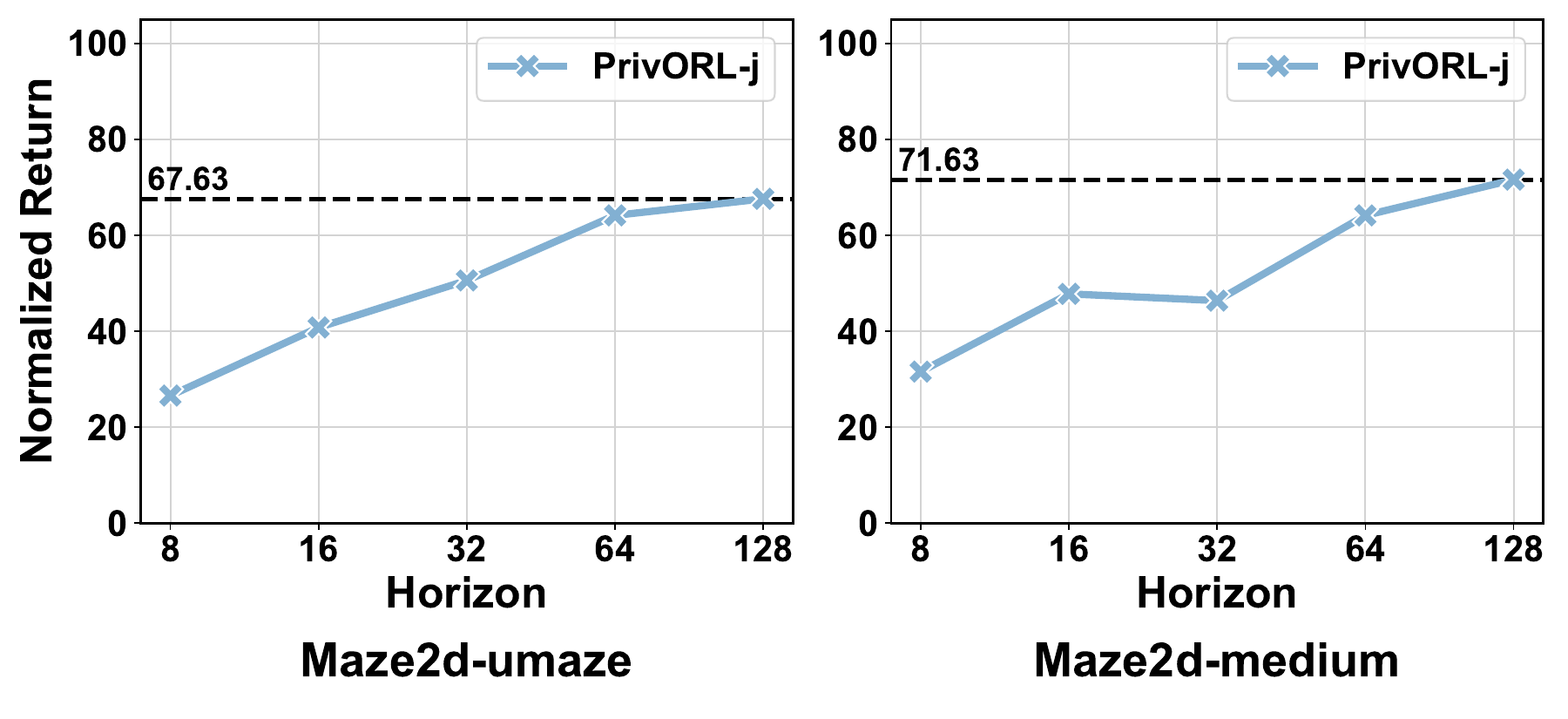}
    \caption{The average normalized returns of agents trained using three offline algorithms on synthetic trajectories using \toolnametraj, varying with the horizon $H$. }
    \label{fig:hyper-parameter-h}
\end{figure}

\subsection{Hyper-parameter Analysis}
\label{supsubsec:hyper}
In this section, we explore the horizon for \toolnametraj, $H=\{8,16,32,64,128\}$. Figure~\ref{fig:hyper-parameter-h} shows the average normalized returns of agents trained
using three algorithms on synthetic trajectories using
PrivORL-j, varying with the horizon $H$. In this figure, we observe that a large value of $H$ helps to increase the performance of synthetic trajectories.

\subsection{\rev{Privacy Accounting Using PRV}}
\label{supsubsec:prv}

\rev{In this paper, we use RDP for fair comparisons with baselines, as RDP is widely adopted for privacy accounting in DP machine learning. We also investigate the use of an alternative privacy accounting method, Privacy Random Variable (PRV)~\cite{PRV}, which provides a tighter analysis of privacy loss compared to RDP. Since PRV and RDP share similar analytical frameworks (both supporting moment accounting and composition across multiple training steps), the integration of PRV into our approach is seamless. Specifically, we replace the original RDP accountant in \toolname with the PRV accountant implemented in the Opacus library~\cite{opacus}.}

\rev{Table~\ref{tab:prv} presents the average normalized returns of agents ($\epsilon = 10$) under two privacy accounting methods (RDP and PRV), evaluated with IQL and TD3PlusBC algorithms. In this table, we observe that although PRV provides a tighter analysis of privacy loss compared to RDP, the PRV’s benefit is marginal in most cases. In particular, for \toolnametran and \toolnametraj, using PRV results in only slight improvements: an average of 0.6 and 0.5 for IQL, and 0.5 and 2.2 for TD3PlusBC. }

\begin{figure}[!t]
\vspace{-2mm}
    \centering
    \setlength{\abovecaptionskip}{0pt}
    \includegraphics[width=1.0\linewidth]{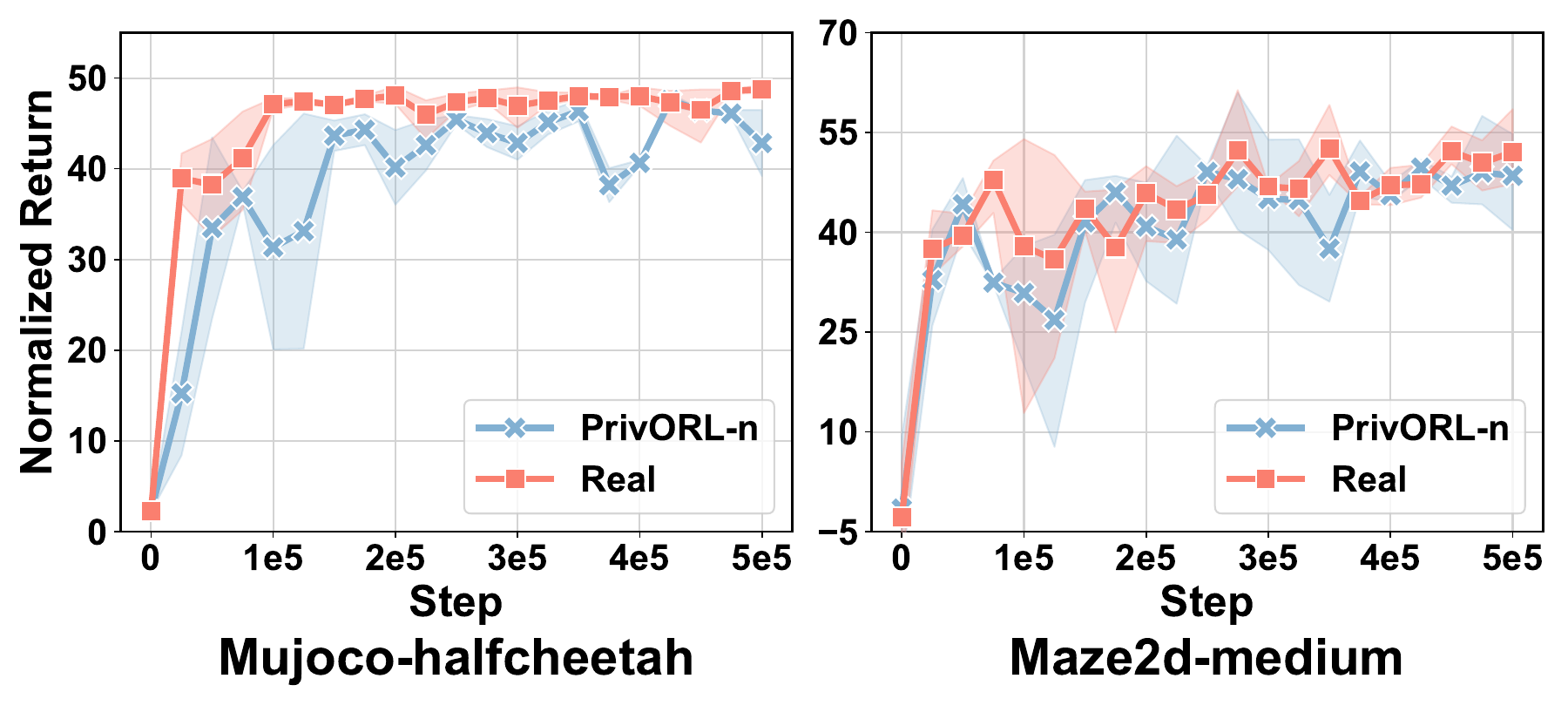}
    \caption{\rev{Training curves of agents on real sensitive datasets and synthetic datasets ($\epsilon = 10$) using the TD3PlusBC algorithm.} }
    \label{fig:curves}
\end{figure}

\subsection{\rev{Training Convergence}}
\label{subapp:curve}

\rev{We analyze the training curves of agents on real sensitive and synthetic datasets under $\epsilon=10$ using the TD3PlusBC algorithm (Figure~\ref{fig:curves}). In {\tt Mujoco-halfcheetah}, agents trained on synthetic data exhibit slightly slower convergence, although their peak return remains comparable to that of real data. In {\tt Maze2d-medium}, convergence behavior is largely similar across real and synthetic datasets, indicating that synthetic data can support effective agent learning. }

\section{More Related works}
\label{app:related}

This section introduces recent works related to our work, including DP dataset synthesis, offline RL with synthetic datasets (non-DP), and broad applications of offline RL.

\vspace{1mm}
\noindent \textbf{Differentially Private Dataset Synthesis.} We group DP dataset synthesis methods into two classes as follows.

\noindent \textit{Marginal-based Synthesis.} A potential approach considers the transition synthesis similar to DP tabular synthesis. Marginal-based synthesis~\cite{privmrf,zhang2021privsyn,PGM}, which is popular in tabular synthesis, focuses on capturing and replicating the marginal distributions of individual variables in a dataset, and the joint distributions of pairs or small groups of variables. However, the DP tabular synthesis method is not adept at handling data with discrete values. They also face challenges in processing data with numerous attributes and large volumes~\cite{zhang2021privsyn}.

\vspace{1mm}
\noindent \textit{Generative model-based Synthesis.} For another direction, we refer to the generative model-based DP synthesizers, primarily focusing on training generative models such as Generative Adversarial Networks (GANs)~\cite{PATE-GAN, yin2022practical}, diffusion models~\cite{ddpm, dp-diffusion}, and large language models~\cite{Privcode} using DP Stochastic Gradient Descent (DP-SGD)~\cite{dpsgd}. PATE-GAN~\cite{PATE-GAN} and DP-CGAN~\cite{DPCGAN}, which apply DP-SGD to GANs, achieve effective synthesis on {\tt MNIST}~\cite{mnist} but perform less well on larger datasets. Recently, diffusion models have emerged as promising generative models surpassing GANs~\cite{dockhorn2023differentially}. DPDM~\cite{dockhorn2023differentially} suggests replacing GANs with diffusion models in DP-CGAN. Inspiration from the success of pre-training and fine-tuning across many challenging tasks~\cite{kumar2023offlinepre}, Sabra et al.~\cite{dp-diffusion} proposed pre-training DP-Diffusion, to first pre-train the diffusion models on a public dataset, and then fine-tune them on the sensitive dataset, achieving impressive outcomes. Based on pre-training DP-Diffusion, Li et al.~\cite{li2023meticulously} introduced Privimage, which reduces the pre-training dataset through semantic queries to resemble sensitive datasets, and achieves state-of-the-art performance.

\noindent \textbf{Offline RL with Synthetic Data.} Diffusion models are powerful tools for augmenting training datasets with synthetic data, widely used in computer vision~\cite{li2023meticulously,yuan2024realfake}. Therefore, the diffusion model is a natural data synthesizer for offline RL datasets, relieving the problematic data scarcity limitation~\cite{offline_survey}. Recent studies~\cite{yu2023scaling, chen2023genaug} have explored augmenting observations in robotic control with a text-guided diffusion model, while keeping the actions unchanged. Building on this, SynthER~\cite{lu2023synthetic} and MTDIFF~\cite{he2023diffusion} proposed using a diffusion model to generate transitions for trained tasks, capable of synthesizing novel actions. Zhao et al.\cite{zhao2024trajsyn} proposed using a transformer to synthesize trajectory datasets. Prior study~\cite{zhu2024trajsynmadiff} has explored dataset synthesis for offline multi-agent systems. Various studies present that diffusion models are vulnerable to MIAs in the image domain~\cite{2023whithmiadiffusion, carlini2023extracting}. This issue extends to the synthesis of offline RL datasets as well. Researchers in offline RL are increasingly focused on addressing privacy leaks in training datasets~\cite{du2023orl} and environmental information~\cite{pan2019you,gomrokchi2022membership}. Our work addresses this concern by proposing training diffusion models with DP guarantees.

\vspace{1mm}
\noindent \textbf{Broad Applications of Offline RL.} Offline RL systems have worked brilliantly on a wide range of real-world fields, including healthcare~\cite{RL4Treatment,RL4BGC}, energy management systems~\cite{RL4Energy,zhang2023mutual}, autonomous driving~\cite{RL4AutonomousVehicles,RL4AutonomousVehicles2}, and recommendation systems~\cite{RL4Recommender,RL4Recommender2}. In some cases, online RL is impractical. For example, experimenting with patients’ health in healthcare poses ethical and practical challenges. Mila et al.~\cite{RL4Treatment} used offline RL methods to develop a policy for recommending diabetes. Besides, Emerson et al.~\cite{RL4BGC} proposed using an offline RL agent to determine the optimal insulin dose.

Offline RL also benefits data utilization efficiency, particularly when data collection is costly~\cite{RL4Energy,RL4AutonomousVehicles}. In the energy industry, users have collected historical datasets and low-fidelity simulation data to train agents operating within safe constraints using the offline RL algorithm~\cite{RL4Energy}. Similarly, in autonomous driving, conducting experiments directly online can result in a large number of crashes and accidents, which are quite costly. Therefore, researchers collect diverse driving behaviors from multiple drivers and subsequently train the agent using offline RL~\cite{RL4AutonomousVehicles, RL4AutonomousVehicles2}. Well-trained offline RL agents can be fine-tuned in an online setting to save resources. Besides, for recommendation systems, Zhang et al.~\cite{RL4Recommender2} exploited offline RL to learn a recommendation policy for long-term user satisfaction. This work provides a tool to address the abuse and privacy concerns of offline RL datasets.

\end{document}